\documentclass[longauth]{aaEC}

\usepackage{graphicx}
\usepackage{natbib}
\usepackage{scalerel}
\usepackage{lastpage}
\usepackage[table]{xcolor}
\usepackage{comment}

\usepackage{txfonts}
\usepackage[pdfencoding=auto,psdextra]{hyperref}
\hypersetup{
    colorlinks=true,
    linkcolor=blue,
    filecolor=magenta,      
    urlcolor=blue,
    citecolor=blue
}
\makeatletter
\renewcommand*\aa@pageof{, page \thepage{} of \pageref*{LastPage}}
\makeatother

\usepackage[utf8]{inputenc}
\usepackage[switch, modulo]{lineno}

\usepackage{euclid}

\usepackage[nameinlink,capitalise]{cleveref}
\crefname{section}{Sect.}{Sects.}
\Crefname{section}{Section}{Sections}
\crefname{figure}{Fig.}{Figs.}
\Crefname{figure}{Figure}{Figures}
\crefname{equation}{Eq.}{Eqs.}
\Crefname{equation}{Equation}{Equations}
\crefname{table}{Table}{Tables}
\crefname{appendix}{Appendix}{Appendices}

\usepackage[nolist,nohyperlinks]{acronym}

\usepackage{orcidlink} 
\newcommand{\orcid}[1]{\orcidlink{#1}}

\begin{document} 

\title{Euclid Quick Data Release (Q1)}
\subtitle{NIR processing and data products}    


\acrodef{AA}{azimuth or $\alpha$ angle}
\acrodef{ADC}{analogue-to-digital converter}
\acrodef{ADU}{analogue-to-digital unit}
\acrodef{AOI}{angle of incidence}
\acrodef{ASIC}{application specific integrated circuit}
\acrodef{BFE}{brighter-fatter effect}
\acrodef{BPM}{bad-pixel mask}
\acrodef{CaLA}{camera-lens assembly}
\acrodef{CCD}{charge-coupled device}
\acrodef{CoLA}{corrector-lens assembly}
\acrodef{CDS}{Correlated Double Sampling}
\acrodef{CFC}{cryo-flex cable}
\acrodef{CFHT}{Canada-France-Hawaii Telescope}
\acrodef{CFRP}{carbon-fibre reinforced plastic}
\acrodef{CGH}{computer-generated hologram}
\acrodef{CME}{coronal mass ejection}
\acrodef{CNES}{Centre National d'Etude Spacial}
\acrodef{CPPM}{Centre de Physique des Particules de Marseille}
\acrodef{CPU}{central processing unit}
\acrodef{CR}{cosmic ray}
\acrodef{CTE}{coefficient of thermal expansion}
\acrodef{CME}{coronal mass ejection}
\acrodef{CTI}{charge-transfer inefficiency}
\acrodef{DCU}{Detector Control Unit}
\acrodef{DES}{Dark Energy Survey}
\acrodef{DNL}{differential nonlinearity}
\acrodef{DP}{Data Product}
\acrodef{DPU}{Data Processing Unit}
\acrodef{DQ}{data quality}
\acrodef{DQC}{data quality control}
\acrodef{DS}{Detector System}
\acrodef{EAS}{Euclid Archive System}
\acrodef{EDS}{Euclid Deep Survey}
\acrodef{EE}{encircled energy}
\acrodef{EPER}{extended pixel-edge response}
\acrodef{ESA}{European Space Agency}
\acrodef{ESP}{Emission of Solar Protons}
\acrodef{ECSS}{European Cooperation for Space Standardization}
\acrodef{EWS}{Euclid Wide Survey}
\acrodef{FDIR}{Fault Detection, Isolation and Recovery}
\acrodef{FGS}{fine guidance sensor}
\acrodef{FOM}{figure of merit}
\acrodef{FOV}{field of view}
\acrodef{FPA}{focal-plane array}
\acrodef{FPR}{false positive rate}
\acrodef{FWA}{filter-wheel assembly}
\acrodef{FWC}{full-well capacity}
\acrodef{FWHM}{full width at half maximum}
\acrodef{GOES}{Geostationary Operational Environmental Satellites}
\acrodef{GCR}{Galactic cosmic ray}
\acrodef{GWA}{grism-wheel assembly}
\acrodef{H2RG}{HAWAII-2RG}
\acrodef{HDU}{header data unit}
\acrodef{HST}{\textit{Hubble} Space Telescope}
\acrodef{IP2I}{Institut de Physique des 2 Infinis de Lyon}
\acrodef{IS}{inverse sensitivity}
\acrodef{JWST}{{\em James Webb} Space Telescope}
\acrodef{IAD}{ion-assisted deposition}
\acrodef{ICU}{Instrument Control Unit}
\acrodef{IPC}{inter-pixel capacitance}
\acrodef{ISES}{International Space Environmental Services}
\acrodef{JWST}{\textit{James Webb} Space Telescope}
\acrodef{LAM}{Laboratoire d'Astrophysique de Marseille}
\acrodef{LE1}[LE\,1]{Level~1 }
\acrodef{LED}{light-emitting diode}
\acrodef{LSB}{low surface brightness}
\acrodef{MACC}{multiple accumulated}
\acrodef{MAD}{median absolute deviation}
\acrodef{MDB}{Mission Database}
\acrodef{MEF}{multi-extension FITS}
\acrodef{MER PF}{MER processing function}
\acrodef{MLI}{multi-layer insulation}
\acrodef{MMU}{Mass Memory Unit}
\acrodef{MPE}{Max-Planck-Institut für extraterrestrische Physik}
\acrodef{MPIA}{Max-Planck-Institut für Astronomie}
\acrodef{NA}{numerical aperture}
\acrodef{NASA}{National Aeronautic and Space Administration}
\acrodef{NEP}{North Ecliptic Pole}
\acrodef{NIEL}{non-ionising energy loss}
\acrodef{NISP}{Near-Infrared Spectrometer and Photometer}
\acrodef{JPL}{NASA Jet Propulsion Laboratory}
\acrodef{MZ-CGH}{multi-zonal computer-generated hologram}
\acrodef{NI-CU}{NISP calibration unit}
\acrodef{NI-OA}{near-infrared optical assembly}
\acrodef{NI-GWA}{NISP Grism Wheel Assembly}
\acrodef{NIR}{near-infrared}
\acrodef{NIR PF}{near-infrared processing function}
\acrodef{NISP}{Near-Infrared Spectrometer and Photometer}
\acrodef{NOAA}{National Oceanic and Atmospheric Administration}
\acrodef{PA}{position angle}
\acrodef{PARMS}{plasma-assisted reactive magnetron sputtering}
\acrodef{PDF}{probability density function}
\acrodef{PE}{processing element}
\acrodef{PLM}{payload module}
\acrodef{PPO}{pipeline processing order}
\acrodef{PRNU}{pixel-response non-uniformity}
\acrodef{PTC}{photon transfer curve}
\acrodef{PTFE}{polytetrafluoroethylene}
\acrodef{PV}{performance verification}
\acrodef{PWM}{pulse-width modulation}
\acrodef{PSF}{point-spread function}
\acrodef{QE}{quantum efficiency}
\acrodef{QF}{quality factor}
\acrodef{RMS}{root mean square}
\acrodef{ROE}{readout electronic block unit}
\acrodef{ROI}{region of interest}
\acrodef{ROIC}{readout-integrated circuit}
\acrodef{ROS}{reference observing sequence}
\acrodef{SAA}{Solar aspect angle}
\acrodef{SCA}{sensor chip array}
\acrodef{SCE}{sensor chip electronic}
\acrodef{SCS}{sensor chip system}
\acrodef{SGS}{Science Ground Segment}
\acrodef{SGPS}{Solar and Galactic Proton Sensor}
\acrodef{SHS}{Shack-Hartmann sensor}
\acrodef{SNR}[S/N]{signal-to-noise ratio}
\acrodef{SED}{spectral energy distribution}
\acrodef{SiC}{silicon carbide}
\acrodef{SEP}{Solar energetic particle}
\acrodef{SSN}{Sunspot number}
\acrodef{STIX}{Spectrometer/Telescope for Imaging X-rays}
\acrodef{SolO}{Solar Orbiter}
\acrodef{SUTR}{sample up-the-ramp}
\acrodef{TP}{trap pumping}
\acrodef{TPR}{true positive rate}
\acrodef{UTR}{up-the-ramp}
\acrodef{SVM}{service module}
\acrodef{VIS}{visible imager}
\acrodef{VIS-PF}{VIS processing function}
\acrodef{WD}{white dwarf}
\acrodef{WCS}{world coordinate system}
\acrodef{WFE}{wavefront error}
\acrodef{ZP}{zero point}


\author{Euclid Collaboration: G.~Polenta\orcid{0000-0003-4067-9196}\thanks{\email{gianluca.polenta@asi.it}}\inst{\ref{aff1}}
\and M.~Frailis\orcid{0000-0002-7400-2135}\inst{\ref{aff2}}
\and A.~Alavi\orcid{0000-0002-8630-6435}\inst{\ref{aff3}}
\and P.~N.~Appleton\orcid{0000-0002-7607-8766}\inst{\ref{aff3},\ref{aff4}}
\and P.~Awad\orcid{0000-0002-0428-849X}\inst{\ref{aff5}}
\and A.~Bonchi\orcid{0000-0002-2667-5482}\inst{\ref{aff1}}
\and R.~Bouwens\orcid{0000-0002-4989-2471}\inst{\ref{aff5}}
\and L.~Bramante\inst{\ref{aff6}}
\and D.~Busonero\orcid{0000-0002-3903-7076}\inst{\ref{aff7}}
\and G.~Calderone\orcid{0000-0002-7738-5389}\inst{\ref{aff2}}
\and F.~Cogato\orcid{0000-0003-4632-6113}\inst{\ref{aff8},\ref{aff9}}
\and S.~Conseil\orcid{0000-0002-3657-4191}\inst{\ref{aff10}}
\and M.~Correnti\orcid{0000-0001-6464-3257}\inst{\ref{aff11},\ref{aff1}}
\and R.~da~Silva\orcid{0000-0003-4788-677X}\inst{\ref{aff11},\ref{aff1}}
\and I.~Das\orcid{0009-0007-7088-2044}\inst{\ref{aff3}}
\and F.~Faustini\orcid{0000-0001-6274-5145}\inst{\ref{aff11},\ref{aff1}}
\and Y.~Fu\orcid{0000-0002-0759-0504}\inst{\ref{aff5},\ref{aff12}}
\and T.~Gasparetto\orcid{0000-0002-7913-4866}\inst{\ref{aff2}}
\and W.~Gillard\orcid{0000-0003-4744-9748}\inst{\ref{aff13}}
\and A.~Grazian\orcid{0000-0002-5688-0663}\inst{\ref{aff14}}
\and S.~Hemmati\orcid{0000-0003-2226-5395}\inst{\ref{aff3}}
\and J.~Jacobson\inst{\ref{aff3}}
\and K.~Jahnke\orcid{0000-0003-3804-2137}\inst{\ref{aff15}}
\and B.~Kubik\orcid{0009-0006-5823-4880}\inst{\ref{aff10}}
\and X.~Liu\inst{\ref{aff16}}
\and C.~Macabiau\inst{\ref{aff10}}
\and E.~Medinaceli\orcid{0000-0002-4040-7783}\inst{\ref{aff9}}
\and P.~W.~Morris\orcid{0000-0002-5186-4381}\inst{\ref{aff16}}
\and K.~Paterson\orcid{0000-0001-8340-3486}\inst{\ref{aff15}}
\and M.~Radovich\orcid{0000-0002-3585-866X}\inst{\ref{aff14}}
\and M.~Schirmer\orcid{0000-0003-2568-9994}\inst{\ref{aff15}}
\and A.~Shulevski\orcid{0000-0002-1827-0469}\inst{\ref{aff17},\ref{aff12},\ref{aff18},\ref{aff19}}
\and H.~I.~Teplitz\orcid{0000-0002-7064-5424}\inst{\ref{aff4}}
\and B.~Venemans\orcid{0000-0001-9024-8322}\inst{\ref{aff5}}
\and N.~Aghanim\orcid{0000-0002-6688-8992}\inst{\ref{aff20}}
\and B.~Altieri\orcid{0000-0003-3936-0284}\inst{\ref{aff21}}
\and A.~Amara\inst{\ref{aff22}}
\and S.~Andreon\orcid{0000-0002-2041-8784}\inst{\ref{aff23}}
\and N.~Auricchio\orcid{0000-0003-4444-8651}\inst{\ref{aff9}}
\and H.~Aussel\orcid{0000-0002-1371-5705}\inst{\ref{aff24}}
\and C.~Baccigalupi\orcid{0000-0002-8211-1630}\inst{\ref{aff25},\ref{aff2},\ref{aff26},\ref{aff27}}
\and M.~Baldi\orcid{0000-0003-4145-1943}\inst{\ref{aff28},\ref{aff9},\ref{aff29}}
\and A.~Balestra\orcid{0000-0002-6967-261X}\inst{\ref{aff14}}
\and S.~Bardelli\orcid{0000-0002-8900-0298}\inst{\ref{aff9}}
\and A.~Basset\inst{\ref{aff30}}
\and P.~Battaglia\orcid{0000-0002-7337-5909}\inst{\ref{aff9}}
\and A.~N.~Belikov\inst{\ref{aff12},\ref{aff31}}
\and R.~Bender\orcid{0000-0001-7179-0626}\inst{\ref{aff32},\ref{aff33}}
\and A.~Biviano\orcid{0000-0002-0857-0732}\inst{\ref{aff2},\ref{aff25}}
\and E.~Branchini\orcid{0000-0002-0808-6908}\inst{\ref{aff34},\ref{aff35},\ref{aff23}}
\and M.~Brescia\orcid{0000-0001-9506-5680}\inst{\ref{aff36},\ref{aff37}}
\and J.~Brinchmann\orcid{0000-0003-4359-8797}\inst{\ref{aff38},\ref{aff39}}
\and A.~Caillat\inst{\ref{aff40}}
\and S.~Camera\orcid{0000-0003-3399-3574}\inst{\ref{aff41},\ref{aff42},\ref{aff7}}
\and G.~Ca\~nas-Herrera\orcid{0000-0003-2796-2149}\inst{\ref{aff43},\ref{aff44},\ref{aff5}}
\and V.~Capobianco\orcid{0000-0002-3309-7692}\inst{\ref{aff7}}
\and C.~Carbone\orcid{0000-0003-0125-3563}\inst{\ref{aff45}}
\and J.~Carretero\orcid{0000-0002-3130-0204}\inst{\ref{aff46},\ref{aff47}}
\and S.~Casas\orcid{0000-0002-4751-5138}\inst{\ref{aff48}}
\and F.~J.~Castander\orcid{0000-0001-7316-4573}\inst{\ref{aff49},\ref{aff50}}
\and M.~Castellano\orcid{0000-0001-9875-8263}\inst{\ref{aff11}}
\and G.~Castignani\orcid{0000-0001-6831-0687}\inst{\ref{aff9}}
\and S.~Cavuoti\orcid{0000-0002-3787-4196}\inst{\ref{aff37},\ref{aff51}}
\and K.~C.~Chambers\orcid{0000-0001-6965-7789}\inst{\ref{aff52}}
\and A.~Cimatti\inst{\ref{aff53}}
\and C.~Colodro-Conde\inst{\ref{aff54}}
\and G.~Congedo\orcid{0000-0003-2508-0046}\inst{\ref{aff55}}
\and C.~J.~Conselice\orcid{0000-0003-1949-7638}\inst{\ref{aff56}}
\and L.~Conversi\orcid{0000-0002-6710-8476}\inst{\ref{aff57},\ref{aff21}}
\and Y.~Copin\orcid{0000-0002-5317-7518}\inst{\ref{aff10}}
\and L.~Corcione\orcid{0000-0002-6497-5881}\inst{\ref{aff7}}
\and A.~Costille\inst{\ref{aff40}}
\and F.~Courbin\orcid{0000-0003-0758-6510}\inst{\ref{aff58},\ref{aff59}}
\and H.~M.~Courtois\orcid{0000-0003-0509-1776}\inst{\ref{aff60}}
\and A.~Da~Silva\orcid{0000-0002-6385-1609}\inst{\ref{aff61},\ref{aff62}}
\and H.~Degaudenzi\orcid{0000-0002-5887-6799}\inst{\ref{aff63}}
\and G.~De~Lucia\orcid{0000-0002-6220-9104}\inst{\ref{aff2}}
\and A.~M.~Di~Giorgio\orcid{0000-0002-4767-2360}\inst{\ref{aff64}}
\and H.~Dole\orcid{0000-0002-9767-3839}\inst{\ref{aff20}}
\and F.~Dubath\orcid{0000-0002-6533-2810}\inst{\ref{aff63}}
\and C.~A.~J.~Duncan\orcid{0009-0003-3573-0791}\inst{\ref{aff55},\ref{aff56}}
\and X.~Dupac\inst{\ref{aff21}}
\and S.~Dusini\orcid{0000-0002-1128-0664}\inst{\ref{aff65}}
\and A.~Ealet\orcid{0000-0003-3070-014X}\inst{\ref{aff10}}
\and S.~Escoffier\orcid{0000-0002-2847-7498}\inst{\ref{aff13}}
\and M.~Fabricius\orcid{0000-0002-7025-6058}\inst{\ref{aff32},\ref{aff33}}
\and M.~Farina\orcid{0000-0002-3089-7846}\inst{\ref{aff64}}
\and R.~Farinelli\inst{\ref{aff9}}
\and S.~Ferriol\inst{\ref{aff10}}
\and F.~Finelli\orcid{0000-0002-6694-3269}\inst{\ref{aff9},\ref{aff66}}
\and S.~Fotopoulou\orcid{0000-0002-9686-254X}\inst{\ref{aff67}}
\and N.~Fourmanoit\orcid{0009-0005-6816-6925}\inst{\ref{aff13}}
\and E.~Franceschi\orcid{0000-0002-0585-6591}\inst{\ref{aff9}}
\and M.~Fumana\orcid{0000-0001-6787-5950}\inst{\ref{aff45}}
\and S.~Galeotta\orcid{0000-0002-3748-5115}\inst{\ref{aff2}}
\and K.~George\orcid{0000-0002-1734-8455}\inst{\ref{aff33}}
\and B.~Gillis\orcid{0000-0002-4478-1270}\inst{\ref{aff55}}
\and C.~Giocoli\orcid{0000-0002-9590-7961}\inst{\ref{aff9},\ref{aff29}}
\and P.~G\'omez-Alvarez\orcid{0000-0002-8594-5358}\inst{\ref{aff68},\ref{aff21}}
\and J.~Gracia-Carpio\inst{\ref{aff32}}
\and B.~R.~Granett\orcid{0000-0003-2694-9284}\inst{\ref{aff23}}
\and F.~Grupp\inst{\ref{aff32},\ref{aff33}}
\and S.~V.~H.~Haugan\orcid{0000-0001-9648-7260}\inst{\ref{aff69}}
\and J.~Hoar\inst{\ref{aff21}}
\and H.~Hoekstra\orcid{0000-0002-0641-3231}\inst{\ref{aff5}}
\and W.~Holmes\inst{\ref{aff70}}
\and I.~M.~Hook\orcid{0000-0002-2960-978X}\inst{\ref{aff71}}
\and F.~Hormuth\inst{\ref{aff72}}
\and A.~Hornstrup\orcid{0000-0002-3363-0936}\inst{\ref{aff73},\ref{aff74}}
\and P.~Hudelot\inst{\ref{aff75}}
\and M.~Jhabvala\inst{\ref{aff76}}
\and E.~Keih\"anen\orcid{0000-0003-1804-7715}\inst{\ref{aff77}}
\and S.~Kermiche\orcid{0000-0002-0302-5735}\inst{\ref{aff13}}
\and A.~Kiessling\orcid{0000-0002-2590-1273}\inst{\ref{aff70}}
\and K.~Kuijken\orcid{0000-0002-3827-0175}\inst{\ref{aff5}}
\and M.~K\"ummel\orcid{0000-0003-2791-2117}\inst{\ref{aff33}}
\and M.~Kunz\orcid{0000-0002-3052-7394}\inst{\ref{aff78}}
\and H.~Kurki-Suonio\orcid{0000-0002-4618-3063}\inst{\ref{aff79},\ref{aff80}}
\and Q.~Le~Boulc'h\inst{\ref{aff81}}
\and A.~M.~C.~Le~Brun\orcid{0000-0002-0936-4594}\inst{\ref{aff82}}
\and D.~Le~Mignant\orcid{0000-0002-5339-5515}\inst{\ref{aff40}}
\and P.~Liebing\inst{\ref{aff83}}
\and S.~Ligori\orcid{0000-0003-4172-4606}\inst{\ref{aff7}}
\and P.~B.~Lilje\orcid{0000-0003-4324-7794}\inst{\ref{aff69}}
\and V.~Lindholm\orcid{0000-0003-2317-5471}\inst{\ref{aff79},\ref{aff80}}
\and I.~Lloro\orcid{0000-0001-5966-1434}\inst{\ref{aff84}}
\and G.~Mainetti\orcid{0000-0003-2384-2377}\inst{\ref{aff81}}
\and D.~Maino\inst{\ref{aff85},\ref{aff45},\ref{aff86}}
\and E.~Maiorano\orcid{0000-0003-2593-4355}\inst{\ref{aff9}}
\and O.~Mansutti\orcid{0000-0001-5758-4658}\inst{\ref{aff2}}
\and S.~Marcin\inst{\ref{aff87}}
\and O.~Marggraf\orcid{0000-0001-7242-3852}\inst{\ref{aff88}}
\and M.~Martinelli\orcid{0000-0002-6943-7732}\inst{\ref{aff11},\ref{aff89}}
\and N.~Martinet\orcid{0000-0003-2786-7790}\inst{\ref{aff40}}
\and F.~Marulli\orcid{0000-0002-8850-0303}\inst{\ref{aff8},\ref{aff9},\ref{aff29}}
\and R.~Massey\orcid{0000-0002-6085-3780}\inst{\ref{aff90}}
\and S.~Maurogordato\inst{\ref{aff91}}
\and H.~J.~McCracken\orcid{0000-0002-9489-7765}\inst{\ref{aff75}}
\and S.~Mei\orcid{0000-0002-2849-559X}\inst{\ref{aff92},\ref{aff93}}
\and M.~Melchior\inst{\ref{aff87}}
\and Y.~Mellier\inst{\ref{aff94},\ref{aff75}}
\and M.~Meneghetti\orcid{0000-0003-1225-7084}\inst{\ref{aff9},\ref{aff29}}
\and E.~Merlin\orcid{0000-0001-6870-8900}\inst{\ref{aff11}}
\and G.~Meylan\inst{\ref{aff95}}
\and A.~Mora\orcid{0000-0002-1922-8529}\inst{\ref{aff96}}
\and M.~Moresco\orcid{0000-0002-7616-7136}\inst{\ref{aff8},\ref{aff9}}
\and L.~Moscardini\orcid{0000-0002-3473-6716}\inst{\ref{aff8},\ref{aff9},\ref{aff29}}
\and R.~Nakajima\orcid{0009-0009-1213-7040}\inst{\ref{aff88}}
\and C.~Neissner\orcid{0000-0001-8524-4968}\inst{\ref{aff97},\ref{aff47}}
\and R.~C.~Nichol\orcid{0000-0003-0939-6518}\inst{\ref{aff22}}
\and S.-M.~Niemi\inst{\ref{aff43}}
\and J.~W.~Nightingale\orcid{0000-0002-8987-7401}\inst{\ref{aff98}}
\and C.~Padilla\orcid{0000-0001-7951-0166}\inst{\ref{aff97}}
\and S.~Paltani\orcid{0000-0002-8108-9179}\inst{\ref{aff63}}
\and F.~Pasian\orcid{0000-0002-4869-3227}\inst{\ref{aff2}}
\and K.~Pedersen\inst{\ref{aff99}}
\and W.~J.~Percival\orcid{0000-0002-0644-5727}\inst{\ref{aff100},\ref{aff101},\ref{aff102}}
\and V.~Pettorino\inst{\ref{aff43}}
\and S.~Pires\orcid{0000-0002-0249-2104}\inst{\ref{aff24}}
\and M.~Poncet\inst{\ref{aff30}}
\and L.~A.~Popa\inst{\ref{aff103}}
\and L.~Pozzetti\orcid{0000-0001-7085-0412}\inst{\ref{aff9}}
\and G.~D.~Racca\inst{\ref{aff43},\ref{aff5}}
\and F.~Raison\orcid{0000-0002-7819-6918}\inst{\ref{aff32}}
\and R.~Rebolo\orcid{0000-0003-3767-7085}\inst{\ref{aff54},\ref{aff104},\ref{aff105}}
\and A.~Renzi\orcid{0000-0001-9856-1970}\inst{\ref{aff106},\ref{aff65}}
\and J.~Rhodes\orcid{0000-0002-4485-8549}\inst{\ref{aff70}}
\and G.~Riccio\inst{\ref{aff37}}
\and E.~Romelli\orcid{0000-0003-3069-9222}\inst{\ref{aff2}}
\and M.~Roncarelli\orcid{0000-0001-9587-7822}\inst{\ref{aff9}}
\and E.~Rossetti\orcid{0000-0003-0238-4047}\inst{\ref{aff28}}
\and B.~Rusholme\orcid{0000-0001-7648-4142}\inst{\ref{aff3}}
\and R.~Saglia\orcid{0000-0003-0378-7032}\inst{\ref{aff33},\ref{aff32}}
\and Z.~Sakr\orcid{0000-0002-4823-3757}\inst{\ref{aff107},\ref{aff108},\ref{aff109}}
\and A.~G.~S\'anchez\orcid{0000-0003-1198-831X}\inst{\ref{aff32}}
\and D.~Sapone\orcid{0000-0001-7089-4503}\inst{\ref{aff110}}
\and B.~Sartoris\orcid{0000-0003-1337-5269}\inst{\ref{aff33},\ref{aff2}}
\and J.~A.~Schewtschenko\orcid{0000-0002-4913-6393}\inst{\ref{aff55}}
\and P.~Schneider\orcid{0000-0001-8561-2679}\inst{\ref{aff88}}
\and M.~Scodeggio\inst{\ref{aff45}}
\and A.~Secroun\orcid{0000-0003-0505-3710}\inst{\ref{aff13}}
\and E.~Sefusatti\orcid{0000-0003-0473-1567}\inst{\ref{aff2},\ref{aff25},\ref{aff26}}
\and G.~Seidel\orcid{0000-0003-2907-353X}\inst{\ref{aff15}}
\and M.~Seiffert\orcid{0000-0002-7536-9393}\inst{\ref{aff70}}
\and S.~Serrano\orcid{0000-0002-0211-2861}\inst{\ref{aff50},\ref{aff111},\ref{aff49}}
\and P.~Simon\inst{\ref{aff88}}
\and C.~Sirignano\orcid{0000-0002-0995-7146}\inst{\ref{aff106},\ref{aff65}}
\and G.~Sirri\orcid{0000-0003-2626-2853}\inst{\ref{aff29}}
\and A.~Spurio~Mancini\orcid{0000-0001-5698-0990}\inst{\ref{aff112}}
\and L.~Stanco\orcid{0000-0002-9706-5104}\inst{\ref{aff65}}
\and J.~Steinwagner\orcid{0000-0001-7443-1047}\inst{\ref{aff32}}
\and P.~Tallada-Cresp\'{i}\orcid{0000-0002-1336-8328}\inst{\ref{aff46},\ref{aff47}}
\and D.~Tavagnacco\orcid{0000-0001-7475-9894}\inst{\ref{aff2}}
\and A.~N.~Taylor\inst{\ref{aff55}}
\and I.~Tereno\inst{\ref{aff61},\ref{aff113}}
\and N.~Tessore\orcid{0000-0002-9696-7931}\inst{\ref{aff114}}
\and S.~Toft\orcid{0000-0003-3631-7176}\inst{\ref{aff115},\ref{aff116}}
\and R.~Toledo-Moreo\orcid{0000-0002-2997-4859}\inst{\ref{aff117}}
\and F.~Torradeflot\orcid{0000-0003-1160-1517}\inst{\ref{aff47},\ref{aff46}}
\and I.~Tutusaus\orcid{0000-0002-3199-0399}\inst{\ref{aff108}}
\and E.~A.~Valentijn\inst{\ref{aff12}}
\and L.~Valenziano\orcid{0000-0002-1170-0104}\inst{\ref{aff9},\ref{aff66}}
\and J.~Valiviita\orcid{0000-0001-6225-3693}\inst{\ref{aff79},\ref{aff80}}
\and T.~Vassallo\orcid{0000-0001-6512-6358}\inst{\ref{aff33},\ref{aff2}}
\and G.~Verdoes~Kleijn\orcid{0000-0001-5803-2580}\inst{\ref{aff12}}
\and A.~Veropalumbo\orcid{0000-0003-2387-1194}\inst{\ref{aff23},\ref{aff35},\ref{aff34}}
\and Y.~Wang\orcid{0000-0002-4749-2984}\inst{\ref{aff4}}
\and J.~Weller\orcid{0000-0002-8282-2010}\inst{\ref{aff33},\ref{aff32}}
\and A.~Zacchei\orcid{0000-0003-0396-1192}\inst{\ref{aff2},\ref{aff25}}
\and G.~Zamorani\orcid{0000-0002-2318-301X}\inst{\ref{aff9}}
\and F.~M.~Zerbi\inst{\ref{aff23}}
\and I.~A.~Zinchenko\orcid{0000-0002-2944-2449}\inst{\ref{aff33}}
\and E.~Zucca\orcid{0000-0002-5845-8132}\inst{\ref{aff9}}
\and V.~Allevato\orcid{0000-0001-7232-5152}\inst{\ref{aff37}}
\and M.~Ballardini\orcid{0000-0003-4481-3559}\inst{\ref{aff118},\ref{aff119},\ref{aff9}}
\and M.~Bolzonella\orcid{0000-0003-3278-4607}\inst{\ref{aff9}}
\and E.~Bozzo\orcid{0000-0002-8201-1525}\inst{\ref{aff63}}
\and C.~Burigana\orcid{0000-0002-3005-5796}\inst{\ref{aff120},\ref{aff66}}
\and R.~Cabanac\orcid{0000-0001-6679-2600}\inst{\ref{aff108}}
\and M.~Calabrese\orcid{0000-0002-2637-2422}\inst{\ref{aff121},\ref{aff45}}
\and A.~Cappi\inst{\ref{aff9},\ref{aff91}}
\and P.~Casenove\orcid{0009-0006-6736-1670}\inst{\ref{aff30}}
\and D.~Di~Ferdinando\inst{\ref{aff29}}
\and J.~A.~Escartin~Vigo\inst{\ref{aff32}}
\and G.~Fabbian\orcid{0000-0002-3255-4695}\inst{\ref{aff122},\ref{aff123}}
\and L.~Gabarra\orcid{0000-0002-8486-8856}\inst{\ref{aff124}}
\and M.~Huertas-Company\orcid{0000-0002-1416-8483}\inst{\ref{aff54},\ref{aff125},\ref{aff126},\ref{aff127}}
\and J.~Mart\'{i}n-Fleitas\orcid{0000-0002-8594-569X}\inst{\ref{aff96}}
\and S.~Matthew\orcid{0000-0001-8448-1697}\inst{\ref{aff55}}
\and N.~Mauri\orcid{0000-0001-8196-1548}\inst{\ref{aff53},\ref{aff29}}
\and A.~A.~Nucita\inst{\ref{aff128},\ref{aff129},\ref{aff130}}
\and A.~Pezzotta\orcid{0000-0003-0726-2268}\inst{\ref{aff131},\ref{aff32}}
\and M.~P\"ontinen\orcid{0000-0001-5442-2530}\inst{\ref{aff79}}
\and C.~Porciani\orcid{0000-0002-7797-2508}\inst{\ref{aff88}}
\and I.~Risso\orcid{0000-0003-2525-7761}\inst{\ref{aff132}}
\and V.~Scottez\inst{\ref{aff94},\ref{aff133}}
\and M.~Sereno\orcid{0000-0003-0302-0325}\inst{\ref{aff9},\ref{aff29}}
\and M.~Tenti\orcid{0000-0002-4254-5901}\inst{\ref{aff29}}
\and M.~Viel\orcid{0000-0002-2642-5707}\inst{\ref{aff25},\ref{aff2},\ref{aff27},\ref{aff26},\ref{aff134}}
\and M.~Wiesmann\orcid{0009-0000-8199-5860}\inst{\ref{aff69}}
\and Y.~Akrami\orcid{0000-0002-2407-7956}\inst{\ref{aff135},\ref{aff136}}
\and I.~T.~Andika\orcid{0000-0001-6102-9526}\inst{\ref{aff137},\ref{aff138}}
\and S.~Anselmi\orcid{0000-0002-3579-9583}\inst{\ref{aff65},\ref{aff106},\ref{aff139}}
\and M.~Archidiacono\orcid{0000-0003-4952-9012}\inst{\ref{aff85},\ref{aff86}}
\and F.~Atrio-Barandela\orcid{0000-0002-2130-2513}\inst{\ref{aff140}}
\and C.~Benoist\inst{\ref{aff91}}
\and P.~Bergamini\orcid{0000-0003-1383-9414}\inst{\ref{aff85},\ref{aff9}}
\and D.~Bertacca\orcid{0000-0002-2490-7139}\inst{\ref{aff106},\ref{aff14},\ref{aff65}}
\and M.~Bethermin\orcid{0000-0002-3915-2015}\inst{\ref{aff141}}
\and L.~Bisigello\orcid{0000-0003-0492-4924}\inst{\ref{aff14}}
\and A.~Blanchard\orcid{0000-0001-8555-9003}\inst{\ref{aff108}}
\and L.~Blot\orcid{0000-0002-9622-7167}\inst{\ref{aff142},\ref{aff139}}
\and S.~Borgani\orcid{0000-0001-6151-6439}\inst{\ref{aff143},\ref{aff25},\ref{aff2},\ref{aff26},\ref{aff134}}
\and A.~S.~Borlaff\orcid{0000-0003-3249-4431}\inst{\ref{aff144},\ref{aff145}}
\and M.~L.~Brown\orcid{0000-0002-0370-8077}\inst{\ref{aff56}}
\and S.~Bruton\orcid{0000-0002-6503-5218}\inst{\ref{aff16}}
\and A.~Calabro\orcid{0000-0003-2536-1614}\inst{\ref{aff11}}
\and B.~Camacho~Quevedo\orcid{0000-0002-8789-4232}\inst{\ref{aff50},\ref{aff49}}
\and F.~Caro\inst{\ref{aff11}}
\and C.~S.~Carvalho\inst{\ref{aff113}}
\and T.~Castro\orcid{0000-0002-6292-3228}\inst{\ref{aff2},\ref{aff26},\ref{aff25},\ref{aff134}}
\and Y.~Charles\inst{\ref{aff40}}
\and R.~Chary\orcid{0000-0001-7583-0621}\inst{\ref{aff4},\ref{aff146}}
\and A.~R.~Cooray\orcid{0000-0002-3892-0190}\inst{\ref{aff147}}
\and O.~Cucciati\orcid{0000-0002-9336-7551}\inst{\ref{aff9}}
\and S.~Davini\orcid{0000-0003-3269-1718}\inst{\ref{aff35}}
\and F.~De~Paolis\orcid{0000-0001-6460-7563}\inst{\ref{aff128},\ref{aff129},\ref{aff130}}
\and G.~Desprez\orcid{0000-0001-8325-1742}\inst{\ref{aff12}}
\and A.~D\'iaz-S\'anchez\orcid{0000-0003-0748-4768}\inst{\ref{aff148}}
\and J.~J.~Diaz\inst{\ref{aff125}}
\and S.~Di~Domizio\orcid{0000-0003-2863-5895}\inst{\ref{aff34},\ref{aff35}}
\and J.~M.~Diego\orcid{0000-0001-9065-3926}\inst{\ref{aff149}}
\and P.-A.~Duc\orcid{0000-0003-3343-6284}\inst{\ref{aff141}}
\and A.~Enia\orcid{0000-0002-0200-2857}\inst{\ref{aff28},\ref{aff9}}
\and Y.~Fang\inst{\ref{aff33}}
\and A.~M.~N.~Ferguson\inst{\ref{aff55}}
\and A.~G.~Ferrari\orcid{0009-0005-5266-4110}\inst{\ref{aff29}}
\and P.~G.~Ferreira\orcid{0000-0002-3021-2851}\inst{\ref{aff124}}
\and A.~Finoguenov\orcid{0000-0002-4606-5403}\inst{\ref{aff79}}
\and A.~Fontana\orcid{0000-0003-3820-2823}\inst{\ref{aff11}}
\and A.~Franco\orcid{0000-0002-4761-366X}\inst{\ref{aff129},\ref{aff128},\ref{aff130}}
\and K.~Ganga\orcid{0000-0001-8159-8208}\inst{\ref{aff92}}
\and J.~Garc\'ia-Bellido\orcid{0000-0002-9370-8360}\inst{\ref{aff135}}
\and V.~Gautard\inst{\ref{aff150}}
\and E.~Gaztanaga\orcid{0000-0001-9632-0815}\inst{\ref{aff49},\ref{aff50},\ref{aff151}}
\and F.~Giacomini\orcid{0000-0002-3129-2814}\inst{\ref{aff29}}
\and F.~Gianotti\orcid{0000-0003-4666-119X}\inst{\ref{aff9}}
\and A.~H.~Gonzalez\orcid{0000-0002-0933-8601}\inst{\ref{aff152}}
\and G.~Gozaliasl\orcid{0000-0002-0236-919X}\inst{\ref{aff153},\ref{aff79}}
\and A.~Gregorio\orcid{0000-0003-4028-8785}\inst{\ref{aff143},\ref{aff2},\ref{aff26}}
\and A.~Gruppuso\orcid{0000-0001-9272-5292}\inst{\ref{aff9},\ref{aff29}}
\and M.~Guidi\orcid{0000-0001-9408-1101}\inst{\ref{aff28},\ref{aff9}}
\and C.~M.~Gutierrez\orcid{0000-0001-7854-783X}\inst{\ref{aff154}}
\and A.~Hall\orcid{0000-0002-3139-8651}\inst{\ref{aff55}}
\and W.~G.~Hartley\inst{\ref{aff63}}
\and C.~Hern\'andez-Monteagudo\orcid{0000-0001-5471-9166}\inst{\ref{aff105},\ref{aff54}}
\and H.~Hildebrandt\orcid{0000-0002-9814-3338}\inst{\ref{aff155}}
\and J.~Hjorth\orcid{0000-0002-4571-2306}\inst{\ref{aff99}}
\and J.~J.~E.~Kajava\orcid{0000-0002-3010-8333}\inst{\ref{aff156},\ref{aff157}}
\and Y.~Kang\orcid{0009-0000-8588-7250}\inst{\ref{aff63}}
\and V.~Kansal\orcid{0000-0002-4008-6078}\inst{\ref{aff158},\ref{aff159}}
\and D.~Karagiannis\orcid{0000-0002-4927-0816}\inst{\ref{aff118},\ref{aff160}}
\and K.~Kiiveri\inst{\ref{aff77}}
\and C.~C.~Kirkpatrick\inst{\ref{aff77}}
\and S.~Kruk\orcid{0000-0001-8010-8879}\inst{\ref{aff21}}
\and V.~Le~Brun\orcid{0000-0002-5027-1939}\inst{\ref{aff40}}
\and J.~Le~Graet\orcid{0000-0001-6523-7971}\inst{\ref{aff13}}
\and L.~Legrand\orcid{0000-0003-0610-5252}\inst{\ref{aff161},\ref{aff123}}
\and M.~Lembo\orcid{0000-0002-5271-5070}\inst{\ref{aff118},\ref{aff119}}
\and F.~Lepori\orcid{0009-0000-5061-7138}\inst{\ref{aff162}}
\and G.~Leroy\orcid{0009-0004-2523-4425}\inst{\ref{aff163},\ref{aff90}}
\and G.~F.~Lesci\orcid{0000-0002-4607-2830}\inst{\ref{aff8},\ref{aff9}}
\and J.~Lesgourgues\orcid{0000-0001-7627-353X}\inst{\ref{aff48}}
\and L.~Leuzzi\orcid{0009-0006-4479-7017}\inst{\ref{aff8},\ref{aff9}}
\and T.~I.~Liaudat\orcid{0000-0002-9104-314X}\inst{\ref{aff164}}
\and S.~J.~Liu\orcid{0000-0001-7680-2139}\inst{\ref{aff64}}
\and A.~Loureiro\orcid{0000-0002-4371-0876}\inst{\ref{aff165},\ref{aff166}}
\and J.~Macias-Perez\orcid{0000-0002-5385-2763}\inst{\ref{aff167}}
\and G.~Maggio\orcid{0000-0003-4020-4836}\inst{\ref{aff2}}
\and M.~Magliocchetti\orcid{0000-0001-9158-4838}\inst{\ref{aff64}}
\and F.~Mannucci\orcid{0000-0002-4803-2381}\inst{\ref{aff168}}
\and R.~Maoli\orcid{0000-0002-6065-3025}\inst{\ref{aff169},\ref{aff11}}
\and C.~J.~A.~P.~Martins\orcid{0000-0002-4886-9261}\inst{\ref{aff170},\ref{aff38}}
\and L.~Maurin\orcid{0000-0002-8406-0857}\inst{\ref{aff20}}
\and C.~J.~R.~McPartland\orcid{0000-0003-0639-025X}\inst{\ref{aff74},\ref{aff116}}
\and M.~Miluzio\inst{\ref{aff21},\ref{aff171}}
\and P.~Monaco\orcid{0000-0003-2083-7564}\inst{\ref{aff143},\ref{aff2},\ref{aff26},\ref{aff25}}
\and A.~Montoro\orcid{0000-0003-4730-8590}\inst{\ref{aff49},\ref{aff50}}
\and C.~Moretti\orcid{0000-0003-3314-8936}\inst{\ref{aff27},\ref{aff134},\ref{aff2},\ref{aff25},\ref{aff26}}
\and G.~Morgante\inst{\ref{aff9}}
\and S.~Nadathur\orcid{0000-0001-9070-3102}\inst{\ref{aff151}}
\and K.~Naidoo\orcid{0000-0002-9182-1802}\inst{\ref{aff151}}
\and P.~Natoli\orcid{0000-0003-0126-9100}\inst{\ref{aff118},\ref{aff119}}
\and A.~Navarro-Alsina\orcid{0000-0002-3173-2592}\inst{\ref{aff88}}
\and S.~Nesseris\orcid{0000-0002-0567-0324}\inst{\ref{aff135}}
\and L.~Nicastro\orcid{0000-0001-8534-6788}\inst{\ref{aff9}}
\and F.~Passalacqua\orcid{0000-0002-8606-4093}\inst{\ref{aff106},\ref{aff65}}
\and L.~Patrizii\inst{\ref{aff29}}
\and A.~Pisani\orcid{0000-0002-6146-4437}\inst{\ref{aff13},\ref{aff172}}
\and D.~Potter\orcid{0000-0002-0757-5195}\inst{\ref{aff162}}
\and S.~Quai\orcid{0000-0002-0449-8163}\inst{\ref{aff8},\ref{aff9}}
\and P.-F.~Rocci\inst{\ref{aff20}}
\and G.~Rodighiero\orcid{0000-0002-9415-2296}\inst{\ref{aff106},\ref{aff14}}
\and S.~Sacquegna\orcid{0000-0002-8433-6630}\inst{\ref{aff128},\ref{aff129},\ref{aff130}}
\and M.~Sahl\'en\orcid{0000-0003-0973-4804}\inst{\ref{aff173}}
\and D.~B.~Sanders\orcid{0000-0002-1233-9998}\inst{\ref{aff52}}
\and E.~Sarpa\orcid{0000-0002-1256-655X}\inst{\ref{aff27},\ref{aff134},\ref{aff26}}
\and C.~Scarlata\orcid{0000-0002-9136-8876}\inst{\ref{aff174}}
\and A.~Schneider\orcid{0000-0001-7055-8104}\inst{\ref{aff162}}
\and D.~Sciotti\orcid{0009-0008-4519-2620}\inst{\ref{aff11},\ref{aff89}}
\and E.~Sellentin\inst{\ref{aff175},\ref{aff5}}
\and F.~Shankar\orcid{0000-0001-8973-5051}\inst{\ref{aff176}}
\and L.~C.~Smith\orcid{0000-0002-3259-2771}\inst{\ref{aff177}}
\and K.~Tanidis\orcid{0000-0001-9843-5130}\inst{\ref{aff124}}
\and C.~Tao\orcid{0000-0001-7961-8177}\inst{\ref{aff13}}
\and G.~Testera\inst{\ref{aff35}}
\and R.~Teyssier\orcid{0000-0001-7689-0933}\inst{\ref{aff172}}
\and S.~Tosi\orcid{0000-0002-7275-9193}\inst{\ref{aff34},\ref{aff35},\ref{aff23}}
\and A.~Troja\orcid{0000-0003-0239-4595}\inst{\ref{aff106},\ref{aff65}}
\and M.~Tucci\inst{\ref{aff63}}
\and C.~Valieri\inst{\ref{aff29}}
\and A.~Venhola\orcid{0000-0001-6071-4564}\inst{\ref{aff178}}
\and D.~Vergani\orcid{0000-0003-0898-2216}\inst{\ref{aff9}}
\and G.~Verza\orcid{0000-0002-1886-8348}\inst{\ref{aff179}}
\and P.~Vielzeuf\orcid{0000-0003-2035-9339}\inst{\ref{aff13}}
\and N.~A.~Walton\orcid{0000-0003-3983-8778}\inst{\ref{aff177}}
\and J.~R.~Weaver\orcid{0000-0003-1614-196X}\inst{\ref{aff180}}
\and M.~Bella\orcid{0000-0002-6406-4789}\inst{\ref{aff108}}
\and D.~Scott\orcid{0000-0002-6878-9840}\inst{\ref{aff181}}}
										   
\institute{Space Science Data Center, Italian Space Agency, via del Politecnico snc, 00133 Roma, Italy\label{aff1}
\and
INAF-Osservatorio Astronomico di Trieste, Via G. B. Tiepolo 11, 34143 Trieste, Italy\label{aff2}
\and
Caltech/IPAC, 1200 E. California Blvd., Pasadena, CA 91125, USA\label{aff3}
\and
Infrared Processing and Analysis Center, California Institute of Technology, Pasadena, CA 91125, USA\label{aff4}
\and
Leiden Observatory, Leiden University, Einsteinweg 55, 2333 CC Leiden, The Netherlands\label{aff5}
\and
Aerospace Logistics Technology Engineering Company, Corso Marche 79, 10146 Torino, Italy\label{aff6}
\and
INAF-Osservatorio Astrofisico di Torino, Via Osservatorio 20, 10025 Pino Torinese (TO), Italy\label{aff7}
\and
Dipartimento di Fisica e Astronomia "Augusto Righi" - Alma Mater Studiorum Universit\`a di Bologna, via Piero Gobetti 93/2, 40129 Bologna, Italy\label{aff8}
\and
INAF-Osservatorio di Astrofisica e Scienza dello Spazio di Bologna, Via Piero Gobetti 93/3, 40129 Bologna, Italy\label{aff9}
\and
Universit\'e Claude Bernard Lyon 1, CNRS/IN2P3, IP2I Lyon, UMR 5822, Villeurbanne, F-69100, France\label{aff10}
\and
INAF-Osservatorio Astronomico di Roma, Via Frascati 33, 00078 Monteporzio Catone, Italy\label{aff11}
\and
Kapteyn Astronomical Institute, University of Groningen, PO Box 800, 9700 AV Groningen, The Netherlands\label{aff12}
\and
Aix-Marseille Universit\'e, CNRS/IN2P3, CPPM, Marseille, France\label{aff13}
\and
INAF-Osservatorio Astronomico di Padova, Via dell'Osservatorio 5, 35122 Padova, Italy\label{aff14}
\and
Max-Planck-Institut f\"ur Astronomie, K\"onigstuhl 17, 69117 Heidelberg, Germany\label{aff15}
\and
California Institute of Technology, 1200 E California Blvd, Pasadena, CA 91125, USA\label{aff16}
\and
ASTRON, the Netherlands Institute for Radio Astronomy, Postbus 2, 7990 AA, Dwingeloo, The Netherlands\label{aff17}
\and
Anton Pannekoek Institute for Astronomy, University of Amsterdam, Postbus 94249, 1090 GE Amsterdam, The Netherlands\label{aff18}
\and
Center for Advanced Interdisciplinary Research, Ss. Cyril and Methodius University in Skopje, Macedonia\label{aff19}
\and
Universit\'e Paris-Saclay, CNRS, Institut d'astrophysique spatiale, 91405, Orsay, France\label{aff20}
\and
ESAC/ESA, Camino Bajo del Castillo, s/n., Urb. Villafranca del Castillo, 28692 Villanueva de la Ca\~nada, Madrid, Spain\label{aff21}
\and
School of Mathematics and Physics, University of Surrey, Guildford, Surrey, GU2 7XH, UK\label{aff22}
\and
INAF-Osservatorio Astronomico di Brera, Via Brera 28, 20122 Milano, Italy\label{aff23}
\and
Universit\'e Paris-Saclay, Universit\'e Paris Cit\'e, CEA, CNRS, AIM, 91191, Gif-sur-Yvette, France\label{aff24}
\and
IFPU, Institute for Fundamental Physics of the Universe, via Beirut 2, 34151 Trieste, Italy\label{aff25}
\and
INFN, Sezione di Trieste, Via Valerio 2, 34127 Trieste TS, Italy\label{aff26}
\and
SISSA, International School for Advanced Studies, Via Bonomea 265, 34136 Trieste TS, Italy\label{aff27}
\and
Dipartimento di Fisica e Astronomia, Universit\`a di Bologna, Via Gobetti 93/2, 40129 Bologna, Italy\label{aff28}
\and
INFN-Sezione di Bologna, Viale Berti Pichat 6/2, 40127 Bologna, Italy\label{aff29}
\and
Centre National d'Etudes Spatiales -- Centre spatial de Toulouse, 18 avenue Edouard Belin, 31401 Toulouse Cedex 9, France\label{aff30}
\and
ATG Europe BV, Huygensstraat 34, 2201 DK Noordwijk, The Netherlands\label{aff31}
\and
Max Planck Institute for Extraterrestrial Physics, Giessenbachstr. 1, 85748 Garching, Germany\label{aff32}
\and
Universit\"ats-Sternwarte M\"unchen, Fakult\"at f\"ur Physik, Ludwig-Maximilians-Universit\"at M\"unchen, Scheinerstrasse 1, 81679 M\"unchen, Germany\label{aff33}
\and
Dipartimento di Fisica, Universit\`a di Genova, Via Dodecaneso 33, 16146, Genova, Italy\label{aff34}
\and
INFN-Sezione di Genova, Via Dodecaneso 33, 16146, Genova, Italy\label{aff35}
\and
Department of Physics "E. Pancini", University Federico II, Via Cinthia 6, 80126, Napoli, Italy\label{aff36}
\and
INAF-Osservatorio Astronomico di Capodimonte, Via Moiariello 16, 80131 Napoli, Italy\label{aff37}
\and
Instituto de Astrof\'isica e Ci\^encias do Espa\c{c}o, Universidade do Porto, CAUP, Rua das Estrelas, PT4150-762 Porto, Portugal\label{aff38}
\and
Faculdade de Ci\^encias da Universidade do Porto, Rua do Campo de Alegre, 4150-007 Porto, Portugal\label{aff39}
\and
Aix-Marseille Universit\'e, CNRS, CNES, LAM, Marseille, France\label{aff40}
\and
Dipartimento di Fisica, Universit\`a degli Studi di Torino, Via P. Giuria 1, 10125 Torino, Italy\label{aff41}
\and
INFN-Sezione di Torino, Via P. Giuria 1, 10125 Torino, Italy\label{aff42}
\and
European Space Agency/ESTEC, Keplerlaan 1, 2201 AZ Noordwijk, The Netherlands\label{aff43}
\and
Institute Lorentz, Leiden University, Niels Bohrweg 2, 2333 CA Leiden, The Netherlands\label{aff44}
\and
INAF-IASF Milano, Via Alfonso Corti 12, 20133 Milano, Italy\label{aff45}
\and
Centro de Investigaciones Energ\'eticas, Medioambientales y Tecnol\'ogicas (CIEMAT), Avenida Complutense 40, 28040 Madrid, Spain\label{aff46}
\and
Port d'Informaci\'{o} Cient\'{i}fica, Campus UAB, C. Albareda s/n, 08193 Bellaterra (Barcelona), Spain\label{aff47}
\and
Institute for Theoretical Particle Physics and Cosmology (TTK), RWTH Aachen University, 52056 Aachen, Germany\label{aff48}
\and
Institute of Space Sciences (ICE, CSIC), Campus UAB, Carrer de Can Magrans, s/n, 08193 Barcelona, Spain\label{aff49}
\and
Institut d'Estudis Espacials de Catalunya (IEEC),  Edifici RDIT, Campus UPC, 08860 Castelldefels, Barcelona, Spain\label{aff50}
\and
INFN section of Naples, Via Cinthia 6, 80126, Napoli, Italy\label{aff51}
\and
Institute for Astronomy, University of Hawaii, 2680 Woodlawn Drive, Honolulu, HI 96822, USA\label{aff52}
\and
Dipartimento di Fisica e Astronomia "Augusto Righi" - Alma Mater Studiorum Universit\`a di Bologna, Viale Berti Pichat 6/2, 40127 Bologna, Italy\label{aff53}
\and
Instituto de Astrof\'{\i}sica de Canarias, V\'{\i}a L\'actea, 38205 La Laguna, Tenerife, Spain\label{aff54}
\and
Institute for Astronomy, University of Edinburgh, Royal Observatory, Blackford Hill, Edinburgh EH9 3HJ, UK\label{aff55}
\and
Jodrell Bank Centre for Astrophysics, Department of Physics and Astronomy, University of Manchester, Oxford Road, Manchester M13 9PL, UK\label{aff56}
\and
European Space Agency/ESRIN, Largo Galileo Galilei 1, 00044 Frascati, Roma, Italy\label{aff57}
\and
Institut de Ci\`{e}ncies del Cosmos (ICCUB), Universitat de Barcelona (IEEC-UB), Mart\'{i} i Franqu\`{e}s 1, 08028 Barcelona, Spain\label{aff58}
\and
Instituci\'o Catalana de Recerca i Estudis Avan\c{c}ats (ICREA), Passeig de Llu\'{\i}s Companys 23, 08010 Barcelona, Spain\label{aff59}
\and
UCB Lyon 1, CNRS/IN2P3, IUF, IP2I Lyon, 4 rue Enrico Fermi, 69622 Villeurbanne, France\label{aff60}
\and
Departamento de F\'isica, Faculdade de Ci\^encias, Universidade de Lisboa, Edif\'icio C8, Campo Grande, PT1749-016 Lisboa, Portugal\label{aff61}
\and
Instituto de Astrof\'isica e Ci\^encias do Espa\c{c}o, Faculdade de Ci\^encias, Universidade de Lisboa, Campo Grande, 1749-016 Lisboa, Portugal\label{aff62}
\and
Department of Astronomy, University of Geneva, ch. d'Ecogia 16, 1290 Versoix, Switzerland\label{aff63}
\and
INAF-Istituto di Astrofisica e Planetologia Spaziali, via del Fosso del Cavaliere, 100, 00100 Roma, Italy\label{aff64}
\and
INFN-Padova, Via Marzolo 8, 35131 Padova, Italy\label{aff65}
\and
INFN-Bologna, Via Irnerio 46, 40126 Bologna, Italy\label{aff66}
\and
School of Physics, HH Wills Physics Laboratory, University of Bristol, Tyndall Avenue, Bristol, BS8 1TL, UK\label{aff67}
\and
FRACTAL S.L.N.E., calle Tulip\'an 2, Portal 13 1A, 28231, Las Rozas de Madrid, Spain\label{aff68}
\and
Institute of Theoretical Astrophysics, University of Oslo, P.O. Box 1029 Blindern, 0315 Oslo, Norway\label{aff69}
\and
Jet Propulsion Laboratory, California Institute of Technology, 4800 Oak Grove Drive, Pasadena, CA, 91109, USA\label{aff70}
\and
Department of Physics, Lancaster University, Lancaster, LA1 4YB, UK\label{aff71}
\and
Felix Hormuth Engineering, Goethestr. 17, 69181 Leimen, Germany\label{aff72}
\and
Technical University of Denmark, Elektrovej 327, 2800 Kgs. Lyngby, Denmark\label{aff73}
\and
Cosmic Dawn Center (DAWN), Denmark\label{aff74}
\and
Institut d'Astrophysique de Paris, UMR 7095, CNRS, and Sorbonne Universit\'e, 98 bis boulevard Arago, 75014 Paris, France\label{aff75}
\and
NASA Goddard Space Flight Center, Greenbelt, MD 20771, USA\label{aff76}
\and
Department of Physics and Helsinki Institute of Physics, Gustaf H\"allstr\"omin katu 2, 00014 University of Helsinki, Finland\label{aff77}
\and
Universit\'e de Gen\`eve, D\'epartement de Physique Th\'eorique and Centre for Astroparticle Physics, 24 quai Ernest-Ansermet, CH-1211 Gen\`eve 4, Switzerland\label{aff78}
\and
Department of Physics, P.O. Box 64, 00014 University of Helsinki, Finland\label{aff79}
\and
Helsinki Institute of Physics, Gustaf H{\"a}llstr{\"o}min katu 2, University of Helsinki, Helsinki, Finland\label{aff80}
\and
Centre de Calcul de l'IN2P3/CNRS, 21 avenue Pierre de Coubertin 69627 Villeurbanne Cedex, France\label{aff81}
\and
Laboratoire d'etude de l'Univers et des phenomenes eXtremes, Observatoire de Paris, Universit\'e PSL, Sorbonne Universit\'e, CNRS, 92190 Meudon, France\label{aff82}
\and
Mullard Space Science Laboratory, University College London, Holmbury St Mary, Dorking, Surrey RH5 6NT, UK\label{aff83}
\and
SKA Observatory, Jodrell Bank, Lower Withington, Macclesfield, Cheshire SK11 9FT, UK\label{aff84}
\and
Dipartimento di Fisica "Aldo Pontremoli", Universit\`a degli Studi di Milano, Via Celoria 16, 20133 Milano, Italy\label{aff85}
\and
INFN-Sezione di Milano, Via Celoria 16, 20133 Milano, Italy\label{aff86}
\and
University of Applied Sciences and Arts of Northwestern Switzerland, School of Engineering, 5210 Windisch, Switzerland\label{aff87}
\and
Universit\"at Bonn, Argelander-Institut f\"ur Astronomie, Auf dem H\"ugel 71, 53121 Bonn, Germany\label{aff88}
\and
INFN-Sezione di Roma, Piazzale Aldo Moro, 2 - c/o Dipartimento di Fisica, Edificio G. Marconi, 00185 Roma, Italy\label{aff89}
\and
Department of Physics, Institute for Computational Cosmology, Durham University, South Road, Durham, DH1 3LE, UK\label{aff90}
\and
Universit\'e C\^{o}te d'Azur, Observatoire de la C\^{o}te d'Azur, CNRS, Laboratoire Lagrange, Bd de l'Observatoire, CS 34229, 06304 Nice cedex 4, France\label{aff91}
\and
Universit\'e Paris Cit\'e, CNRS, Astroparticule et Cosmologie, 75013 Paris, France\label{aff92}
\and
CNRS-UCB International Research Laboratory, Centre Pierre Binetruy, IRL2007, CPB-IN2P3, Berkeley, USA\label{aff93}
\and
Institut d'Astrophysique de Paris, 98bis Boulevard Arago, 75014, Paris, France\label{aff94}
\and
Institute of Physics, Laboratory of Astrophysics, Ecole Polytechnique F\'ed\'erale de Lausanne (EPFL), Observatoire de Sauverny, 1290 Versoix, Switzerland\label{aff95}
\and
Aurora Technology for European Space Agency (ESA), Camino bajo del Castillo, s/n, Urbanizacion Villafranca del Castillo, Villanueva de la Ca\~nada, 28692 Madrid, Spain\label{aff96}
\and
Institut de F\'{i}sica d'Altes Energies (IFAE), The Barcelona Institute of Science and Technology, Campus UAB, 08193 Bellaterra (Barcelona), Spain\label{aff97}
\and
School of Mathematics, Statistics and Physics, Newcastle University, Herschel Building, Newcastle-upon-Tyne, NE1 7RU, UK\label{aff98}
\and
DARK, Niels Bohr Institute, University of Copenhagen, Jagtvej 155, 2200 Copenhagen, Denmark\label{aff99}
\and
Waterloo Centre for Astrophysics, University of Waterloo, Waterloo, Ontario N2L 3G1, Canada\label{aff100}
\and
Department of Physics and Astronomy, University of Waterloo, Waterloo, Ontario N2L 3G1, Canada\label{aff101}
\and
Perimeter Institute for Theoretical Physics, Waterloo, Ontario N2L 2Y5, Canada\label{aff102}
\and
Institute of Space Science, Str. Atomistilor, nr. 409 M\u{a}gurele, Ilfov, 077125, Romania\label{aff103}
\and
Consejo Superior de Investigaciones Cientificas, Calle Serrano 117, 28006 Madrid, Spain\label{aff104}
\and
Universidad de La Laguna, Departamento de Astrof\'{\i}sica, 38206 La Laguna, Tenerife, Spain\label{aff105}
\and
Dipartimento di Fisica e Astronomia "G. Galilei", Universit\`a di Padova, Via Marzolo 8, 35131 Padova, Italy\label{aff106}
\and
Institut f\"ur Theoretische Physik, University of Heidelberg, Philosophenweg 16, 69120 Heidelberg, Germany\label{aff107}
\and
Institut de Recherche en Astrophysique et Plan\'etologie (IRAP), Universit\'e de Toulouse, CNRS, UPS, CNES, 14 Av. Edouard Belin, 31400 Toulouse, France\label{aff108}
\and
Universit\'e St Joseph; Faculty of Sciences, Beirut, Lebanon\label{aff109}
\and
Departamento de F\'isica, FCFM, Universidad de Chile, Blanco Encalada 2008, Santiago, Chile\label{aff110}
\and
Satlantis, University Science Park, Sede Bld 48940, Leioa-Bilbao, Spain\label{aff111}
\and
Department of Physics, Royal Holloway, University of London, TW20 0EX, UK\label{aff112}
\and
Instituto de Astrof\'isica e Ci\^encias do Espa\c{c}o, Faculdade de Ci\^encias, Universidade de Lisboa, Tapada da Ajuda, 1349-018 Lisboa, Portugal\label{aff113}
\and
Department of Physics and Astronomy, University College London, Gower Street, London WC1E 6BT, UK\label{aff114}
\and
Cosmic Dawn Center (DAWN)\label{aff115}
\and
Niels Bohr Institute, University of Copenhagen, Jagtvej 128, 2200 Copenhagen, Denmark\label{aff116}
\and
Universidad Polit\'ecnica de Cartagena, Departamento de Electr\'onica y Tecnolog\'ia de Computadoras,  Plaza del Hospital 1, 30202 Cartagena, Spain\label{aff117}
\and
Dipartimento di Fisica e Scienze della Terra, Universit\`a degli Studi di Ferrara, Via Giuseppe Saragat 1, 44122 Ferrara, Italy\label{aff118}
\and
Istituto Nazionale di Fisica Nucleare, Sezione di Ferrara, Via Giuseppe Saragat 1, 44122 Ferrara, Italy\label{aff119}
\and
INAF, Istituto di Radioastronomia, Via Piero Gobetti 101, 40129 Bologna, Italy\label{aff120}
\and
Astronomical Observatory of the Autonomous Region of the Aosta Valley (OAVdA), Loc. Lignan 39, I-11020, Nus (Aosta Valley), Italy\label{aff121}
\and
School of Physics and Astronomy, Cardiff University, The Parade, Cardiff, CF24 3AA, UK\label{aff122}
\and
Kavli Institute for Cosmology Cambridge, Madingley Road, Cambridge, CB3 0HA, UK\label{aff123}
\and
Department of Physics, Oxford University, Keble Road, Oxford OX1 3RH, UK\label{aff124}
\and
Instituto de Astrof\'isica de Canarias (IAC); Departamento de Astrof\'isica, Universidad de La Laguna (ULL), 38200, La Laguna, Tenerife, Spain\label{aff125}
\and
Universit\'e PSL, Observatoire de Paris, Sorbonne Universit\'e, CNRS, LERMA, 75014, Paris, France\label{aff126}
\and
Universit\'e Paris-Cit\'e, 5 Rue Thomas Mann, 75013, Paris, France\label{aff127}
\and
Department of Mathematics and Physics E. De Giorgi, University of Salento, Via per Arnesano, CP-I93, 73100, Lecce, Italy\label{aff128}
\and
INFN, Sezione di Lecce, Via per Arnesano, CP-193, 73100, Lecce, Italy\label{aff129}
\and
INAF-Sezione di Lecce, c/o Dipartimento Matematica e Fisica, Via per Arnesano, 73100, Lecce, Italy\label{aff130}
\and
INAF - Osservatorio Astronomico di Brera, via Emilio Bianchi 46, 23807 Merate, Italy\label{aff131}
\and
INAF-Osservatorio Astronomico di Brera, Via Brera 28, 20122 Milano, Italy, and INFN-Sezione di Genova, Via Dodecaneso 33, 16146, Genova, Italy\label{aff132}
\and
ICL, Junia, Universit\'e Catholique de Lille, LITL, 59000 Lille, France\label{aff133}
\and
ICSC - Centro Nazionale di Ricerca in High Performance Computing, Big Data e Quantum Computing, Via Magnanelli 2, Bologna, Italy\label{aff134}
\and
Instituto de F\'isica Te\'orica UAM-CSIC, Campus de Cantoblanco, 28049 Madrid, Spain\label{aff135}
\and
CERCA/ISO, Department of Physics, Case Western Reserve University, 10900 Euclid Avenue, Cleveland, OH 44106, USA\label{aff136}
\and
Technical University of Munich, TUM School of Natural Sciences, Physics Department, James-Franck-Str.~1, 85748 Garching, Germany\label{aff137}
\and
Max-Planck-Institut f\"ur Astrophysik, Karl-Schwarzschild-Str.~1, 85748 Garching, Germany\label{aff138}
\and
Laboratoire Univers et Th\'eorie, Observatoire de Paris, Universit\'e PSL, Universit\'e Paris Cit\'e, CNRS, 92190 Meudon, France\label{aff139}
\and
Departamento de F{\'\i}sica Fundamental. Universidad de Salamanca. Plaza de la Merced s/n. 37008 Salamanca, Spain\label{aff140}
\and
Universit\'e de Strasbourg, CNRS, Observatoire astronomique de Strasbourg, UMR 7550, 67000 Strasbourg, France\label{aff141}
\and
Center for Data-Driven Discovery, Kavli IPMU (WPI), UTIAS, The University of Tokyo, Kashiwa, Chiba 277-8583, Japan\label{aff142}
\and
Dipartimento di Fisica - Sezione di Astronomia, Universit\`a di Trieste, Via Tiepolo 11, 34131 Trieste, Italy\label{aff143}
\and
NASA Ames Research Center, Moffett Field, CA 94035, USA\label{aff144}
\and
Bay Area Environmental Research Institute, Moffett Field, California 94035, USA\label{aff145}
\and
University of California, Los Angeles, CA 90095-1562, USA\label{aff146}
\and
Department of Physics \& Astronomy, University of California Irvine, Irvine CA 92697, USA\label{aff147}
\and
Departamento F\'isica Aplicada, Universidad Polit\'ecnica de Cartagena, Campus Muralla del Mar, 30202 Cartagena, Murcia, Spain\label{aff148}
\and
Instituto de F\'isica de Cantabria, Edificio Juan Jord\'a, Avenida de los Castros, 39005 Santander, Spain\label{aff149}
\and
CEA Saclay, DFR/IRFU, Service d'Astrophysique, Bat. 709, 91191 Gif-sur-Yvette, France\label{aff150}
\and
Institute of Cosmology and Gravitation, University of Portsmouth, Portsmouth PO1 3FX, UK\label{aff151}
\and
Department of Astronomy, University of Florida, Bryant Space Science Center, Gainesville, FL 32611, USA\label{aff152}
\and
Department of Computer Science, Aalto University, PO Box 15400, Espoo, FI-00 076, Finland\label{aff153}
\and
Instituto de Astrof\'\i sica de Canarias, c/ Via Lactea s/n, La Laguna 38200, Spain. Departamento de Astrof\'\i sica de la Universidad de La Laguna, Avda. Francisco Sanchez, La Laguna, 38200, Spain\label{aff154}
\and
Ruhr University Bochum, Faculty of Physics and Astronomy, Astronomical Institute (AIRUB), German Centre for Cosmological Lensing (GCCL), 44780 Bochum, Germany\label{aff155}
\and
Department of Physics and Astronomy, Vesilinnantie 5, 20014 University of Turku, Finland\label{aff156}
\and
Serco for European Space Agency (ESA), Camino bajo del Castillo, s/n, Urbanizacion Villafranca del Castillo, Villanueva de la Ca\~nada, 28692 Madrid, Spain\label{aff157}
\and
ARC Centre of Excellence for Dark Matter Particle Physics, Melbourne, Australia\label{aff158}
\and
Centre for Astrophysics \& Supercomputing, Swinburne University of Technology,  Hawthorn, Victoria 3122, Australia\label{aff159}
\and
Department of Physics and Astronomy, University of the Western Cape, Bellville, Cape Town, 7535, South Africa\label{aff160}
\and
DAMTP, Centre for Mathematical Sciences, Wilberforce Road, Cambridge CB3 0WA, UK\label{aff161}
\and
Department of Astrophysics, University of Zurich, Winterthurerstrasse 190, 8057 Zurich, Switzerland\label{aff162}
\and
Department of Physics, Centre for Extragalactic Astronomy, Durham University, South Road, Durham, DH1 3LE, UK\label{aff163}
\and
IRFU, CEA, Universit\'e Paris-Saclay 91191 Gif-sur-Yvette Cedex, France\label{aff164}
\and
Oskar Klein Centre for Cosmoparticle Physics, Department of Physics, Stockholm University, Stockholm, SE-106 91, Sweden\label{aff165}
\and
Astrophysics Group, Blackett Laboratory, Imperial College London, London SW7 2AZ, UK\label{aff166}
\and
Univ. Grenoble Alpes, CNRS, Grenoble INP, LPSC-IN2P3, 53, Avenue des Martyrs, 38000, Grenoble, France\label{aff167}
\and
INAF-Osservatorio Astrofisico di Arcetri, Largo E. Fermi 5, 50125, Firenze, Italy\label{aff168}
\and
Dipartimento di Fisica, Sapienza Universit\`a di Roma, Piazzale Aldo Moro 2, 00185 Roma, Italy\label{aff169}
\and
Centro de Astrof\'{\i}sica da Universidade do Porto, Rua das Estrelas, 4150-762 Porto, Portugal\label{aff170}
\and
HE Space for European Space Agency (ESA), Camino bajo del Castillo, s/n, Urbanizacion Villafranca del Castillo, Villanueva de la Ca\~nada, 28692 Madrid, Spain\label{aff171}
\and
Department of Astrophysical Sciences, Peyton Hall, Princeton University, Princeton, NJ 08544, USA\label{aff172}
\and
Theoretical astrophysics, Department of Physics and Astronomy, Uppsala University, Box 515, 751 20 Uppsala, Sweden\label{aff173}
\and
Minnesota Institute for Astrophysics, University of Minnesota, 116 Church St SE, Minneapolis, MN 55455, USA\label{aff174}
\and
Mathematical Institute, University of Leiden, Einsteinweg 55, 2333 CA Leiden, The Netherlands\label{aff175}
\and
School of Physics \& Astronomy, University of Southampton, Highfield Campus, Southampton SO17 1BJ, UK\label{aff176}
\and
Institute of Astronomy, University of Cambridge, Madingley Road, Cambridge CB3 0HA, UK\label{aff177}
\and
Space physics and astronomy research unit, University of Oulu, Pentti Kaiteran katu 1, FI-90014 Oulu, Finland\label{aff178}
\and
Center for Computational Astrophysics, Flatiron Institute, 162 5th Avenue, 10010, New York, NY, USA\label{aff179}
\and
Department of Astronomy, University of Massachusetts, Amherst, MA 01003, USA\label{aff180}
\and
Department of Physics and Astronomy, University of British Columbia, Vancouver, BC V6T 1Z1, Canada\label{aff181}}    


   \date{Received Now; accepted Never}

\abstract{
This paper describes the \ac{NIR PF} that processes near-infrared images from the \ac{NISP} instrument onboard the \Euclid satellite. \ac{NIR PF} consists of three main components: (i) a common pre-processing stage for both photometric (NIR) and spectroscopic (SIR) data to remove instrumental effects; (ii) astrometric and photometric calibration of NIR data, along with catalogue extraction; and (iii) resampling and stacking. The necessary calibration products are generated using dedicated pipelines that process observations from both the early \ac{PV} phase in 2023 and the nominal survey operations. After outlining the pipeline’s structure and algorithms, we demonstrate its application to \Euclid Q1 images. For Q1, we achieve an astrometric accuracy of 9--15\,mas, a relative photometric accuracy of 5\,mmag, and an absolute flux calibration limited by the 1\% uncertainty of the \ac{HST} CALSPEC database. We characterise the \ac{PSF} that we find very stable across the focal plane, and we discuss current limitations of \ac{NIR PF} that will be improved upon for future data releases.}



   \titlerunning{\Euclid Quick Data Release (Q1): NIR processing and data products}
   \authorrunning{Euclid Collaboration: G.~Polenta et al.}

   \maketitle
   

\section{Introduction}
\label{sec:intro}

   \Euclid was designed to study the dark Universe through a photometric and spectroscopic survey of the extragalactic sky in the visible and near-infrared bands. An overview of this \ac{ESA} mission, the survey, the data products, and its science programme is presented in \citet{EuclidSkyOverview}. The Euclid Wide Survey \citep[EWS,][]{Scaramella-EP1} will cover about $14\,000\,\rm{deg}^2$, providing tens of thousands of images to extract positions, shapes, and photometric redshifts for 1.5 billion galaxies, as well as 35 million spectroscopic redshifts. 
   To manage and process these data in a largely automated manner, the Euclid Consortium has built a number of software pipelines and tools constituting the \ac{SGS}, which has been designed, developed, and validated through a series of data challenges and simulations \citep{EuclidSkyFlagship} during the last 10 years, in parallel with the development of the instruments.
   
   In this paper, we describe the \ac{NIR PF}, which produces calibrated near-infrared images for the \citet{Q1cite}, starting from raw photometric exposures acquired by NISP \citep[see \cref{sec:nisp} and][]{EuclidSkyNISP}. This is achieved through three main steps (\cref{sec:pf}): (i) pre-processing to account for instrumental and detector effects that are common to both photometric and spectroscopic exposures \citep{Q1-TP006}; (ii) image astrometric and photometric calibration, and catalogue extraction; and (iii) resampling and stacking. The latter step will be described for a future data release, since NIR stacks and associated source catalogues are not part of the Q1 release. The relevant calibration products have been generated through dedicated pipelines designed to process specific calibration observations taken during ground calibrations, \ac{PV} phase, and nominal science operations. An additional post-processing pipeline has been developed to evaluate the quality of the processing (\cref{sec:dqc}), thus allowing us to mark valid data products and trigger further analyses on invalid ones. 
   The resulting NIR data products, publicly available through the \Euclid archive for Q1 \citep{Q1-TP001} are presented in \cref{sec:products}, and their basic performance, characterisation, and current limitations are discussed in \cref{sec:imageaccuracy,sec:conclusions}.


\section{The NISP Instrument}
\label{sec:nisp}

NISP is the Near-Infrared Spectrometer and Photometer \citep{EuclidSkyNISP} working in parallel to the optical VIS imager \citep{EuclidSkyVIS} by means of a dichroic beam splitter. Two wheels located after a correction lens select either the photometric (NISP-P) or spectroscopic mode (NISP-S), with their corresponding bandpasses. A camera-lens assembly then focusses the light on the 16 \ac{H2RG} detectors of 2048\,$\times$\,2048 pixels, each. Light from five calibration \acp{LED} with different wavelengths can directly illuminate the \ac{FPA} without passing through any optical element. The detectors are operated in an non-destructive, \ac{MACC} \ac{SUTR} scheme, with several groups of exposures being averaged. In particular, photometric exposures are taken using MACC(4,16,4)\footnote{The acquisition mode is indicated as MACC($n_{\rm g}$,$n_{\rm f}$,$n_{\rm d}$) where $n_g$ is the number of groups, $n_f$ is the number of frames being averaged within a group, and $n_d$ is the number of dropped frames between two consecutive groups.}, that is 4 groups of 16 averaged
frames, interleaved by 4 dropped frames between groups.
The slope of the signal between groups is fitted on-board \citep{Kubiketal2016}, and only this slope and a quality image is transmitted to Earth. The details of NISP's layout, operations, and calibration approach are described in \citet{EuclidSkyNISP}.

NISP-S can utilise one of three grisms in the grism wheel, covering the 926\,--\,1892\,nm wavelength range. NISP-P can select from three broadband filters -- \YE, \JE, and \HE\ -- in the 950\,nm and 2021\,nm range \citep{Schirmer-EP18}. A closed position in the filter wheel allows calibration observations without sky, either for dark frames or \ac{LED} flats \citep{EuclidSkyNISPCU}.

\Euclid observes both science fields and most calibration targets using a standardised reference observation sequence \citep[ROS,][]{Scaramella-EP1}. This approach ensures instrument stability and maintains high uniformity across science and calibration data. Special sequences are used only for specific calibration tasks. All scientific and calibration data are processed by the \ac{SGS}.

%



\section{The NIR Processing Function}
\label{sec:pf}
The \ac{NIR PF} is designed to provide fully calibrated NISP images accounting for all instrumental effects, and it consists of numerous \acp{PE} for specific tasks. These comprise both the construction of master calibration files from dedicated calibration observations, and their application to the science images to remove the instrumental fingerprints.


\subsection{Common NIR--SIR pre-processing}
\label{sec:preprocessing}
\begin{figure}
\centering
\resizebox{\hsize}{!}{\includegraphics{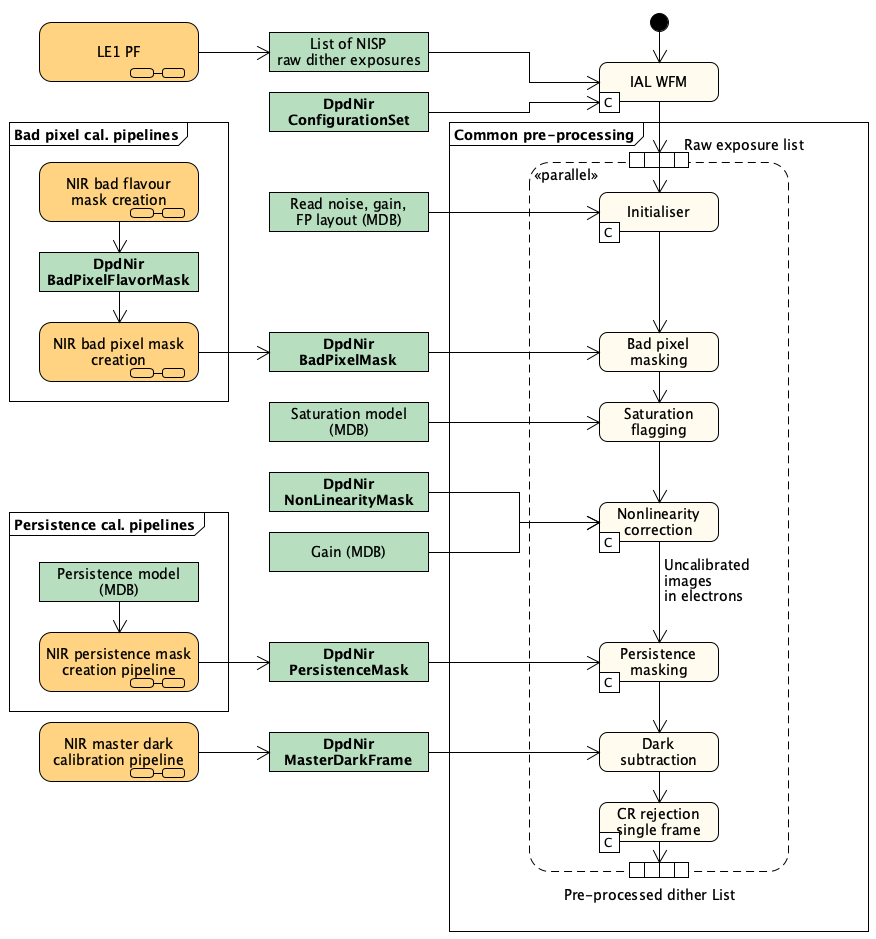}}
\caption{\ac{NIR PF} pre-processing. Input products are shown as green boxes on the left. Some of them are taken from the \ac{MDB}, while other products are generated by dedicated processing pipelines represented as orange boxes.}
\label{fig:preproc}
\end{figure}

In this section we summarise the early \acp{PE} in \ac{NIR PF} shown in \cref{fig:preproc}. These correct effects mostly at the detector-level, common to both the NISP-P and NISP-S channels.


\subsubsection{Detector baseline}
\label{sec:baseline}

The detector baseline is a pedestal set to ensure positive values in the \ac{ADC} output, as well as an optimal use of the detectors' dynamic range in a linear manner. The baseline \ac{PE} does not manipulate or calibrate NIR \ac{LE1} data. 
Instead, based on the results obtained from the NISP on-ground test campaign \citep[Kubik et al., in prep.]{Barbier2018}, it is used to validate the correct baseline setting. This is essential for guaranteeing accurate signal measurements, that is to operate the detectors within the \ac{ADC} linear range.

After an initial setting done during \Euclid's commissioning phase \citep{Cogato2024}, the detectors’ baseline was verified at a nominal operational temperature of 94\,K. To this end, 64 consecutive dark exposures were acquired with the MACC(1,16,1) mode \citep{Kubiketal2016}, that is a single group of 16 co-added frames. From these, the baseline map is estimated as the median signal over the image stack after correcting raw images for reference pixel drift \citep[Kubik et al., in prep.]{Kubik2014, Medinaceli2020}.

\begin{figure}
\centering
\resizebox{\hsize}{!}{\includegraphics{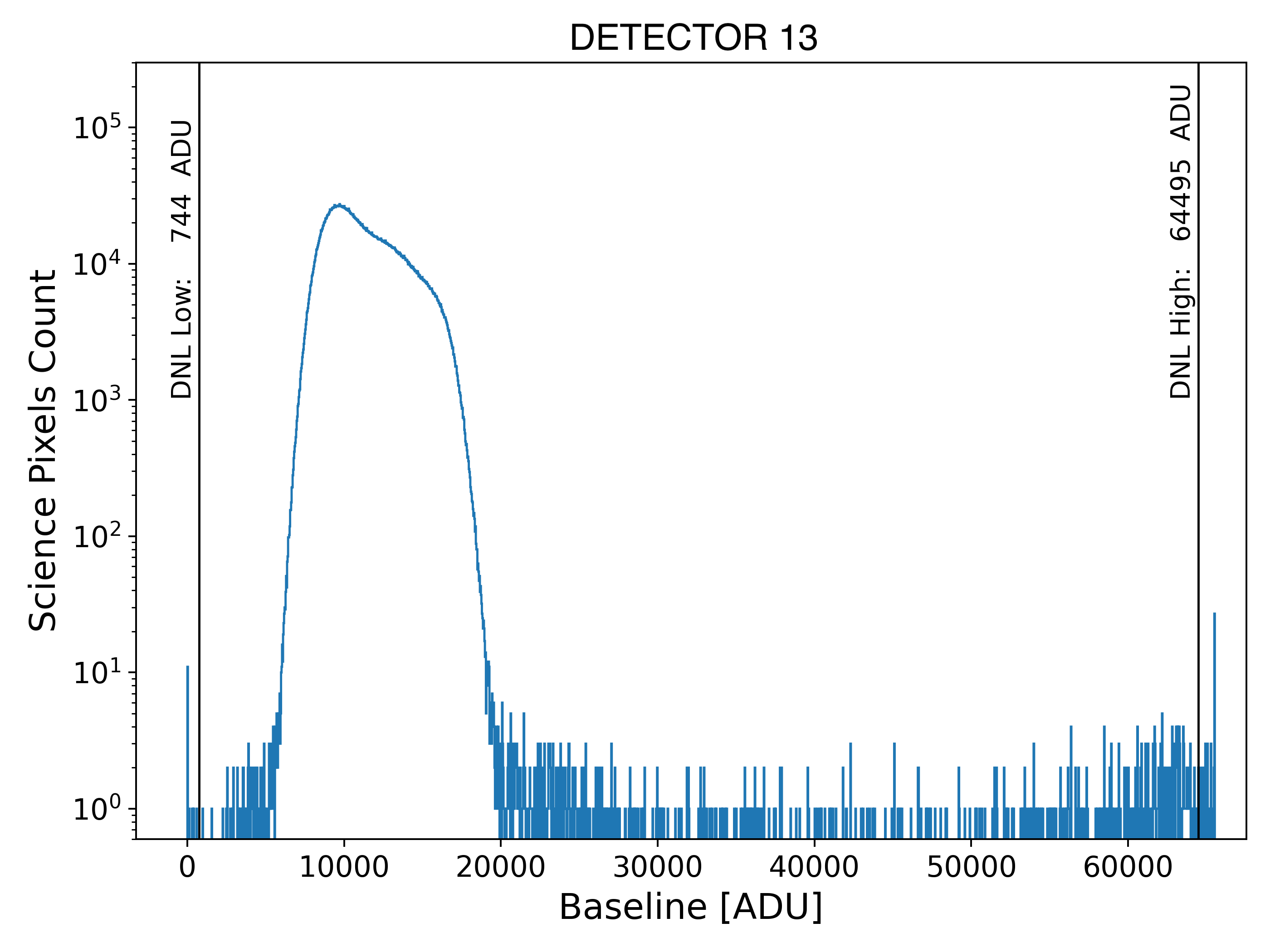}}
\caption{Example of the science pixels' baseline distribution for one of the detectors (DET13) measured during \ac{PV}. Solid black vertical lines show the boundary of the \ac{DNL} range defined during the 
ground-test campaign.}
\label{fig:baseline}
\end{figure}

The baseline maps are verified by comparison with the ground-based  \ac{DNL} range for reference and science pixel distributions. The baseline maps are stored in the \Euclid archive and the \ac{DNL} outliers are included in the bad-pixel map (see \cref{sec:badpix} and \cref{fig:baseline}). 
As a by-product of this \ac{PE}, the so-called kTC noise map is estimated by computing the signal standard deviation across the dark image stack (Kubik et al., in prep.). The kTC noise originates from the thermal fluctuations in the pixel’s capacitance during reset.

\subsubsection{Unstable channels}
\label{sec:unstablechannels}

\Euclid \acp{H2RG} detectors operate in 32-output readout mode. This means that they are divided into 32 channels of 64 2048-pixel columns, and the channels are read out in parallel. Over the 16 detectors, a few channels show a top-bottom gradient, with various amplitudes and frequency of occurrence. This effect was identified on the ground for some channels, but the set of channels affected by this has changed slightly in flight. Two types of gradients have been identified.

First, most cases show a linear slope from top to bottom. Those cases can be detected and corrected using the reference pixels.
Each detector channel has four rows of reference pixels at the top and at the bottom of the array. Reference pixels are not connected to the detector photodiodes and they allow subtraction of the 
frame-to-frame bias and temperature drift (Kubik et al., in prep.). In flight we have access to all frames at the top and bottom of two reference lines, used to compute a gradient from top to bottom and thus identify unstable channels. The amplitude of this slope can be used to compute a correction. This procedure is currently applied to the 6 channels for which the amplitude can be detected with high significance above the noise floor.

Second, other cases show a quadratic slope, where the top and bottom parts of all columns of the channel are brighter while the middle part is darker. This happens  usually with the longer integration times of the 
MACC(15,16,11) spectroscopic exposures, and cannot be detected with reference pixels because they do not detect the lower values near the channel's mid-point.


\subsubsection{Bad-pixel identification and masking}
\label{sec:badpix}

We use ground test data of the \acp{H2RG} as the primary resource to identify bad pixels for Q1. These pre-launch evaluations of the pixel performance were taken over three main campaigns: acceptance testing and ranking at the NASA Goddard Space Flight Center's Detector Characterisation Laboratory (DCL) in 2016; individual component tests in 2019
at the CNRS-IN2P3 Center for Particle Physics in Marseille (CPPM); and fully-integrated flight-module tests in space-like conditions (thermal vacuum and thermal balance, or TBTV) at \ac{LAM} in 2019--2020. The TBTV and final pre-launch \ac{PLM} tests with NISP integrated into the telescope are summarised in \citet{medinaceli2022}. 

In addition, baseline calibration measurements were taken during commissioning and incorporated into the bad-pixel maps for \ac{NIR PF} (see \cref{sec:badpix}). Calibration observations obtained during the \ac{PV} Phase are not reflected in Q1.

Pixels revealed to be permanently inoperable for science, or exhibiting noteworthy performance characteristics that may affect operability, are contained in calibration data products for the NISP Level 2 pipeline. A map has been created for each pixel condition (or `flavour') by detector, assigned a 16-bit unsigned integer 2$^n$, $n$ > 1, and merged into a master \ac{BPM}  that can be used by \ac{NIR PF}. The \ac{BPM} application updates the \ac{DQ} layer accompanying each of the 16 science frames, with no distinction between imaging and spectroscopy modes. \ac{DQ} values are the bit-wise sum for pixels of more than one flavour.   

The BPM is not restricted to scientifically inoperable pixels. To distinguish unusable from informational flavours, the bit 0 is reserved for the unusable cases.  This `INVALID' bit is summed only once, even when a pixel has been flagged with more than one fatal condition, as is often the case. Thus {\tt INVALID} pixels will always have an odd \ac{DQ} value. Further details, including bit assignments, can be found in \citet{TNflags} and \citet{EuclidDpdd}. Statistics for the non-transient -- that is static -- pixel masks are summarised in \cref{table:badpixdarks}, and in the following we review the most important flags.

\begin{figure}
\centering
\begin{minipage}[t]{0.48\textwidth}
\centering
\resizebox{\hsize}{!}{\includegraphics{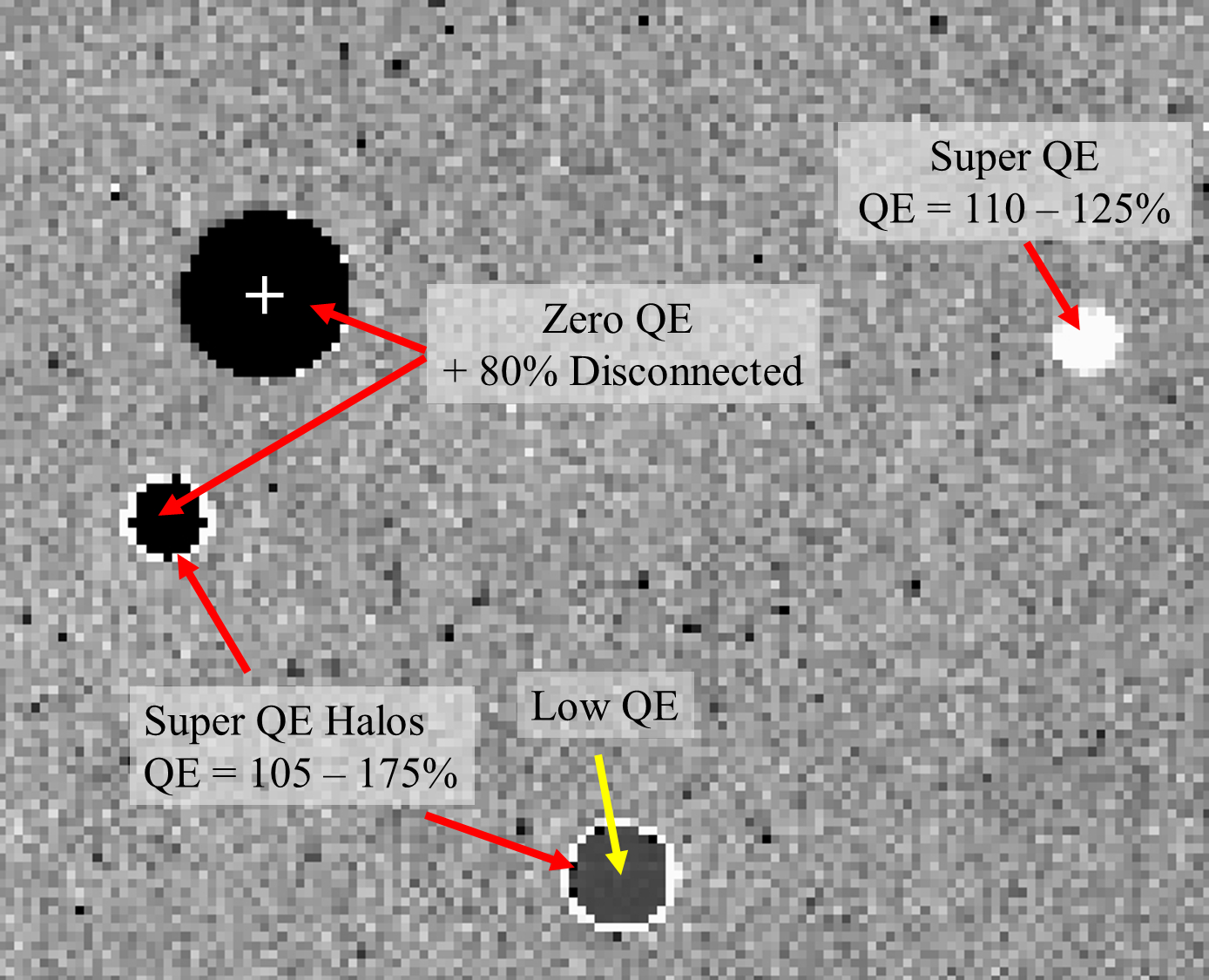}}
\end{minipage}
\caption{Representative flagged QE `flavours' in Q1, detector 44 in this example. Grey background pixels are in the range of 5--100\% QE. Zero and Super QE pixels are {\tt INVALID} in \ac{NIR PF}, while Low QE pixels are flagged for monitoring.  Super QE pixels often appear as a 1-pixel wide bright halo around groups of disconnected pixels, although the largest group in this example does not exhibit a halo -- the QEs of neighbouring pixels are within the nominal range to within the 5\% measurement uncertainty. For reference, `+' indicates pixel [$x$,$y$] = [1458,1336] in Level 2 2040\,$\times$\,2040 array coordinates. The labelled QE ranges are wavelength-averaged.} 
\label{figure:badpix_QE}
\end{figure}

{\bf Invalid flags.} Pixels in Q1 data are flagged invalid if: (i) they are unresponsive due to disconnection from the readout-integrated circuit; (ii) their quantum efficiency is very low or poorly known; and (iii) if their baseline after bias reset is outside the photometrically calibratable range.

\paragraph{DISCONNECTED} pixels have absent or non-working indium bumps for connection to the photosensitive diodes. They terminate at the capacitor and have no detectable photo-sensitivity. Some positive residual charge is generally present for them in the readout electronics. Measurements were done under specific laboratory conditions at ambient room temperature and thus could not be repeated in the same fashion after launch. 

\paragraph{ZERO QUANTUM EFFICIENCY} pixels were identified during detector 
qualification tests at the DCL. Such pixels are flagged when the measured \ac{QE} is less than 5\% (corresponding to the low-end measurement uncertainty) averaged over wavelengths $920\,{\rm nm} \leq \lambda \leq1120\,{\rm nm}$. Most disconnected pixels have zero \ac{QE}.

\paragraph{SUPER QE} pixels have wavelength-averaged \acp{QE} greater than 110\%.  Such pixels generally surround clusters of disconnected pixels, creating a bright, 1-pixel wide halo under illumination as photocurrent flows from the disconnected pixels to the nearest photosensitive neighbours with usable well depths. \Cref{figure:badpix_QE} shows an example. Thus the measured \acp{QE} become artificially high -- reaching as much as 200\% in some cases -- and unrecoverable.  

\paragraph{BAD BASELINE} pixels have baselines that are either too high or too low after bias reset and fall into a range of excessive kTC noise.
Because the ground-based baseline data were obtained at a lower temperature than where the NISP \ac{FPA} operates on orbit by about 6\,K, the baselines were remeasured during commissioning (see \cref{sec:baseline}) and updated in the \ac{BPM}. This is the only flavour that  has been updated with in-flight data for the Q1 release.

{\bf Informational flags.} Certain flags have been reserved for monitoring purposes.  In Q1, hot and low QE pixels are tracked but are not set to {\tt INVALID}, pending further evaluation of their operability for science with flight data.

\paragraph{HOT} pixels have been defined differently in the literature by the 
sensor-chip manufacturers and the intended application. For NISP the signal of a hot pixel is 3\,$\sigma$ above the most probable value of the detector's measured dark current. Pixels with elevated dark currents are not necessarily {\tt INVALID}, as long as the unilluminated output signal is stable and below the saturation limit.  Hot pixels were identified from ground-based dark current measurements, and are updated and checked for stability in flatfield measurements taken during \ac{PV} and in the routine phase with the \acp{LED} over a range of fluences.

\paragraph{LOW \ac{QE}} pixels are flagged when nominal requirements set by NASA are not met to within the 5\% RMS uncertainty for measurements at $920\,{\rm nm} \leq \lambda \leq2020\,{\rm nm}$,
\begin{itemize}
\item QE $\geq$ 74\% for $\lambda\in[1120, 2020]$\,nm, and
\item $\mathrm{QE} \geq \left[64 + (\lambda/\mathrm{nm} - 920)/20 \right]\%$ for $\lambda\in[920;1120]$\,nm.
\end{itemize}
The data from such pixels are expected to be calibratable, pending further evaluation with flight data, but will generally exhibit reduced signal-to-noise levels compared to pixels with nominal \ac{QE}.

\begin{table*}
\caption{NISP Detector Non-Transient DQ Flags and Dark Currents in Q1}             
\label{table:badpixdarks}      
\centering          
\begin{tabular}{c r r r r r r c c}     \hline\hline       
\noalign{\vskip 1pt}

DetID\tablefootmark{a} & \multicolumn{4}{c}{Invalid pixels\tablefootmark{b}} & \multicolumn{2}{c}{Informational\tablefootmark{b}} & \multicolumn{2}{c}{Dark MPV\tablefootmark{c}} \\ 
\hline
\noalign{\vskip 1pt}
 & DISCONN & ZEROQE & SUPERQE & BADBASE & LOWQE & HOT & Photo & Spectro \\ 
\hline   
\noalign{\vskip 1pt}                 
   11 & 135  & 747  & 290  & 27 & 591 & 630 & 0.007 & 0.016 \\  
   12 & 3223 & 5399 & 7038 & 25 & 953 & 3056 & 0.000 & 0.015 \\  
   13 & 589   & 1261 & 551 & 74 & 219 & 644 & 0.014 & 0.024 \\  
   14 & 252   & 1858 & 579 & 183 & 597 & 2794 & 0.018 & 0.029 \\  
   21 & 313    & 1275 & 16 & 171 & 292 & 1060 & 0.009 & 0.012 \\  
   22 & 2580 & 3510 & 1313 & 2127 & 331 & 1645 & 0.003 & 0.014 \\  
   23 & 1512 & 1992 & 2373 & 75 & 605 & 754 & 0.009 & 0.015 \\  
   24 & 146    & 698 & 302 & 15 & 216 & 420 & 0.000 & 0.010 \\  
   31 & 1420  & 2054 & 992 & 72 & 725 & 1743 & 0.004 & 0.014 \\  
   32 & 269   &  895 & 307 & 76 & 420 & 1161 & 0.004 & 0.013 \\  
   33 & 262   & 1288 & 644 & 95 & 402 & 1535 & 0.003 & 0.014 \\  
   34 & 970   & 2072 & 249 & 105 & 390 & 1159 & 0.002 & 0.013 \\  
   41 & 198   & 1858 & 339 & 63 & 585 & 631 & 0.000 & 0.009 \\  
   42 & 334    & 847 & 404 & 143 & 390 & 807 & 0.006 & 0.018 \\  
   43 & 222   & 1129 & 401 & 17 & 278 & 369 & 0.007 & 0.017 \\  
   44 & 255   & 7420 & 558 & 707 & 1542 & 4076 & 0.010 & 0.017 \\  

  \hline                  
\end{tabular}
\tablefoot{
\tablefoottext{a}{Detector slot position in the NISP \ac{FPA}; see Fig.~A.1 in \citet{EuclidSkyNISP}.}
\tablefoottext{b}{Pixel flag column names correspond to metadata keywords in the \ac{DQ} layer with each L2 science product (see \cref{table:flag_descriptions}).}
\tablefoottext{c}{Dark current most 
probable value in e$^{-}$\,s$^{-1}$ estimated from a Gaussian-profile fit to the signal distribution excluding invalid pixels.}
}
\end{table*}


\subsubsection{Nonlinearity correction}
\label{sec:nonlinearity}

Nonlinearity in \acp{H2RG} is primarily caused by the behaviour of the photodiode's junction capacitance. As charge accumulates in a pixel during exposure, the depletion region of the photodiode narrows, leading to an increase in junction capacitance. This change affects the conversion of accumulated charge to voltage, causing deviations from the ideal linear response. Additionally, other components in the signal chain, such as the source follower in the readout integrated circuit and the \ac{ADC} process, can introduce further nonlinearities. These combined effects result in a nonlinear relationship between the incident photon flux and the measured signal, with the nonlinearity becoming increasingly pronounced as the flux level rises, independent of wavelength \citep[see e.g.,][]{Plazas2017,Barbier2018}.

We calibrate the pixel-level response and nonlinearity with custom-designed \acp{LED}
\citep{EuclidSkyNISPCU,EuclidSkyNISP}. Each \ac{LED} has a narrow emission spectrum and can illuminate the \ac{FPA} with varying fluxes. For nonlinearity calibration -- conducted on-ground and in-flight -- two key measurements are required: the fitted slope image \citep{Kubiketal2016} that represents the nonlinear response, and the individual MACC group images that are used to approximate the linear flux \( F_{\text{Linear}} \) \citep[e.g.,][]{Vacca2004}. For both photometry and spectroscopy readout modes, MACC(4,16,4) and MACC(15,16,11) respectively, five fluence levels are selected to span the dynamic range expected for each mode, accounting for their different exposure times. These measurements are then combined to derive nonlinearity calibration coefficients for each pixel.

The linear flux estimate, \( F_{\text{Linear}} \), is derived by modelling the pixel response across various fluence levels using orthonormal polynomials. The first term of this model provides an approximation of the linear flux, as fully described in Kubik et al.\ (in prep.). Orthonormal polynomials were chosen for their practical advantages, including computational efficiency and manageable memory usage, and this approach has been successfully adopted for correcting \acp{H2RG} at several facilities. In the long term, more sophisticated models, such as a principal-component analysis, will be 
evaluated to enhance nonlinearity corrections \citep[see also][]{Rauscher2019}. Before fitting, the ramps are corrected for reference pixels using sliding windows of four pixels. To optimise memory and CPU usage, the ramps, observed multiple times at each fluence level, are averaged for each run. Additionally, frames exceeding $60\,000$\,ADU are rejected to avoid saturation effects.

Following preprocessing, the nonlinearity per pixel is expressed as a 4th-degree  polynomial relationship between the estimate of the linearised flux \( F_{\text{Linear}} \) and the measured nonlinear flux \( F_{\text{Nonlinear}} \)
\begin{equation}
F_{\text{Linear}} = \sum_{k=0,4} a_k F_{\text{Nonlinear}}^k\,.
\end{equation}
The degree of the polynomial is constrained by the number of available fitting flux levels, and for consistency purposes we enforce the linear flux to be zero for a zero nonlinear flux. 

Fitting these polynomials to the thermal vacuum (TV1) calibration data, we measured the nonlinearity coefficients and their covariances, which are subsequently used in the \ac{NIR PF} nonlinearity correction for Q1. The validity range is dictated by the dynamic range covered by the \ac{LED} fluences (1000\,ADU to 30\,000\,ADU), below which nonlinearity effects are minimal and above which corrections are applied by linearly extending the polynomial fit rather than using the higher-order coefficients. It should be noted that the values above already include baseline removal. Considering the baseline level shown in \cref{fig:baseline}, we expect saturation to occur between 40\,000 and 50\,000\,ADU for most of the pixels. For the first data release DR1, expected in October 2026, additional fluences from \ac{PV} and calibration observations are expected to enlarge the validity range. The correction step also incorporates gain conversion, and a flagging of pixels where the fit is unreliable. In addition, pixels whose signal exceeds the saturation threshold defined from the ground-test campaign are identified by this \ac{PE} and flagged using both {\tt SATURATION} and {\tt INVALID} bits.


\subsubsection{Persistence masking}
\label{sec:persistence}

\paragraph{Persistence model - calibration product from ground characterisation }
Charge persistence is the effect of electrons being trapped in detector defects, and being released again on time scales of minutes and hours \citep[see e.g.,][]{smith2008,Leisenring2016,tulloch2018}. This leads to ghost images of previous exposures 
in subsequent exposures, affecting photometric and spectroscopic measurements. \ac{NIR PF}  masks persistence using an effective persistence model calibrated on ground-characterisation data \citep[cf.\ Kubik et al., in prep.,][]{Kubik_SPIE_2024} for illuminations below saturation.

During the TV1 characterisation campaign, a standard protocol for persistence calibration was used. It consisted of a set of flat field exposures of 87\,s with increasing fluences. Each of the flat exposures was followed by a dark exposure of 430\,s. 

The signal in the flat fields was estimated by the linear fit to the first 50 frames of the ramp. The subsequent frames were discarded from the fit to avoid nonlinearity effects. Dark exposures were used to measure the persistence signal accumulated in exposure times that match those in the  ROS \citep{EuclidSkyNISP}. The first 26 or 175 frames of the dark exposures were discarded to account for the persistence decay during the dither time or slew time, respectively, between two consecutive exposures in the \ac{ROS}. The remaining frames were fitted using the on-board signal estimator of \cite{Kubiketal2016}. The photometric readout mode, MACC(4,16,4), was used to estimate persistence in photometric exposures. To estimate persistence in spectroscopic exposures MACC(7,16,1) with 16 frames per group was used, that was the maximum possible, since the dark ramps during TV1 persistence testing were too short to compute persistence in the spectroscopic exposure time.  In each case, the obtained slopes were multiplied by the corresponding total exposure time, converted to electrons using the gain per channel map. The dark-current  was also subtracted . All coefficients are estimated per pixel, since the persistence amplitudes have high spatial variability \citep{Kubik_SPIE_2024}. 

The calibrated effective persistence model yields the persistence charge $Q_\mathrm{P}$ as a function of the previous illumination amplitude $S$ using the expression
\begin{equation}\label{eq:pers}
Q_\mathrm{P}(S) = \left( \alpha + \beta S \right)\left( 1 - \mathrm{e}^{-S/\gamma} \right) \,.
\end{equation}
Separate sets of coefficients $\alpha,\beta,and \gamma$ are needed to describe persistence in photometric and spectroscopic exposures, due to their different exposure times.

This model is effective in the sense that the persistence decay during one exposure is built in the coefficients of the model, but there is no explicit time dependence of the persistence decay. 
Other shortcomings of this model are that  it does not take into account the superposition of persistence from multiple bright exposures, and it does not account for charge trapping, resulting in a potential lack of signal for bright sources on pixels that have previously seen a low-background exposure.

\paragraph{Persistence masking procedure}

\begin{figure*}
\centering
\includegraphics[width=\linewidth]{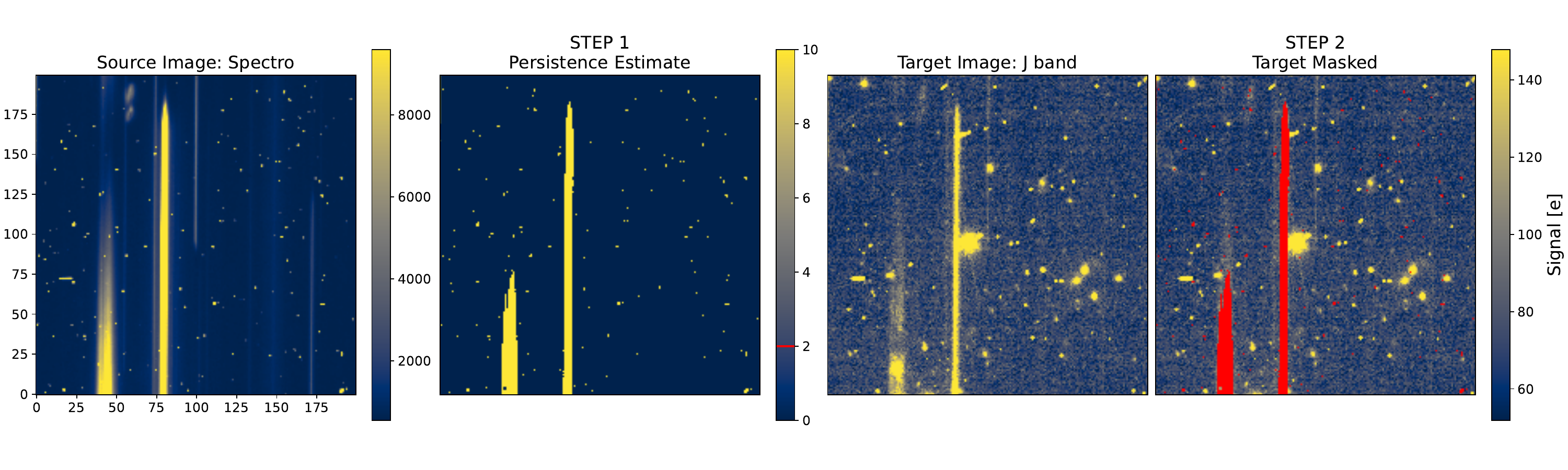}
\caption{Persistence masking scheme. In the first step, the persistence estimate is calculated using the persistence model applied to prior exposures (source images) for each target exposure. In the second step, the persistence estimates are thresholded to identify and mask pixels where the persistence charge exceeds 2\,e$^-$ (indicated in red in the right image). Typically, the percentage of masked pixels does not exceed 2\%.} 
\label{figure:persis}
\end{figure*}

There are two separate steps to mask persistence, presented schematically in \cref{figure:persis}.
The first step consists of the estimation of persistence charge contributing to any of the NISP scientific exposures. It is a separate pipeline executed before the proper NIR data reduction, in which for each exposure the persistence charge originating from previous images is estimated and stored as a separate product (persistence estimate image). 
Taking into account the limitations of the effective model, for each photometric exposure only the persistence charge from one previous spectroscopic image is computed.  To account for the persistence decay over one dither, the model computes $Q_\mathrm{P}$ using the calibrated set of parameters in \cref{eq:pers} -- this is the prediction of the persistence from a spectroscopic image to the \JE-band image 
-- and a rescaling factor is applied to estimate the persistence contribution to the \HE- and \YE-band images taken next in time. We note that for Q1 processing the persistence contribution originating from photometric exposures is not taken into account because, given the \Euclid \ac{ROS}, that would result in masking most of the bright sources. A more accurate model to allow for a correction procedure is being developed. Likewise, the computation and masking of persistence in spectroscopic exposures is deactivated for Q1.

The second step is the persistence masking executed during the NIR scientific pipeline. In this task, for each photometric exposure, the corresponding persistence image is analysed. If the persistence charge amplitude in a pixel exceeds 2\,e$^-$, then a {\tt PERSIST} and {\tt INVALID} flag is raised in the \ac{DQ} layer of the science exposure (\cref{table:flag_descriptions}). For some pixels the persistence coefficients could not be calibrated, and consequently the persistence charge could not be estimated. In that case, the {\tt PERMODFAIL} flag is raised in the \ac{DQ} layer of the corresponding image.


\subsubsection{Dark current}
\label{sec:darkcurrent}

The master dark frames were updated from ground-based (TBTV) characterisation (Kubik et al., in prep.) with two blocks of measurements taken during \ac{PV} at 87\,s and 548\,s integration times, corresponding to the imaging and spectroscopic science modes, respectively. We refer to these as photo- and spectro-darks. The darks are taken after a sufficiently long pause to allow for the decay of previously built-up persistence charge. With the filter wheel in its closed position, the dependence of the measured dark current is thus purely on exposure time. Sets of 500 photo- and 100 spectro-darks were acquired to obtain a sufficient \ac{SNR}. The \ac{PE} that generates the master dark uses an iterative procedure to reduce the impact of spurious detections from the active space weather near the maximum of the 25th Solar cycle.

To estimate pixel noise performances, the \ac{MAD} is measured on the array of pixel values $a_i$ (the 0th axis of the datacube of slope images), where the \ac{MAD} is defined as the median (med) of $\abs{a - {\rm{med(}}a{\rm{)}}}/c$.  The 
normalisation constant is $c \approx 0.6745$.  The \ac{MAD} is used to set the noise threshold, outside of which values are masked from the array $a$; that is $a >$ med($a$) $\pm$ $n_{\rm MAD} \times\,$MAD are masked.  However, They are not removed since this is only a noise-performance-selection step needed for the subsequent dark detection. Simulations using ground-based dark currents including readout noise and varied \ac{CR} fluences in the Earth-Sun L2 orbit set the optimum value of $n_{\rm MAD}$ at 3.0. 

After masking to the \ac{MAD}-based threshold, the median of $a$ is taken again, and the result is evaluated for detection of the dark current above the \ac{SNR} given by the two conditions $\abs{{\rm{med(}}a{\rm{)}}} \leq$ RMS/$\sqrt{N_{\rm{fr}}}$ and RMS/$\sqrt{N_{\rm{fr}}} \leq$ 0.3$\,\times\,\sqrt{480/N_{\rm{fr}}}$\,, where RMS is the standard deviation and $N_{\rm{fr}}$ is the number of measurements in each pixel meeting its initial threshold; for example, $N_{\rm{fr}} \leq 500$ in the case of the \ac{PV} photo-darks.  The final median dark currents and their variances comprise the master dark product, along with a \ac{DQ} flag set for pixels whose dark currents could not be detected to within the two noise conditions.  This flag, {\tt NODARKDET}, is informational and does not necessarily represent an invalid condition. This is because some pixels in \acp{H2RG} have such a low dark current  that the slopes computed are dominated by the readout noise, and for example, also a floor of weak \ac{CR}-induced signal not completely compensated by the masking 
steps -- an issue demonstrated with simulations. Strictly speaking, however, such pixels cannot be distinguished from those with sufficiently high noise arising from multiple sources that their dark currents are hidden from detection.  Thus, for all flagged pixels the dark current is set to zero and these are excluded from the detector dark-current performance 
statistics; see \cref{table:badpixdarks} for the most probable values of the dark current applicable to Q1. 


\subsubsection{Cosmic-ray identification}
\label{sec:cr}

\paragraph{\textit{Single-frame photometric mode}}

The first step in the identification of \acp{CR} in the NISP photometric data is performed by the module {\tt NIR\_CrRejectionSingleFrame}. This module utilises the {\tt LAcosmic}\footnote{\url{https://github.com/cmccully/lacosmicx}} algorithm \citep{vandokkum2001} to locate \acp{CR} in single photometric exposures using  Laplacian edge detection. The algorithm has several input parameters that strongly depend on the instrument and conditions of the observations. A balance has to be found between identifying a large fraction of the \acp{CR} without masking real (point) sources. This is especially important for NISP, which strongly undersamples the \ac{PSF} \citep{EuclidSkyNISP}. For each of the three NISP filters, the optimal set of input parameters were found based on simulated images, for which the ground truth is known. In this tuning, the priority was to avoid masking real objects, since \acp{CR} not identified in this module can be picked up by the multi-frame \ac{CR}-rejection module (see below). The efficiency of this module is higher for the redder filter \HE\ compared to the bluest filter \YE, since undersampling is stronger in the bluer passband. 

Pixels identified by {\tt LAcosmic} as \ac{CR}s are masked by adding the flags {\tt COSMIC} (bit 16) and {\tt INVALID} (bit 0) to the \ac{DQ} layer. We also mask all eight neighbouring pixels to account for \ac{IPC} that spreads charge from a pixel to its neighbors, which then also need to be masked in the case of a \ac{CR}. For details about \ac{IPC} see \cite{LeGraet22} and \cite{EuclidSkyNISP}.

\paragraph{\textit{Multi-frame photometric mode}}

This is the second step in identifying and flagging \acp{CR}, using information from all available dithers simultaneously and implemented in the {\tt NIR\_CrRejectionMultiFrame} module. This task requires precise astrometric calibration (see \cref{sec:astrometry}). Each position in the sky, corresponding to a pixel on the array, is observed with $n$ dithers. In the EWS, the \ac{ROS} dither pattern results in fractions of 1\%, 8\%, 40\%, and 42\% for $n=1$ to 4, respectively \citep[see table 2 in][]{Scaramella-EP1}. We use pull-clipping for outlier detection that is robust on such small samples ($n\geq3$ unflagged data points). More specifically, we reject pixels with the pull $p_i\ge10$, where $p_i$ is defined as 
\begin{equation}
p_i = \frac{(x_i - \Bar{x_i})}{\sqrt{\sigma_i^2 - \Bar{\sigma}_i^2}}\,.    
\end{equation}
Here, \(x_i\) and \(\sigma_i\) are the respective flux and measurement error of dither $i$, and \(\Bar{x}_i\) and \(\Bar{\sigma}_i\) are the inverse-variance weighted average and the associated sample error without point $i$. We note that for the monthly visits of \Euclid's self-calibration field \citep[see Sect.~4.2.3 in][]{EuclidSkyOverview} near the \ac{NEP} we have coverage of up to $n=76$, and that for $n<3$ we fall back to the single-frame mode explained above.

To counter false positive \ac{CR} detections -- for example stars erroneously identified as \acp{CR} -- we compute the sum of pixel values in the 3\,$\times$\,3 pixel region around the candidate \ac{CR}. The same sum is computed for the other dithers. The \ac{CR}-detection is rejected if the sums agree to within 3\%. All detected \acp{CR} are flagged in the \ac{DQ} layer.

For testing purposes, or in the case of a malfunction in specific scenarios, standard
sigma-clipping or Grubb's test can be enabled in \ac{NIR PF}. This estimates the $z$-score, 
\begin{equation}
z_i = \frac{(x_i - \Bar{x})}{s}\,,    
\end{equation} 
where \(\Bar{x}\) and $s$ are the sample mean and standard deviation, respectively. For small $n$, this method has the disadvantages that it includes outliers in the sample estimates, and that it does not fully account for individual measurement errors \(\sigma_i\). A comparison with the pull-statistic for small sample sizes is given in \cite{YCopin}.
Since $n\leq4$ for the EWS, we adopted the pull clipping as default statistic for the \ac{NIR PF}.

\paragraph{\textit{Single-frame spectroscopic mode}}

\begin{figure}
\centering
\resizebox{\hsize}{!}{\includegraphics{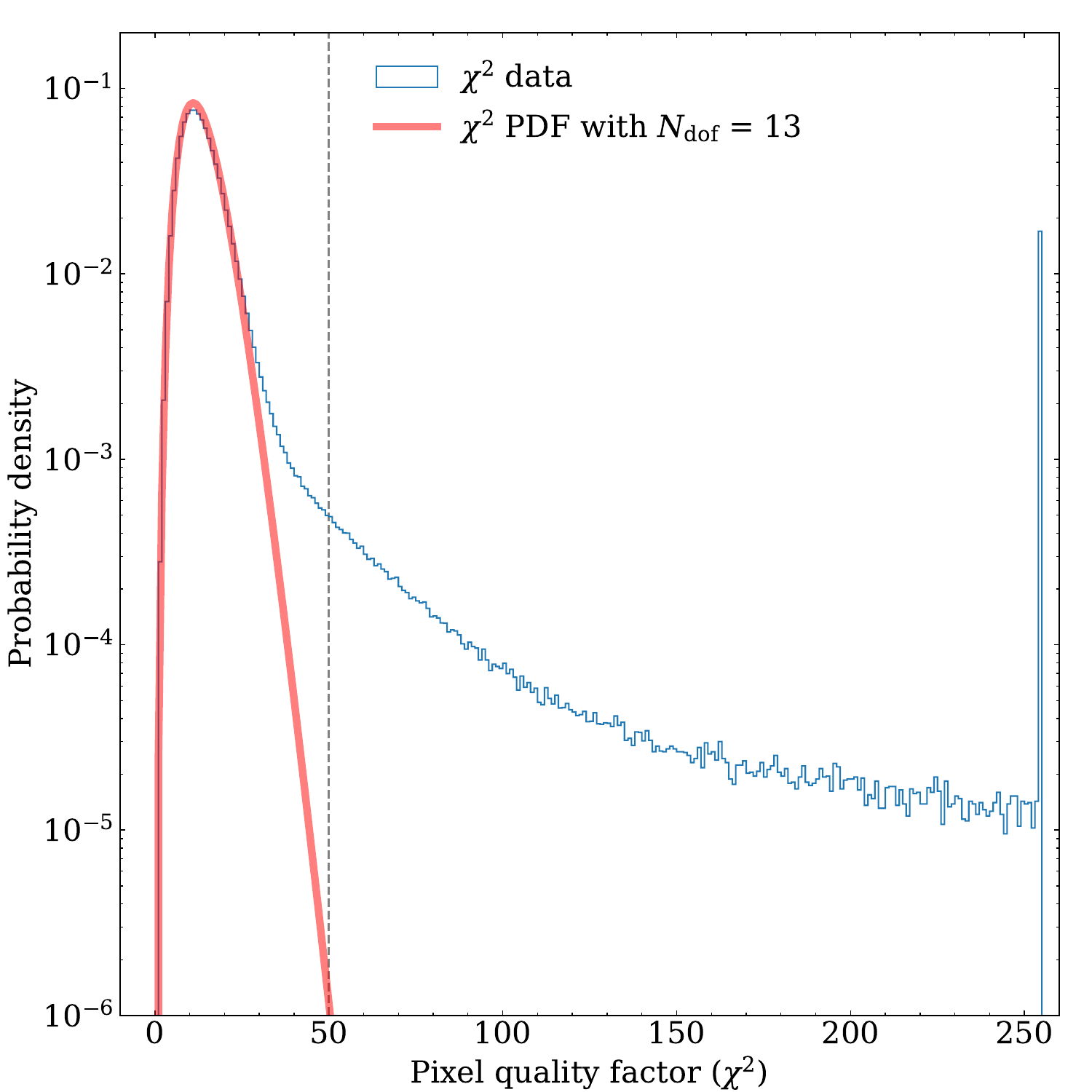}}
\caption{Comparison between the \ac{PDF} of the pixel quality factors ($\chi^2$) for a spectroscopic MACC(15,16,11) exposure for one of the NISP detectors (blue histogram), and the \ac{PDF} calculated for a $\chi^2$ distribution with 
${N_{\rm dof}}=13$ (red line). The threshold $\chi^2=50$ (dashed vertical line) distinguishes good- from bad-quality pixels. The $\chi^2$ value is encoded on-board as an 8-bit integer and thus has an upper limit of 255. Hence, the last bin corresponding to $\chi^2=255$ includes not only \acp{CR}, but all other pixels with a poor slope fit, such as bad pixels.}
\label{figure:chi2}
\end{figure}

Each MACC(15,16,11) grism exposure is associated with an 8-bit image of quality factors containing the $\chi^2$-statistics of the on-board slope-fit \citep[see][]{Kubiketal2016}. The current version of the {\tt SIR\_CrRejectionSingleFrame} module in \ac{NIR PF} uses this $\chi^2$ value to identify \acp{CR}. \Cref{figure:chi2} shows the $\chi^2$ distribution for one detector plotted together with the \ac{PDF}  with degree of freedom ${\rm DOF}=13$. According to \citet{Kubiketal2016}, $N_{\rm dof} = n_{\rm g}-2$, where $n_{\rm g}=15$ is the number of groups 
in the nominal \ac{MACC} spectro-mode. \Cref{figure:chi2} also indicates the threshold (50 counts) to detect bad pixels. From simulated data we found that the $\chi^2$ implementation identifies at least 99\% of the \acp{CR}.

Alternatively to using $\chi^2$, the single-frame spectroscopic mode can also run {\tt LAcosmic} as implemented for photometry in {\tt NIR\_CrRejectionSingleFrame}, albeit with different configuration parameters. We find that combining {\tt LAcosmic} with the $\chi^2$ approach does not increase the completeness fraction of detected \acp{CR}, and therefore {\tt SIR\_CrRejectionSingleFrame} defaults to the $\chi^2$ method alone.


\subsection{Astrometric and photometric calibration of NISP images}
\label{sec:astrophoto}

\begin{figure*}
\centering
\resizebox{\hsize}{!}{\includegraphics{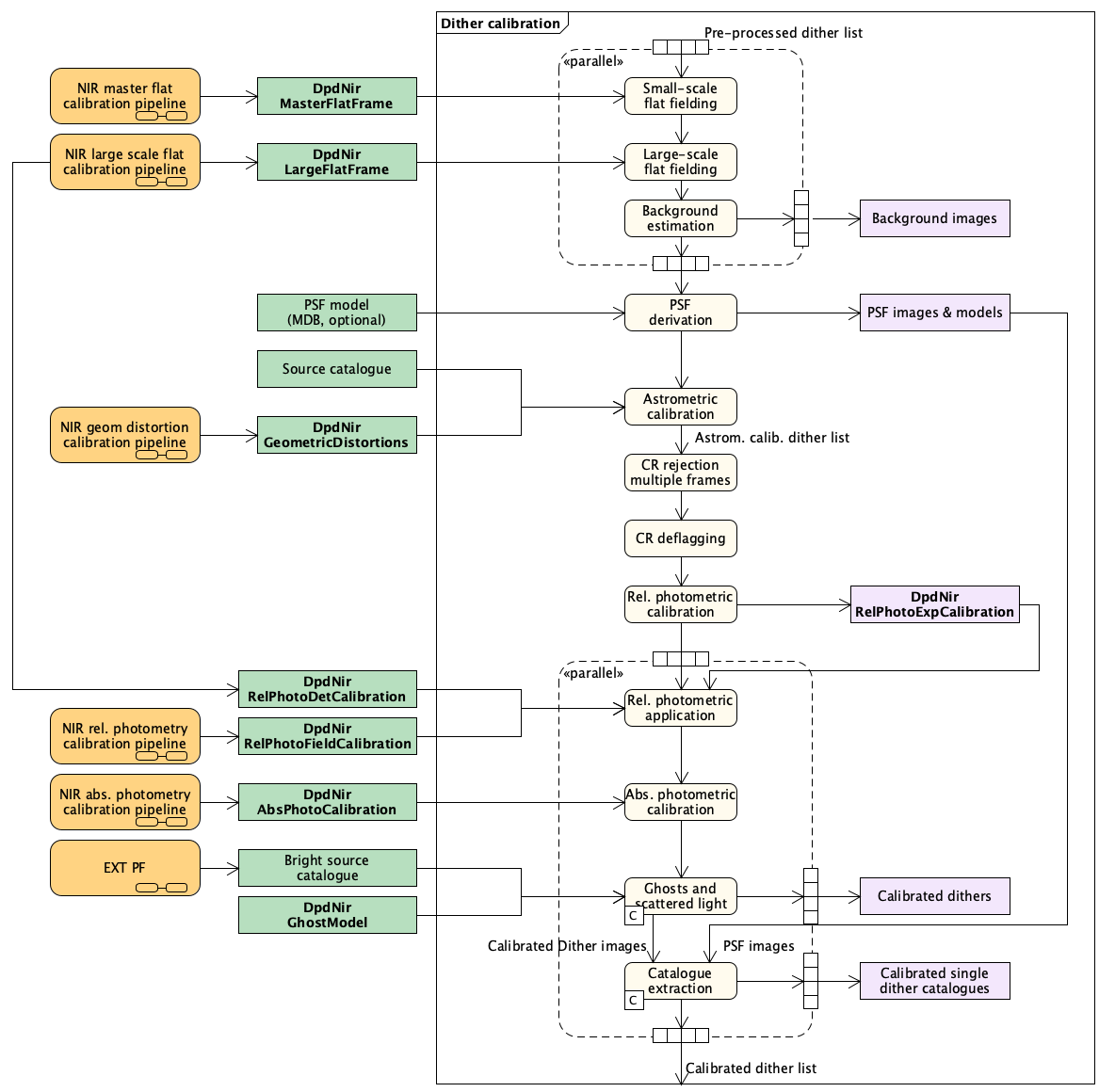}}
\caption{\ac{NIR PF} astrometric and photometric calibration \ac{PE}s. Input products are shown as green boxes on the left. Some of them are taken from the \ac{MDB}, while other products are generated by dedicated processing pipelines represented as orange boxes.}
\label{fig:astrophotocalib}
\end{figure*}

In this section we describe the \ac{PE}s in NIR PF that derive the astrometric and photometric calibrations (\cref{fig:astrophotocalib}).

\subsubsection{PRNU correction using LED flat fields}
\label{sec:flat}

\begin{figure}
    \centering
    \includegraphics[width=0.5\textwidth]{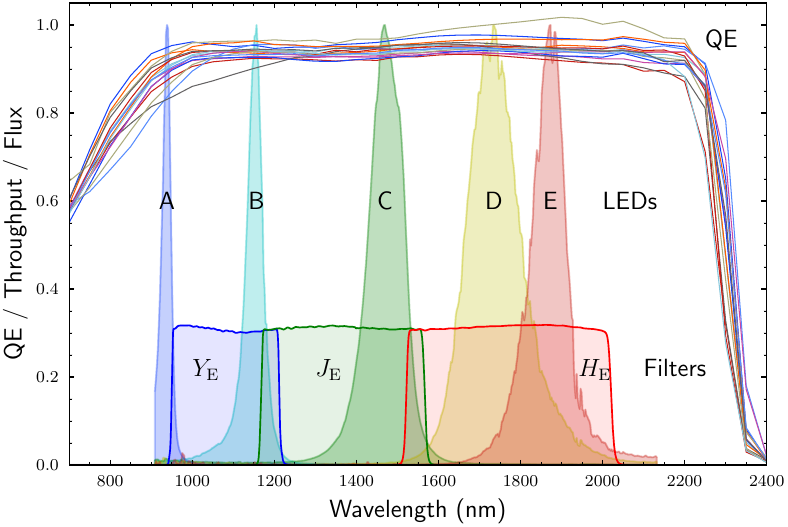}
    \caption{Comparison of the average \ac{QE} curves of the 16 NISP detectors, 
    the normalised \acp{SED} of the \acp{LED}, and the \YE, \JE, and \HE\ passbands (reduced by a factor 3 in height to avoid cluttering near the top of the plot).}
    \label{fig:led_spectra}
\end{figure}

The detectors' \ac{PRNU} is corrected with flat fields. The ideal \ac{PRNU} correction would be given by the relative detector response to a uniform illumination from a continuum white-light source. NISP, however, uses \acp{LED} with a near-monochromatic \ac{SED} in the near-infrared calibration unit \citep[NI-CU;][]{EuclidSkyNISPCU}. NI-CU incorporates five \acp{LED} A to E,  ordered by ascending central wavelengths of their spectra (\cref{fig:led_spectra}). It illuminates the \ac{FPA} directly from a small off-axis angle, without passing through a filter \citep[see section~2.2. of][for more details]{EuclidSkyNISPCU}. The spatial illumination approximately follows a Lambertian cosine profile \citep{EuclidSkyNISPCU}. 

A single \ac{LED} flat alone thus does not represent an ideal \ac{PRNU} correction. Therefore, after going through all pre-processing steps listed in \cref{sec:preprocessing}, we reconstruct filter-dependent flats from the individual \ac{LED} flats. Denoting the \ac{LED} flat as ${\rm FLAT_{\rm LED}}(\mathbf{x})$, with $\mathbf{x}$ being the pixel position, we compute the filter-dependent flat as
\begin{equation}
    {\rm FLAT_{Filter}}(\mathbf{x}) = {\rm FLAT_{\rm LED}}(\mathbf{x})\; P(\mathbf{x})\;.
    \label{eq:flatreconstruction}
\end{equation}
Here,
\begin{equation}
  P(\mathbf{x}) = \frac{\int_{\lambda_1}^{\lambda_2} \mathrm{QE}(\lambda,\mathbf{x})\;T(\lambda,\mathbf{x})\;{\rm d}\lambda}
  {\int_{\lambda_3}^{\lambda_4} \mathrm{QE}(\lambda,\mathbf{x})\;{\rm d}\lambda}
  \label{eq:propagator}
\end{equation}
propagates the \ac{LED}'s \ac{SED} across the detector's \ac{QE} to reconstruct a broad-band flat from a near-monochromatic light source. $T(\lambda,\mathbf{x})$ is the filter transmission curve, $[\lambda_1,\lambda_2]$ is the passband of the filter, and $[\lambda_3,\lambda_4]$ is a representative wavelength coverage of the \ac{LED} spectrum. Notably, due to the \acp{LED}' narrow \ac{SED}, $P$ is independent of the \ac{LED} spectrum. The mathematical derivation of \cref{eq:flatreconstruction,eq:propagator} is given in \cref{app:filter_flats}. The data for \ac{QE} and $T(\lambda)$ used for the Q1 data processing were measured on the ground \citep{Schirmer-EP18}. We note that $P$ is in principle time-dependent over mission-duration due to radiation damage and molecular contamination \citep{Schirmer-EP29}, but in the context of the singular Q1 observations this dependence can be neglected. 

Of the five \acp{LED} available, only \acp{LED} B to D are used to construct the broad-band flats. Given the mutual wavelength coverages, we use \ac{LED} B for filter \YE, C for \JE, and D for \HE in \cref{eq:flatreconstruction}. In principle, several \acp{LED} can be combined to reduce uncertainties from the propagator $P$, but the \ac{LED}-flat \ac{SNR} is so high that there is currently no need to do so, mainly because the \ac{QE} curves are very flat across the passbands, and the \ac{SNR} in the \ac{QE} maps themselves is also very high. 

Each of the three filter flats is computed from 30 \ac{LED} flats. In addition to the bad pixels identified by the pre-processing steps, the filter flats require further masking. The first type is vignetting from a baffle near the \ac{FPA}, which affects the edges of detectors \texttt{11}, \texttt{14}, \texttt{24}, \texttt{34}, and \texttt{44} for \ac{LED} frames \citep[see section~2.3. of][]{EuclidSkyNISPCU}. Pixels in the vignetting regions are flagged as {\tt VIGNET} and their values are set to 1.0 in the master flat. The second type of pixels masked are `flower' pixels, which typically present a cross-shaped pattern with a faint core (`flower centre') and bright adjacent pixels (`flower petals'). The same type is found in \ac{JWST} images, where they are referred to as `open' and `adjacent to open' pixels \citep{2014PASP..126..739R}. Flower pixels are masked as {\tt FLOWER} and {\tt INVALID}. We detect these pixels in the flat exposures using a Laplacian kernel.
Finally, pixels with values that deviate from the median of the 30 frames by over 50\% are flagged as {\tt FLATLH}, indicating too-low or too-high values without further discrimination.

\begin{figure*}
\centering
\begin{minipage}[t]{0.98\textwidth}
\centering
\includegraphics[width=5.9cm]{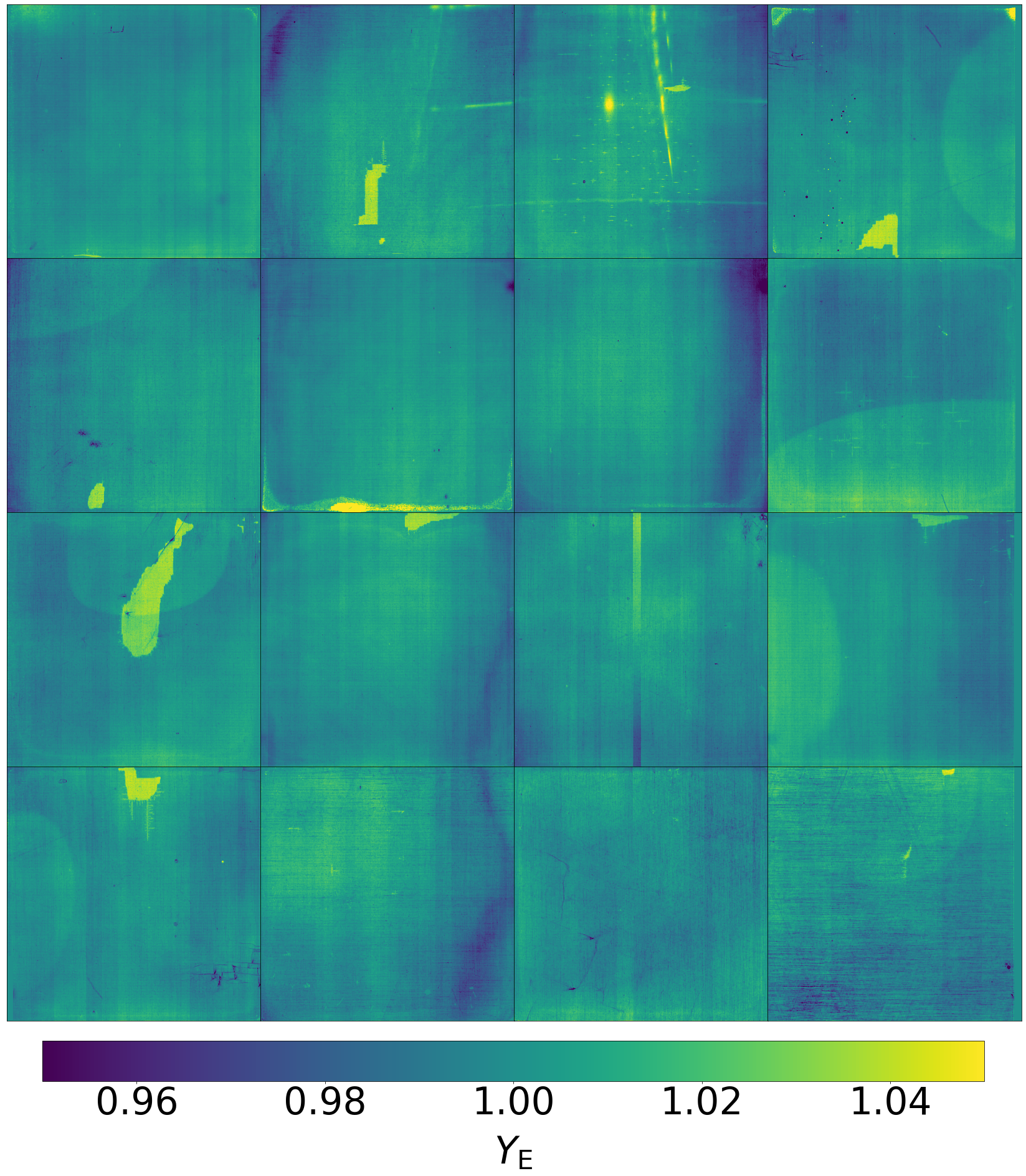}
\includegraphics[width=5.9cm]{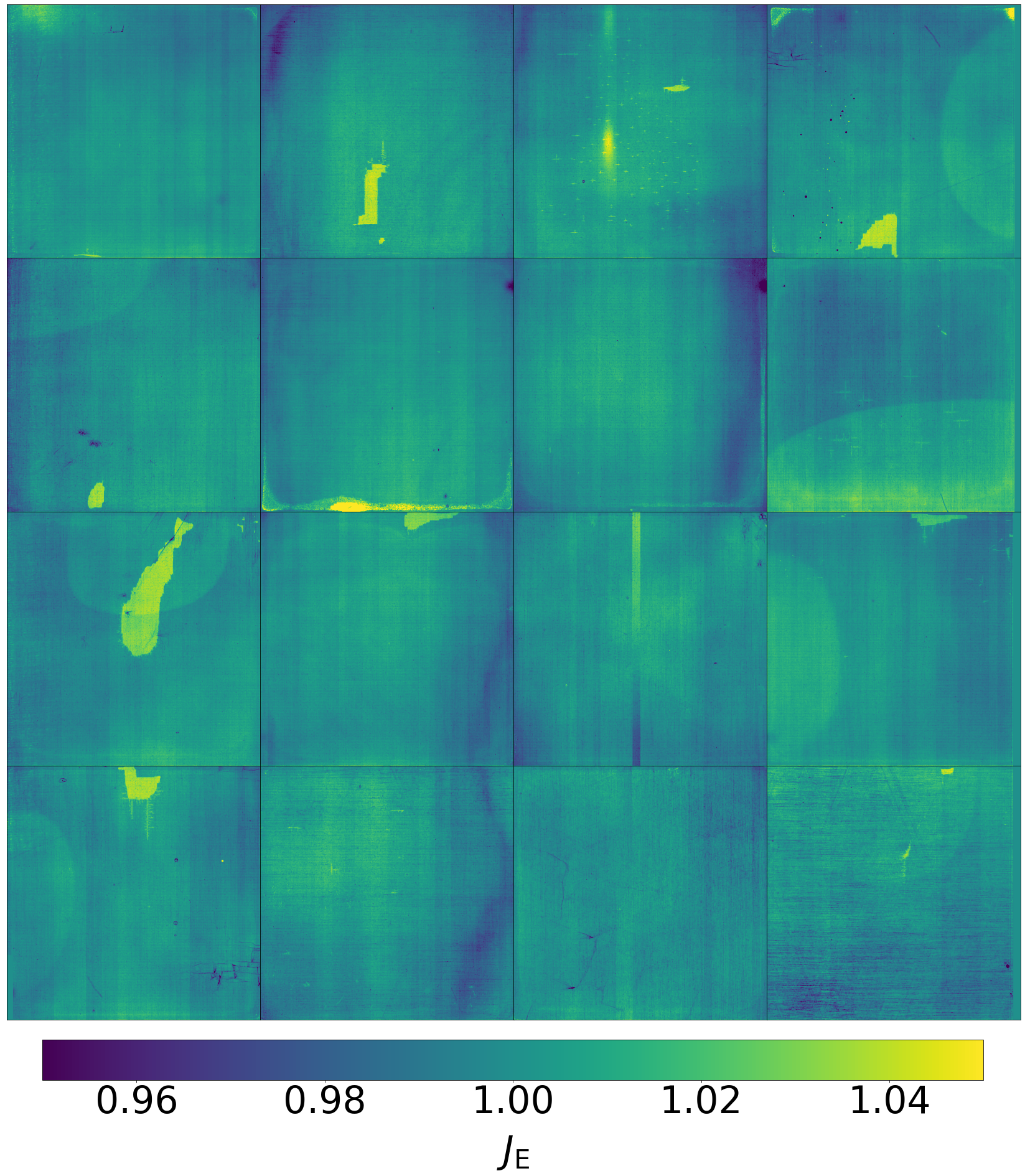}
\includegraphics[width=5.9cm]{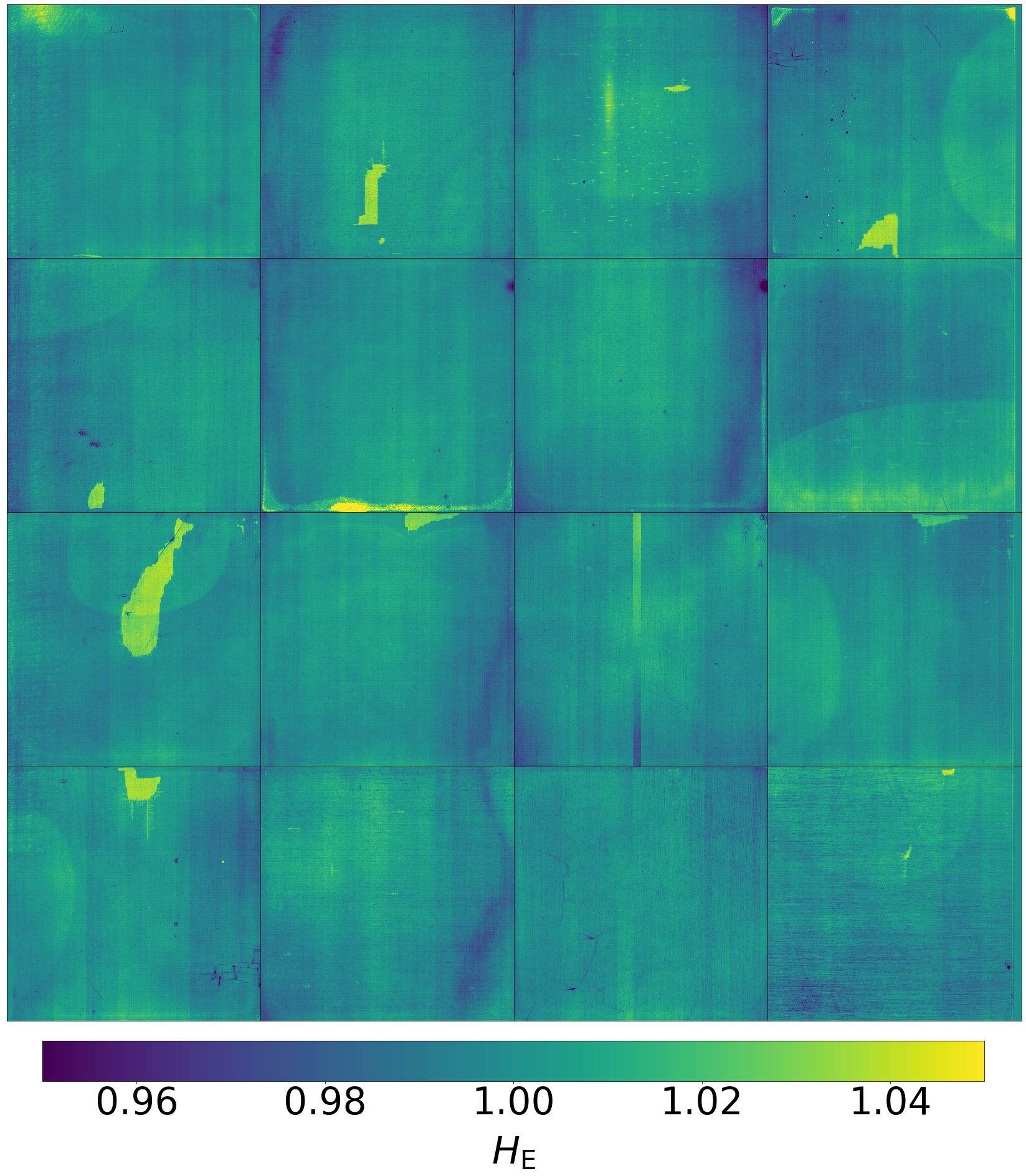}
\end{minipage}
\caption{Master \ac{PRNU} flats for \Euclid \YE, \JE, and \HE bands used for the processing of Q1 images. The uniform \ac{QE} is evident from these images.
} 
\label{figure:masterflats}
\end{figure*}

The master flats of the three NISP bands used for the Q1 processing are shown in \cref{figure:masterflats}. The \ac{PRNU} correction is done by dividing the science frames by the master flat, without prior removal of NI-CU's Lambertian illumination profile.

\subsubsection{Large-scale illumination correction}
\label{sec:largescale}

The \ac{LED} illumination profile is not removed prior to \ac{PRNU} correction (\cref{sec:flat}), introducing a multiplicative flux change across the \ac{PRNU}-corrected images on the order of 10\%. While \Euclid's telescopic beam is free of vignetting, non-uniformities in the illumination of the \ac{FPA} still arise. For example due to the cosine-fourth-power-law of illumination \citep{reiss1945}, modified by \Euclid's off-axis optical design \citep{EuclidSkyOverview}, and by angle-of-incidence dependencies of the mirror reflectances. Jointly, these are expected to be on the order of 1--2\%.

A photometric flat is required to correct for these non-uniformities on scales larger than about 100\,pixels; below that scale, the \ac{LED} flats can be assumed to be uniform apart from \ac{PRNU}. The photometric flat, or large-scale flat, is computed by measuring how the fluxes of stars change when they move to different \ac{FPA} locations in the 76 widely-dithered exposures of \Euclid's self-calibration field. In \ac{NIR PF}, the {\tt NIR\_SelfCalib} processing element performs this computation, and also applies it to the science images. This also includes a determination of the relative, average detector-to-detector sensitivity variations.

We note that the internal computations are performed in magnitudes, not in linear fluxes. Specifically, we consider observations of the same source in different locations on the \ac{FPA}, and assume the observed magnitude is the sum of three contributions: the source actual apparent magnitude; an additional term due to the detector non-flat sensitivity; and a third term due to the detector average sensitivity,
\begin{equation}
    m_{i,d}(x,y) = m_i + {\rm LSF}_d(x,y) + s_d\,.
    \label{eq:LSF_model}    
\end{equation}
Here, $m_{i,d}(x,y)$ is the magnitude of the $i$-th source as observed on position $(x,y)$ of detector $d$, $m_i$ is the magnitude of the source as observed by a \ac{PRNU}-corrected detector, ${\rm LSF}_d(x,y)$ is the detectors's large-scale flat, and $s_d$ its average sensitivity. In \cref{eq:LSF_model}, the quantities on the right represent the model to be fit to the empirical data on the left, within the uncertainties. In particular, $m_i$ are nuisance parameters necessary for the fit but not interesting for our purposes, while both ${\rm LSF}_d(x,y)$ and $s_d$ are relevant to apply the large-scale-flat correction. We note that ${\rm LSF}_d$ and $s_d$ are strongly degenerate, and should be determined from the same analysis and data set. All quantities on the right-hand side are constrained up to a constant term, hence we normalise
${\rm LSF}_d$ and $s_d$ such that they have zero mean across the \ac{FPA}.

\begin{figure*}[hbt!]
\centering
\includegraphics[width=.45\textwidth]{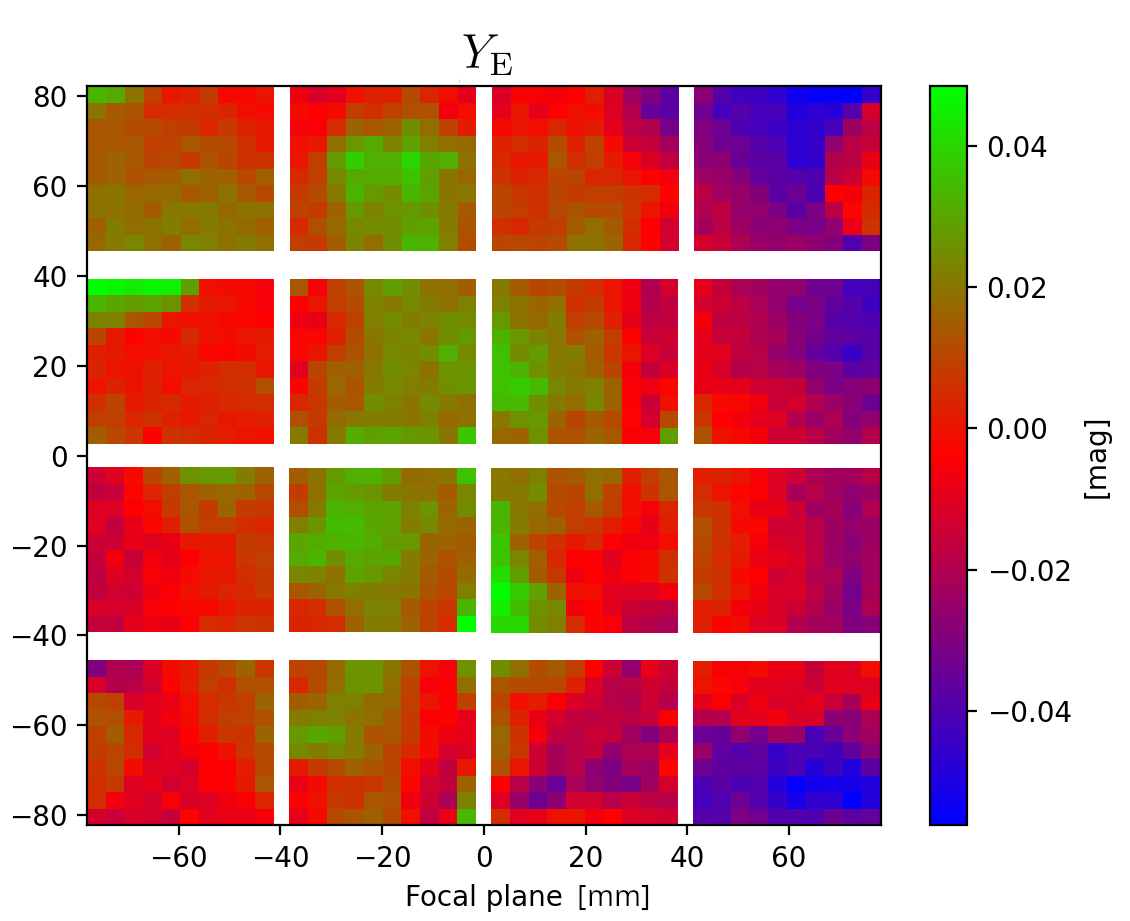}
\includegraphics[width=.45\textwidth]{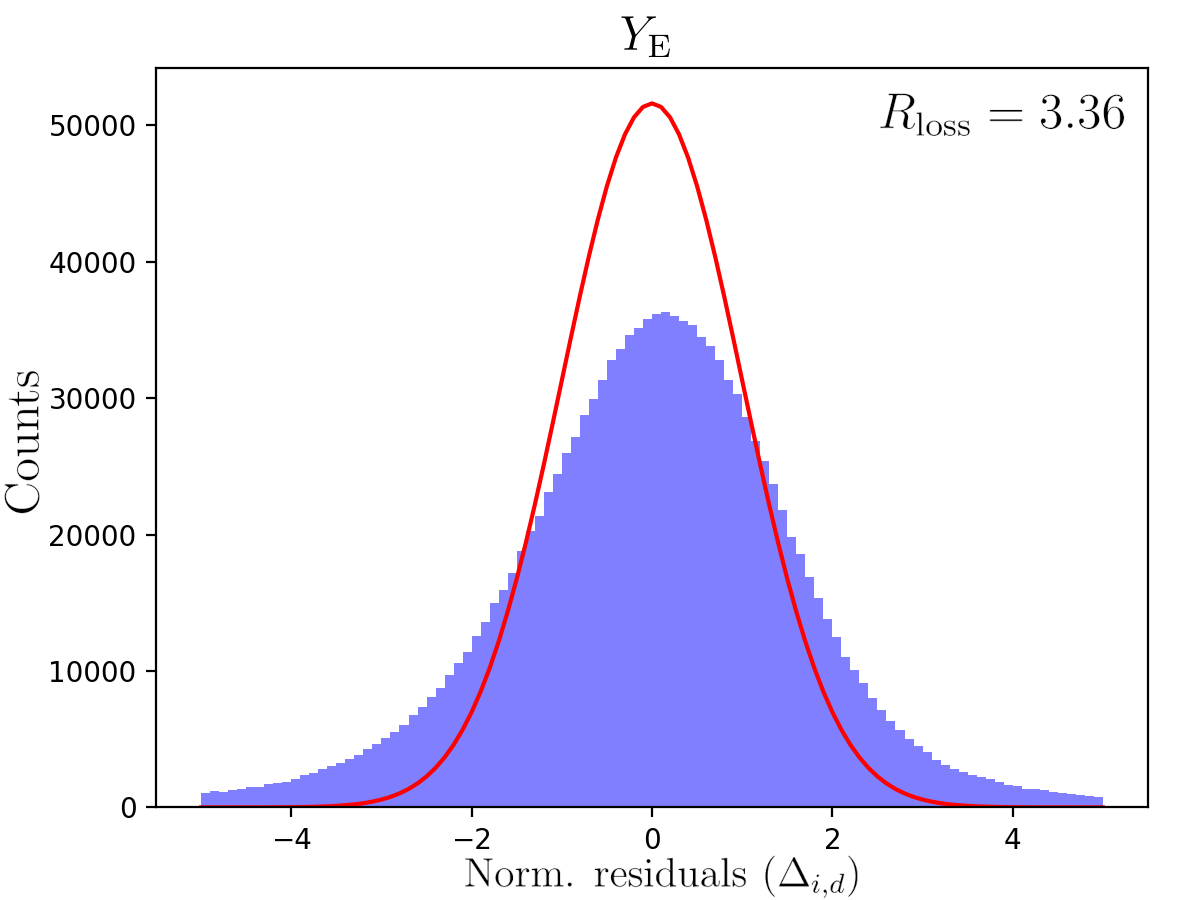}\\
\includegraphics[width=.45\textwidth]{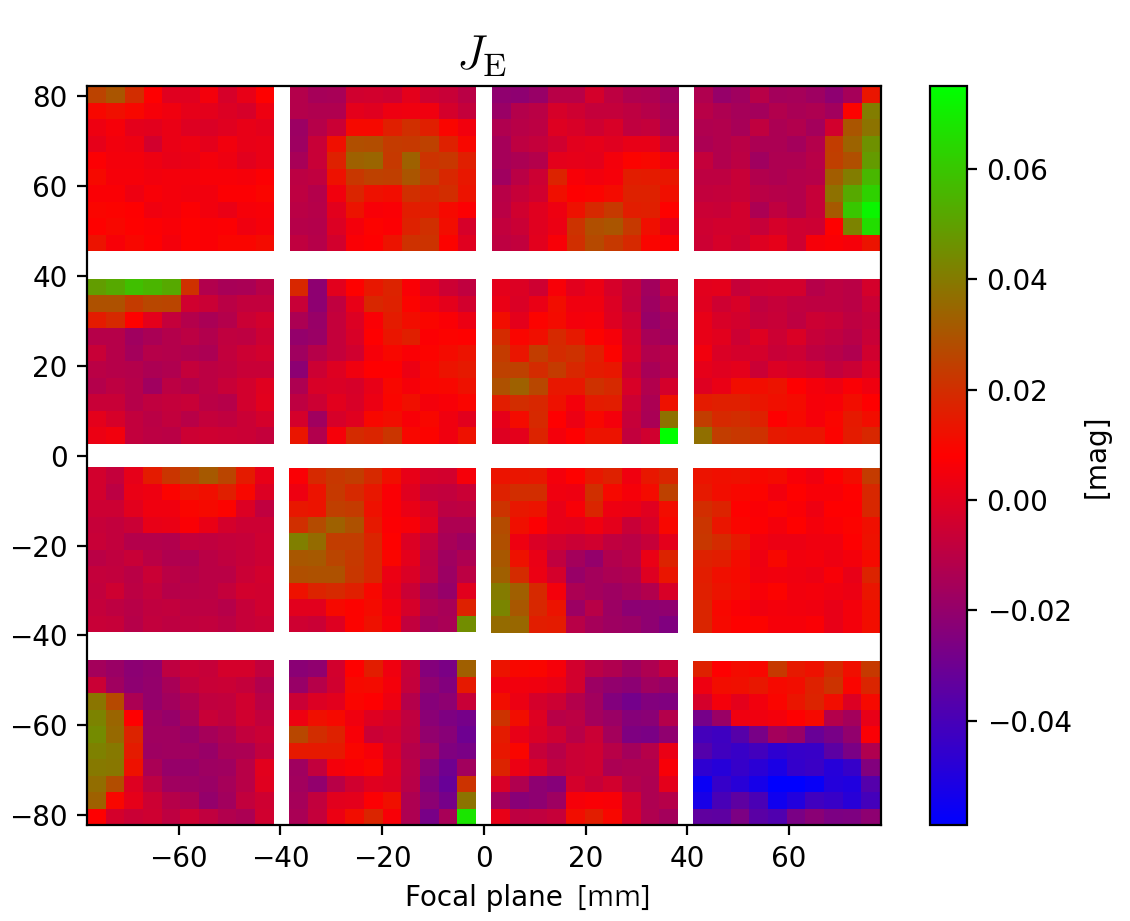}
\includegraphics[width=.45\textwidth]{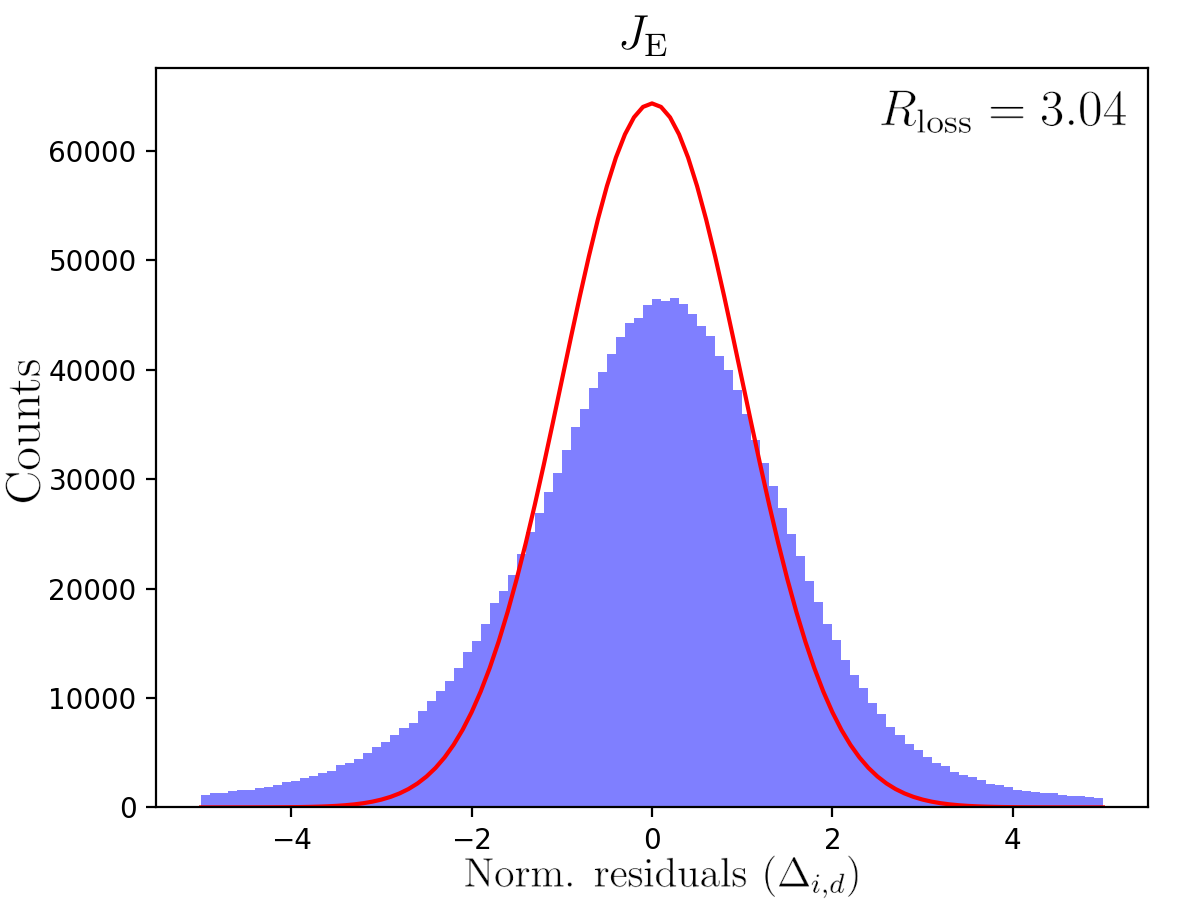}\\
\includegraphics[width=.45\textwidth]{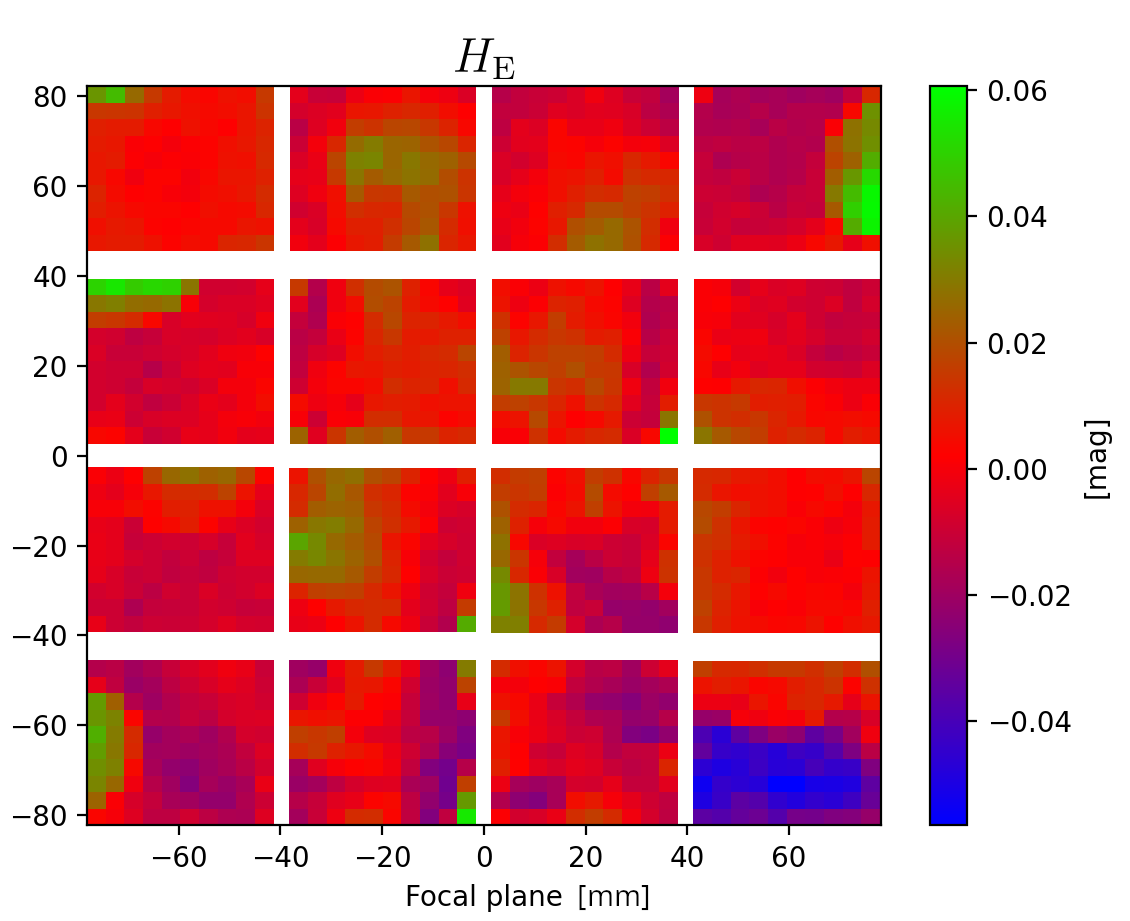}
\includegraphics[width=.45\textwidth]{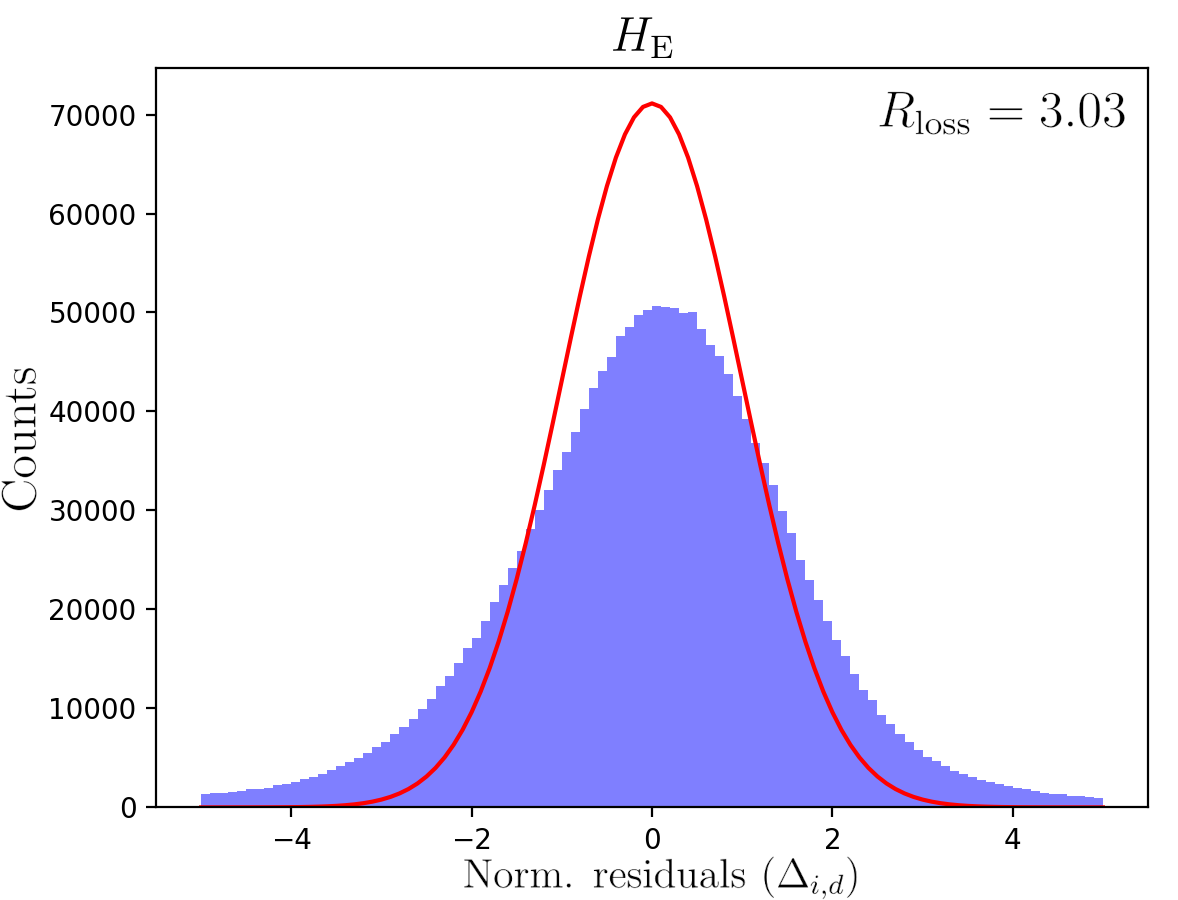}\\
\caption{
\textit{Left column}: best-fit models of the \YE, \JE, and \HE large-scale flats, ${\rm LSF}_d(x, y)$, as in \cref{eq:LSF_model}. \textit{Right column}: histograms of the normalised residuals of $\Delta_{i,d}(x,y)$ from \cref{eq:LSF_normresid}, compared to a normal distribution with mean zero and variance one (red curve). The quantity $R_{\rm loss}$ from \cref{eq:LSF_redloss} is reported in the plots.}
\label{figure:LSF_results}
\end{figure*}

To estimate a goodness of fit for the model with respect to a specific data set, we introduce the normalised residual, defined as
\begin{equation}
    \label{eq:LSF_normresid}
    \Delta_{i,d}(x,y) = \frac{m_{i,d}(x,y) - [m_i + {\rm LSF}_d(x,y) + s_d]}{\delta m_{i,d}(x,y)}\,,
\end{equation}
where $\delta m_{i,d}(x,y)$ is the uncertainty of $m_{i,d,x,y}$. For a representative model, we expect $\Delta_{i,d}(x,y)$ to be normally distributed with mean zero and variance one. Thus, a measure of the goodness-of-fit is given by the reduced loss
\begin{equation}
    \label{eq:LSF_redloss}
    R_{\rm loss} = \frac{\sum_{i,d} \Delta_{i,d}(x,y)^2}{N_{\rm obs}}\,,
\end{equation}
where $N_{\rm obs}$ is the number of observations considered. A value $R_{\rm loss}\approx1$ represents a good fit, while considerably larger or smaller 
values indicate that the model is not flexible enough (underfitting), or that it is too flexible (overfitting), respectively. \Cref{figure:LSF_results} shows the large-scale flats, ${\rm LSF}_d$, for the three NISP filters for the 
self-calibration visit on 14 June 2024, together with the histograms of $\Delta_{i,d}$ and the $R_{\rm loss}$ values.

\begin{figure}[hbt!]
\centering
\includegraphics[width=.49\textwidth]{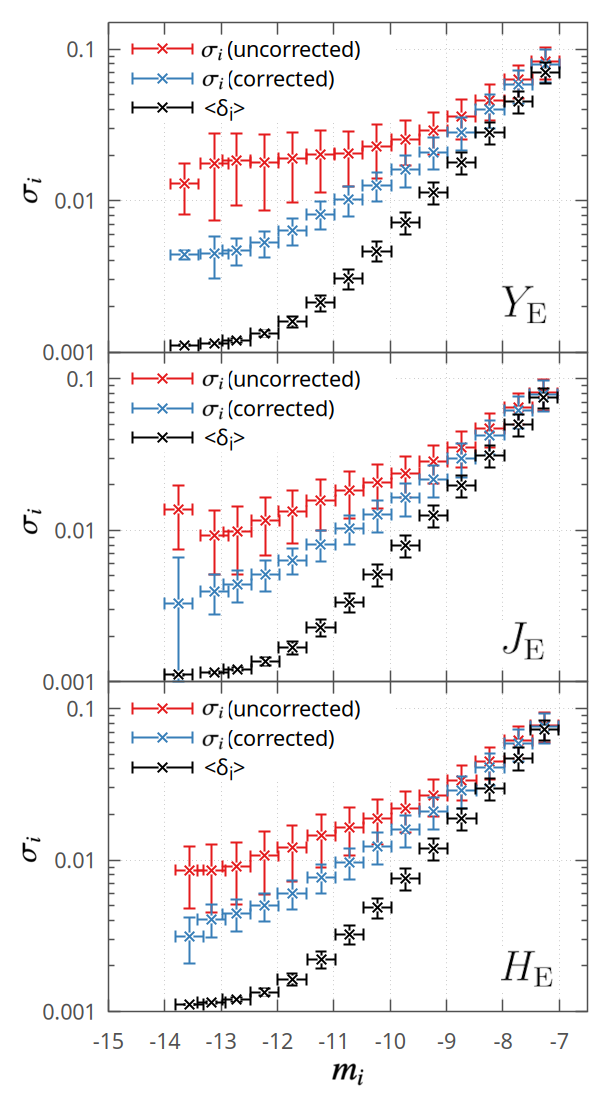}
\caption{
Comparison of the scatter in magnitude from the repeated observations of the same source (`uncorrected' $\sigma_i$, red symbols), with the same scatter after the large-scale flat ($\sigma_i$ from \cref{eq:LSF_scatter}, blue symbols), as a function of the instrumental magnitude $m_i$ for the \YE, \JE, and \HE filters.  Also shown are the average of the photometric uncertainties for references (black symbols).  The scatter introduced by the large-scale non-uniformities is represented by the separation between red and black symbols, while the blue symbols represent the residual scatter after the application of the large-scale flat calibration.}
\label{figure:LSF_corrections}
\end{figure}

An alternative measure of the goodness-of-fit is given by the standard deviation of the observed magnitudes, once corrected for the large-scale flat and average sensitivity. That is, for the $i$-th source,
\begin{equation}
    \label{eq:LSF_scatter}
    \sigma_i = {\rm std} \left[ m_{i,d}(x,y) - {\rm LSF}_d(x,y) - s_d\right]\,,
\end{equation}
where the standard deviation is calculated over all $d$, $x$ and $y$ for which we have an observation of the $i$-th source. If the LSF and average sensitivity corrections are good enough, that is if $R_{\rm loss}\approx$\,1, we expect $\sigma_i$ to be approximately equal to the average of the magnitude uncertainties for the same 
$i$-th source, that is $\sigma_i \sim \langle \delta m_{i,d}(x,y) \rangle$.  On the other hand, if the corrections are suboptimal and $R_{\rm loss}>1$, we expect $\sigma_i > \langle \delta m_{i,d}(x,y) \rangle$. \Cref{figure:LSF_corrections} shows the distributions of $\sigma_i$ versus the corresponding $m_i$ values for the three filters (in blue). We also show two more quantities. First, the distributions of the `uncorrected' $\sigma_i$, namely the standard deviation of the $m_{i,d}(x,y)$ alone (in red), which are always larger than the `corrected' $\sigma_i$, 
otherwise the use of the large-scale correction would be detrimental. And second, the distributions of the average magnitude uncertainties $\delta m_{i,d}(x,y)$ (in black), which are always smaller than the `corrected' $\sigma_i$, since the large-scale flat algorithm cannot recover the random errors intrinsic to the flux measurements.


The separation between the blue and red symbols for the brightest sources provides a hint of the photometric offsets due to the ${\rm LSF}_d$ and $s_d$ contributions, which is of the order of 0.01\,mag in all filters. These offsets are partially corrected by the application of the large-scale flat, by 0.006--0.007\,mag for the brightest sources. At instrumental magnitudes $m_i \gtrsim -9$ the photometric uncertainties become comparable to the `uncorrected' $\sigma_i$, that is, the large-scale-flat correction is negligible for the faintest sources.

We started this section by motivating the large-scale flat with the illumination properties of the telescope, and the \acp{LED}' Lambertian cosine profile, which are purely optical and geometrical effects. One would thus expect a fairly uniform 
pattern, whereas the derived large-scale flats shown in \cref{figure:LSF_results} look rather feature-rich. Some of the features, such as in the upper and lower right detectors, show almost discrete jumps between resolution elements ($200\times200$ NISP pixels). This indicates variations on scales smaller than what the large-scale flat can resolve, consistent with underfitting suggested by $R_{\rm loss}\approx3$. The inability of the model to fully resolve these variations is also evident when considering the histograms of the normalised residuals, $\Delta_{i,d}$. They are not Gaussian with a variance equal to one. \Euclid's optical design precludes such effects, hence the large-scale flat algorithm is picking up systematics in the photometry that are not fully corrected for by the preceding calibration efforts. 

We argue that these features are due to low-level, diffuse charge persistence (\cref{sec:persistence}) that is not captured by our masking efforts. Plausibly, this is caused by the alternating higher and lower zodiacal background registered by the detectors as we cycle through different filters and grisms in the \ac{ROS}. Indeed, these features have a remarkable resemblance to those seen in our  persistence-calibration data. This in itself is a remarkable confirmation that our large-scale flat implementation is working very well, because at no point is the algorithm informed about charge-persistence effects. Hence the large-scale flat correction does exactly what it is supposed to do: it sweeps up all residual effects that are left uncorrected -- or imperfectly corrected -- by the preceding calibration steps, regardless of their multiplicative, additive, or more complicated nature.

\Cref{figure:LSF_corrections} shows that the large-scale flat corrects about 6--7\,mmag of the joint systematics in the photometry. It also shows that we still lack the remaining 2--3\,mmag to bring the blue symbols closer to the black ones, also for the brightest magnitudes. We conclude that the relative photometric accuracy of the Q1 data after the large-scale flat correction has a floor of at least 2--3\,mmag. A persistence correction -- as opposed to masking in Q1 -- in a future data release should considerably improve the photometry further.


\subsubsection{Background estimation}
\label{sec:background}

We capture background variations while simultaneously minimising the influence of bright objects on the background estimates. The background maps are estimated in several steps using {\tt SourceExtractor} \citep{BertinArnouts1996}. First, a local background is calculated in a rectangular area by iteratively clipping the histogram of the pixel values until the value converges at 3\,$\sigma$ around the median. The background value is set to the mode of the histogram defined by

\begin{equation}
  \label{eq:sextractor_background}
  \rm{mode} = 2.5 \, {\rm median} - 1.5 \, {\rm mean}\,.
\end{equation}

The size of the rectangular area in the background-subtraction task is set to $64\,\times\,64$ pixels. The background grid computed this way is subsequently median filtered using a $3\,\times\,3$ box to eliminate high background estimates due to bright stars. The final background image is given by a spline interpolation over the grid values. This is not subtracted from the corresponding scientific image, and it is provided as a companion file (\cref{sec:products}).

While this method provides consistent results for the compact and faint objects at the core of the \Euclid\ science programme, it results in considerable over-subtraction for extended, bright galaxies, limiting for example studies of low-surface brightness features and other faint, diffuse targets. A more refined method will be implemented for DR1.


\subsubsection{PSF modelling from physical optics}
\label{sec:psfMDB}

The `\ac{MDB} mode' of the {\tt NIR\_PointSpreadFunction} module can derive  \ac{PSF} images for single exposures only. The \ac{PSF} images are based on a library of \ac{PSF} models that were derived from a physical-optics model of the telescope and instrument stored in the \ac{MDB}. The \ac{PSF} models account for different \ac{FPA} positions and wavelengths, while also incorporating pixelisation and broadening effects, such as intra-pixel sensitivity and \ac{IPC} \citep{LeGraet22}. 

The \ac{MDB} mode cannot produce a \ac{PSF} model for stacked images. This is because it is agnostic to: (i) the accuracy of the astrometric solver (\cref{sec:astrometry}) that corrects for dither offsets and optical distortions; and (ii) the resampling methods and performance of the stacking software (out-of-scope for this paper).

\subsubsection{PSF modelling from data}
\label{sec:psf}

Contrary to the \ac{MDB} mode, the `Inflight mode' of the {\tt NIR\_PointSpreadFunction} module derives the \ac{PSF} model directly from the data. This is the standard mode used for survey-data processing. It can derive a \ac{PSF} model from single images, a series of dithered single images, and also from stacked images. The derivation of \ac{PSF} models for stacked images is out of the scope for this paper, because \ac{NIR PF} stacks are not included in Q1. Stacks are created by the \ac{MER PF} that is described in a separate Q1 paper \citep{Q1-TP004}.

The Inflight mode utilises {\tt SourceExtractor} and {\tt PSFEx} \citep{Bertin2011} to detect and extract the sources in the images, and then to derive the \ac{PSF} models from these data. A relevant parameter given to {\tt SourceExtractor} is {\tt VIGNET} that defines the size of the image stamp centred on each detected source. {\tt PSFEx} uses these stamps to derive the \ac{PSF} models. The larger the size of the stamps, the more extended the derived \ac{PSF} model can be.

When run in the single-image mode, {\tt SourceExtractor} produces a \ac{MEF} catalogue with one extension per detector. When several images are provided, such as all dithers from the \ac{ROS} in one photometric filter, a single extension in the \ac{MEF} catalogue contains the sources from all dithers for that detector. In this way, the source density is increased proportionally to the number of exposures to improve the \ac{PSF} model. The \ac{PSF} comprises three components: (i) the optical \ac{PSF} from the telescope and the instrument's fore-optics; (ii) a detector-electronic \ac{PSF} that describes charge-sharing processes between pixels during charge-generation and readout; and (iii) a mechanical \ac{PSF} from the spacecraft's tracking performance. While optical and detector \ac{PSF}s are unchanged between dithers, the mechanical \ac{PSF} may vary. The precondition for using the all-dither mode is therefore that the spacecraft's tracking is uniform across dithered exposures. 

This paper focuses on the all-dithers \ac{PSF} models, which is also the default mode in \ac{NIR PF}. Once the catalogues are created, they are passed to {\tt PSFEx} that preselects suitable sources for \ac{PSF} modelling. Only unflagged point sources can enter, that is \acp{CR}, saturated stars, and other compromised sources are excluded (see \cref{sec:preprocessing}). {\tt PSFex} then: (i) defines the size and sampling of the derived models; (ii) accounts for variability of the \ac{PSF} shape across the \ac{FPA}; and (iii) sets the level of details of the models. In particular, the \ac{PSF} variability can be accounted for at the level of individual detectors, or for the \ac{FPA} as a whole. Currently, the \ac{PSF} shape is considered variable among detectors for all three NIR passbands. In addition, for the \YE\ band, we include variations inside the detectors, which result in a better \ac{PSF} reconstruction.

The detail in the \ac{PSF} model depends on how accurately it reconstructs the shape of an observed unsaturated point source. {\tt PSFex} models the \ac{PSF}  either directly from the image pixels, or as a linear combination of basis vectors. For this paper, we chose Gauss--Laguerre basis vectors \citep[also known as polar shapelets;][]{MasseyRefregier2005} that have two free parameters configurable for {\tt PSFEx}: {\tt BASIS\_SCALE} affects the width of the \ac{PSF} profile; and {\tt BASIS\_NUMBER} defines the level of detail in the reconstructed core and wings. We tuned a set of optimal values by minimising the difference between \ac{PSF} models from the Inflight and the \ac{MDB} modes. The respective best values for {\tt BASIS\_SCALE} and {\tt BASIS\_NUMBER} currently used by \ac{NIR PF} are 1.4 and 5 for \YE\ band, 1.38 and 9 for \JE\ band, and 1.7 and 5 for \HE\ band.

\begin{figure*}
\centering
\begin{minipage}[t]{0.33\textwidth}
\centering
\resizebox{\hsize}{!}{\includegraphics{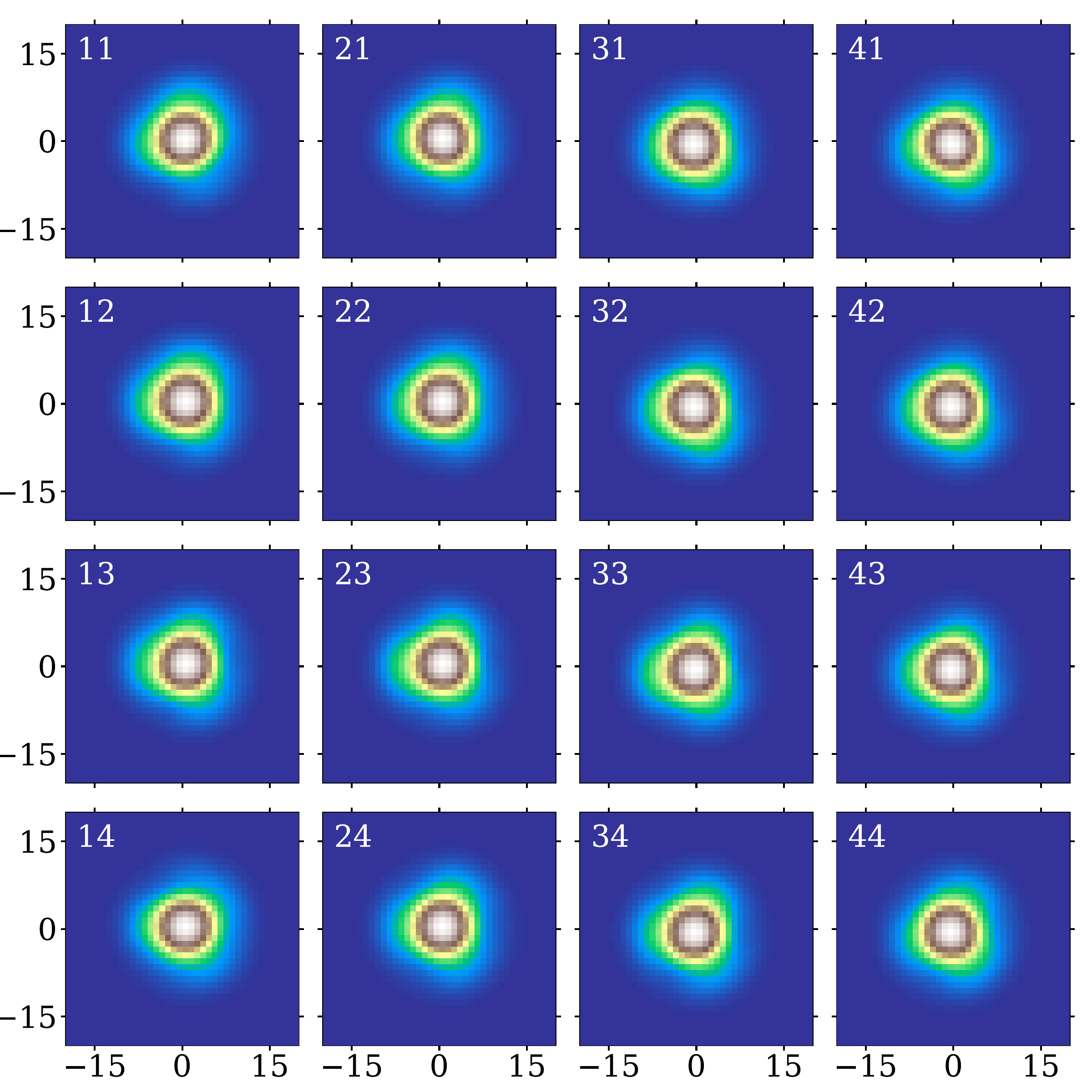}}
\end{minipage}
\begin{minipage}[t]{0.33\textwidth}
\centering
\resizebox{\hsize}{!}{\includegraphics{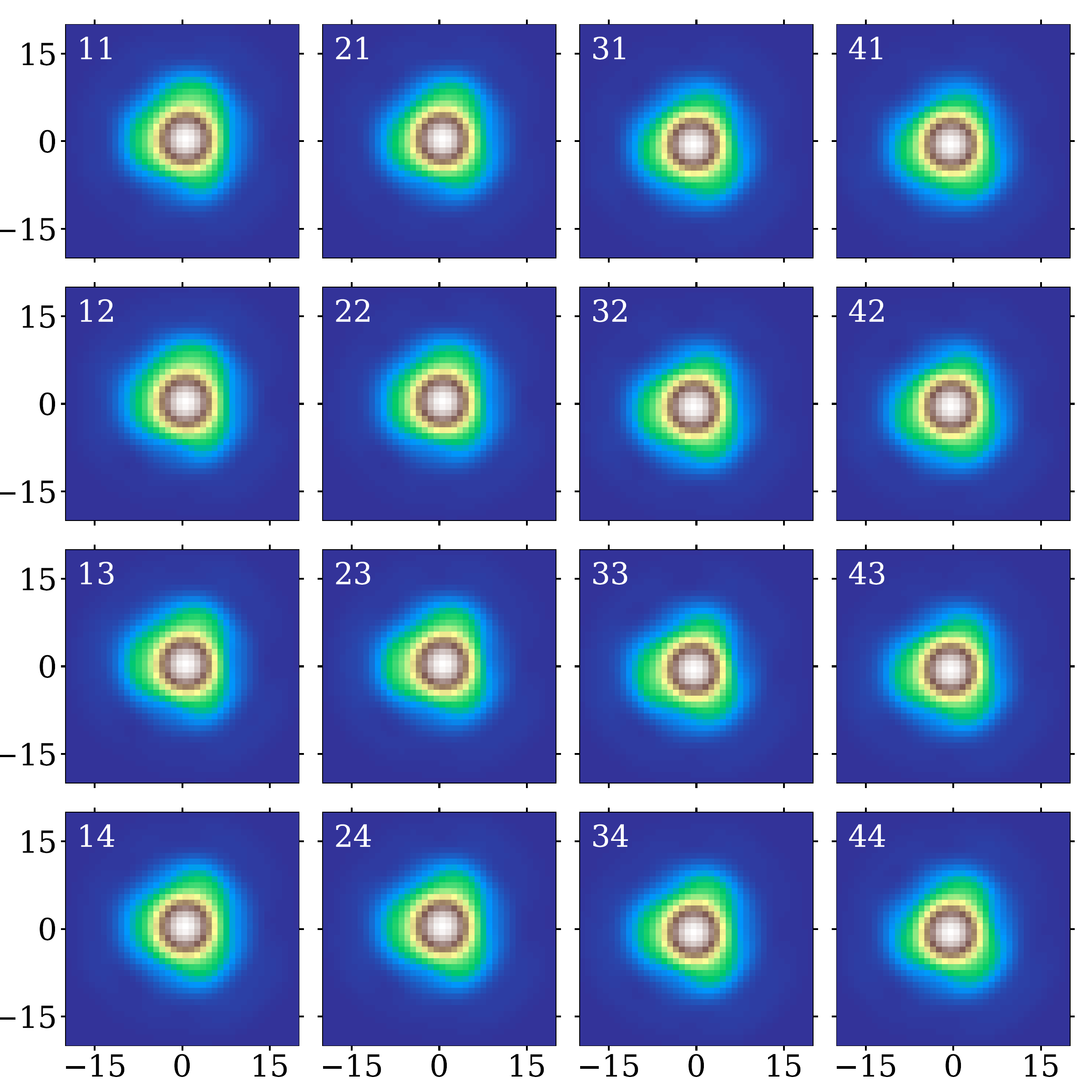}}
\end{minipage}
\begin{minipage}[t]{0.33\textwidth}
\centering
\resizebox{\hsize}{!}{\includegraphics{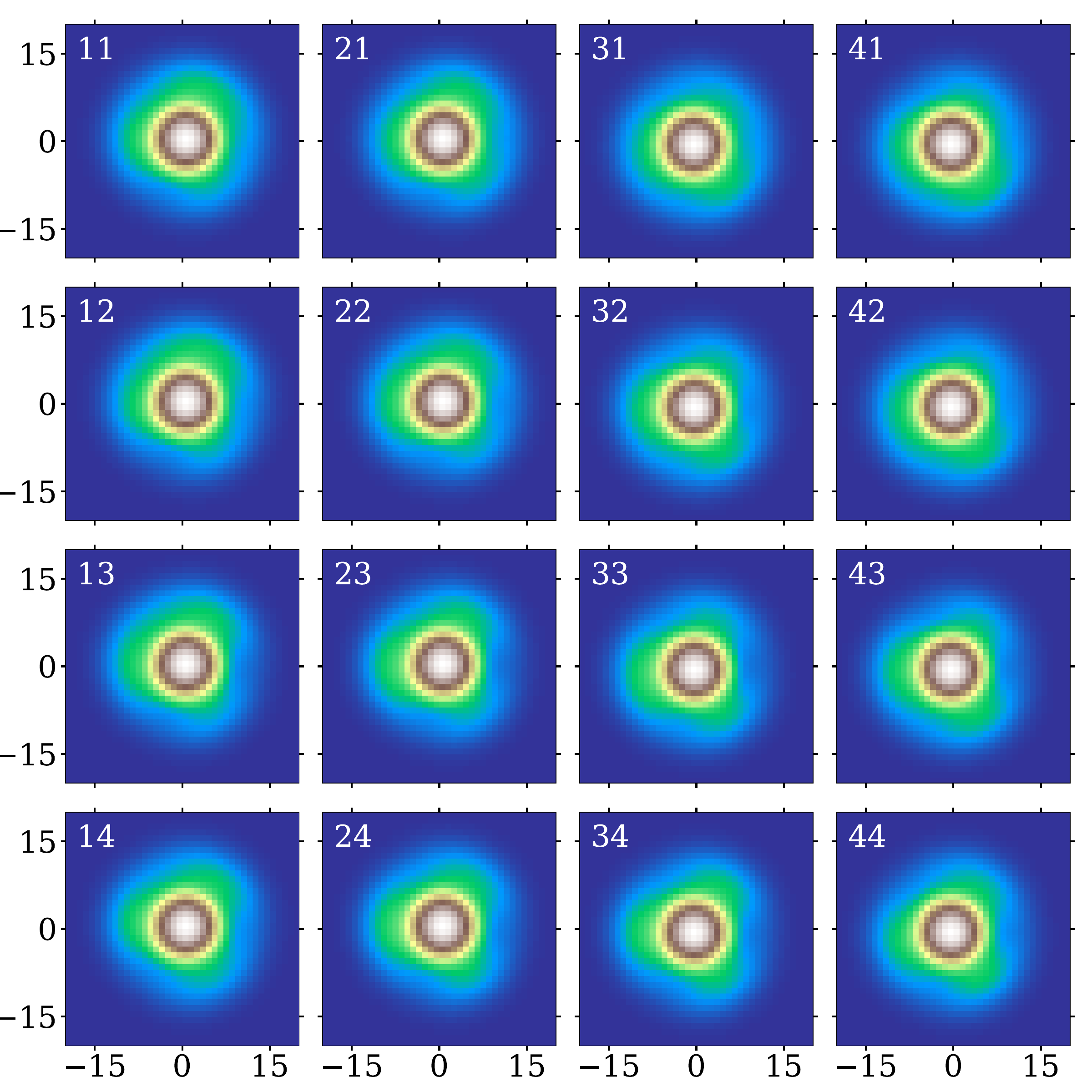}}
\end{minipage}
\caption{2D NISP \ac{PSF} models derived from a typical EWS observation. The different panels are for \YE\ (left), \JE\ (middle), and \HE\ (right) passbands and show the models (40$\times$40 pixel stamps) derived for each of the 16 NISP detectors. The resolution here is \ang{;;0.05}, subsampling the NISP pixel scale by a factor of 6.}
\label{figure:psf_models}
\end{figure*}

\Cref{figure:psf_models} shows the 2D \ac{PSF} models at the centre of each detector and for all passbands, using data from four dithered exposures of the EWS ({\tt ObservationId} 3800). The number of sources effectively used by {\tt PSFEx} ranges from 30 to 40 for each dither, depending on the detector.  The sampling is \ang{;;0.05}, the same as the \ac{MDB} models and corresponding to 1/6th of the NISP pixel size. The \acp{PSF} models recover the known trefoil shape \citep{EuclidSkyOverview} of \Euclid's optics, and show that the \ac{PSF} is very stable 
across the field of view.

\begin{figure}
\centering
\resizebox{\hsize}{!}{\includegraphics{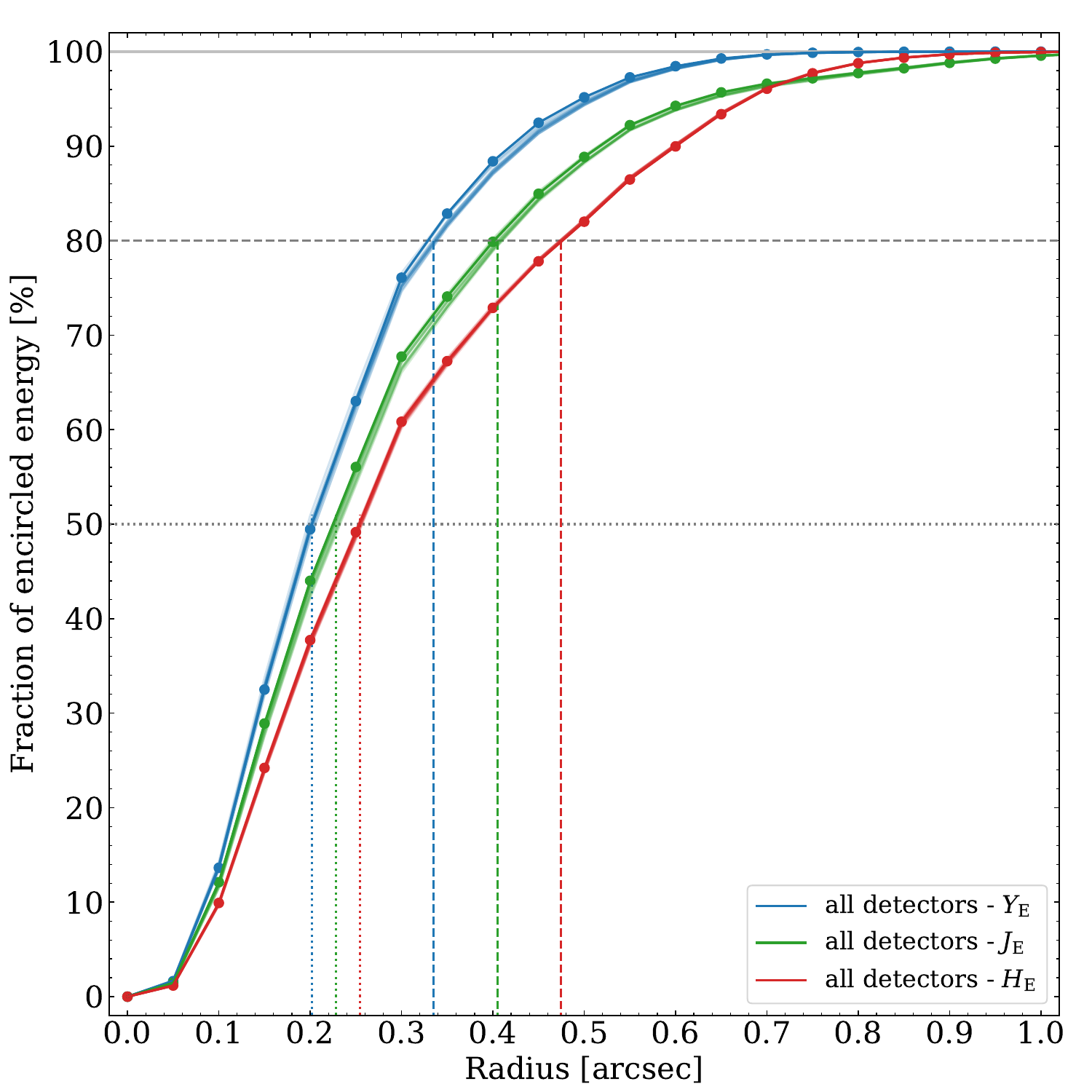}}
\caption{Fraction of encircled energy as a function of the distance to the peak of the \ac{PSF} profile. These curves were calculated using the same models shown in \cref{figure:psf_models} for each of the 16 detectors of the three NISP passbands. The vertical and horizontal lines indicate the average radii encircling 50\% (rEE50, dotted lines) and 80\% (rEE80, dashed lines).}
\label{figure:ree}
\end{figure}

\Cref{figure:ree} shows the encircled-energy curves for all detectors in each passband. The radii encircling 50\% (rEE50) and 80\% (rEE80) of energy must fulfill two mission requirements on the compactness of the \ac{PSF}. Specifically, ${\rm rEE50}<\ang{;;0.4}$ and ${\rm rEE80}<\ang{;;0.7}$ for all wavelengths below 1486\,nm, meaning all of the \YE\ band and most of \JE\ band. The figure shows that even the reddest passband, \HE, easily meets these requirements, a testament to the excellent NISP optics \citep[see also][]{grupp2019}.

\subsubsection{Quality of the PSF model from PSF-fitting photometry}
\label{sec:PSFfittingphot}

\begin{figure*}
\centering
\begin{minipage}[t]{\textwidth}
\centering
\resizebox{\hsize}{!}{\includegraphics{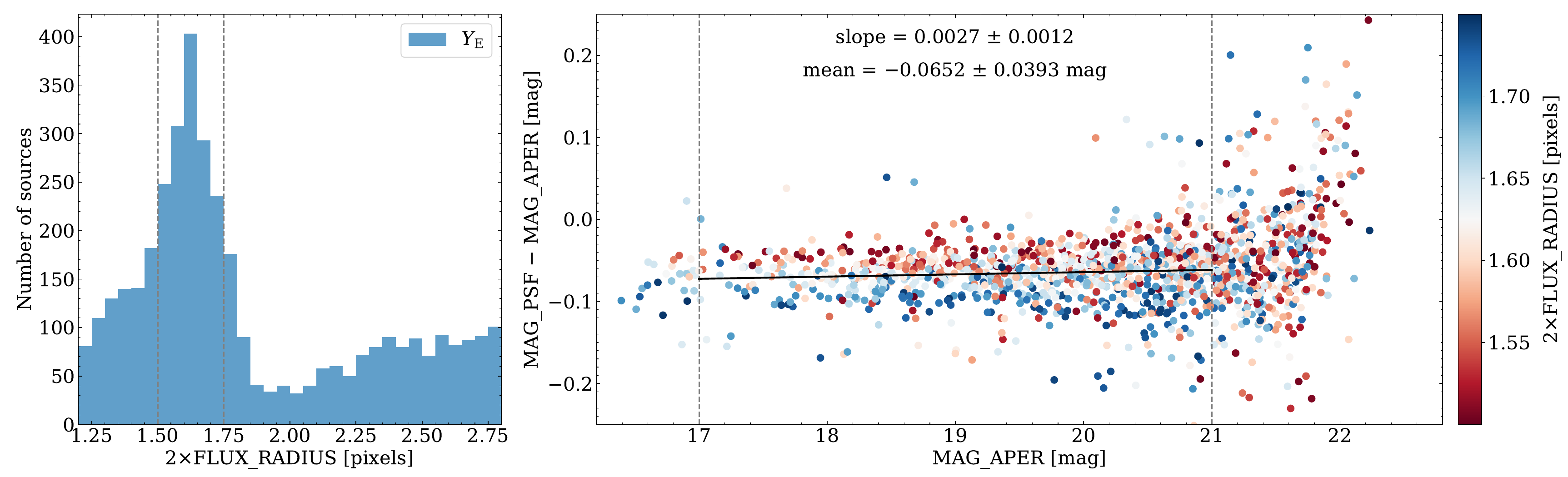}}
\end{minipage} \\
\begin{minipage}[t]{\textwidth}
\centering
\resizebox{\hsize}{!}{\includegraphics{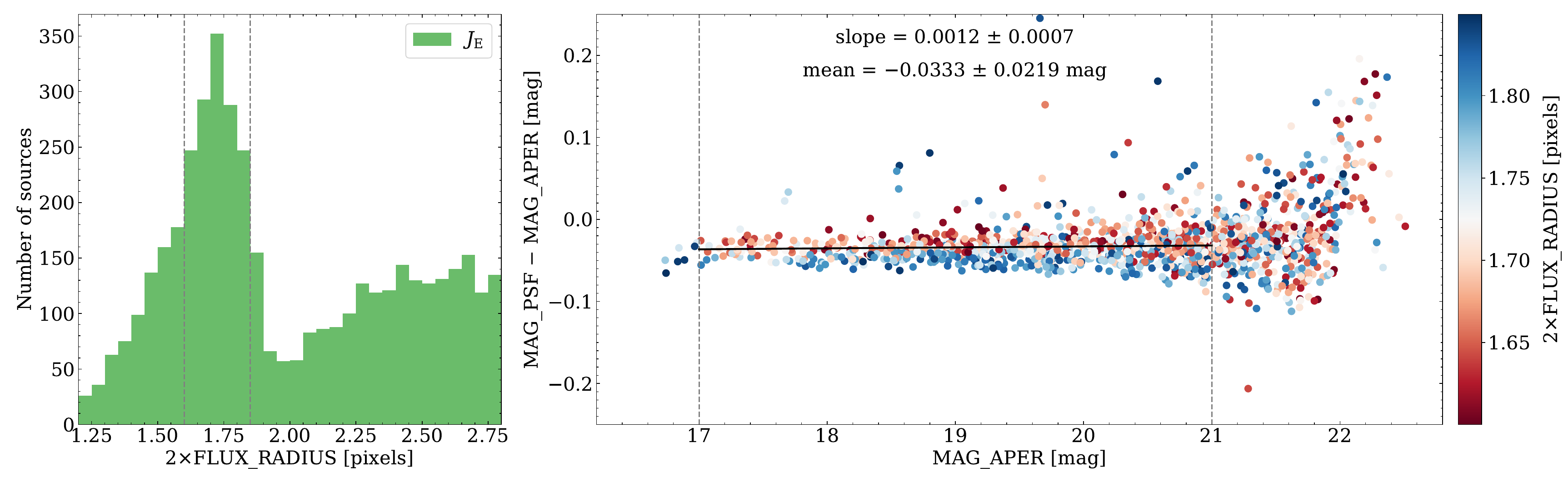}}
\end{minipage} \\
\begin{minipage}[t]{\textwidth}
\centering
\resizebox{\hsize}{!}{\includegraphics{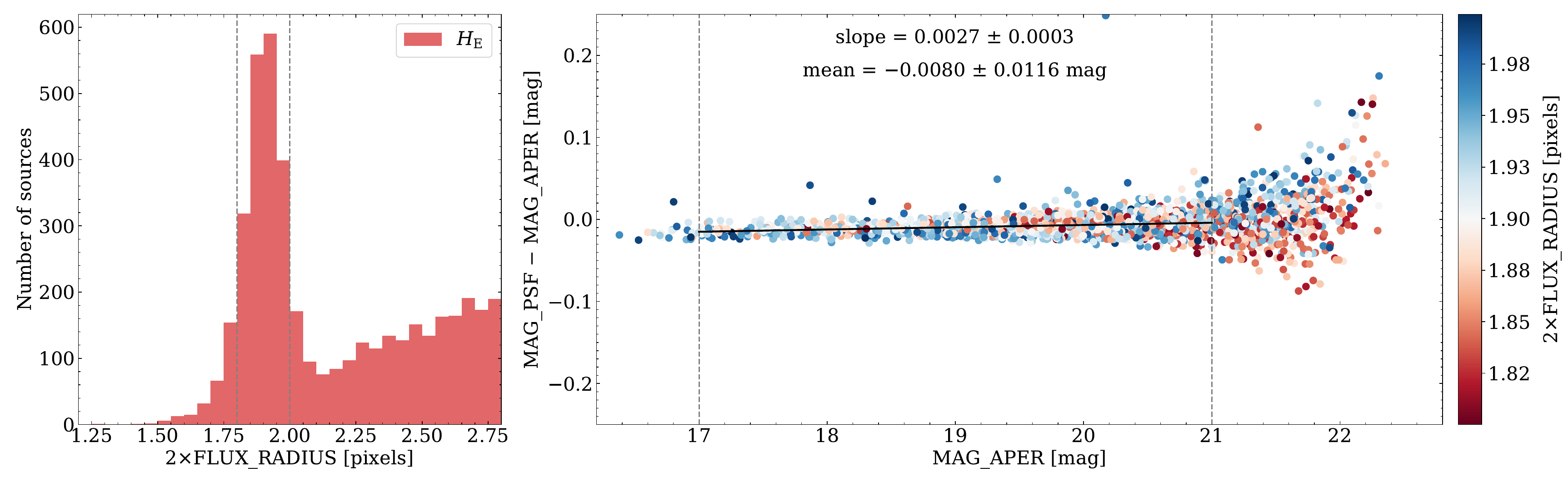}}
\end{minipage}
\caption{Comparison between \ac{PSF}-fitting and aperture photometry.
{\it Left column:} Histograms of the half-light diameter for the three NISP passbands. The dashed lines indicate the ranges of point-sources plotted in the right panels and used by {\tt PSFex} for \ac{PSF} modelling. {\it Right column:} Difference between \texttt{MAG\_PSF} and \texttt{MAG\_APER}, colour-coded according to the half-light diameter. The mean difference and the slope of the linear regression (black line) for magnitudes between 17 and 21 (dashed lines) are also shown.}
\label{figure:hist_mag_psf_aper}
\end{figure*}

The quality of the \ac{PSF} models can be further investigated by \ac{PSF}-fitting photometry. To this end we pass the \ac{PSF} model to {\tt SourceExtractor}, which in turn fits the models to every source extracted from the images. \Cref{figure:hist_mag_psf_aper} (right panels) compares the \ac{PSF} magnitudes ({\tt MAG\_PSF}) to aperture magnitudes ({\tt MAG\_APER}) with a diameter of 6 pixels, the default in \ac{NIR PF}. This aperture then contains -- for the current models -- about $99.99$\% for \YE, $98.8$\% for \JE, and 
$99.7$\% for \HE of the total flux. Aperture corrections compensate for these offsets during absolute photometric calibration (\cref{sect:photom_calib}). The three panels in \cref{figure:hist_mag_psf_aper} are colour-coded according to the {\tt FLUX\_RADIUS}, which contains a certain fraction of a source's total flux. To obtain the half-light radius encompassing half of the total flux, one would set ${\rm {\tt PHOT\_FLUXFRAC}}=0.5$.
\Cref{figure:hist_mag_psf_aper} (left panels) shows the distribution of the half-light diameter, or two times the {\tt FLUX\_RADIUS}, including both point- and extended sources. For each passband, we highlight the range in which the point sources are located. These are the \texttt{FLUX\_RADIUS} ranges of the sources passed to {\tt PSFex} to derive the \ac{PSF} models (the same sources plotted in the right panels). 

The right column in \cref{figure:hist_mag_psf_aper} shows that the \ac{PSF}-fitting photometry for \HE\ band agrees to better than 1\% with the aperture photometry. For the
\JE\ band we see that PSF magnitudes are systematically brighter by about 3\%. For the \YE\ band, the offset increases to more than 6\%, there is a slight dependence on magnitude, and the scatter is considerably larger than for the other two bands. For all three bands we also find a positive correlation between \texttt{FLUX\_RADIUS} on the one hand, and the offset and the scatter on the other hand. For the \HE\ band, in particular, the slope of the linear regression is more prominent for more extended sources. There is ongoing work to improve the \ac{PSF} models. In section~5.1 of \cite{EuclidSkyNISP}, we show preliminary results of \ac{PSF} models derived using a larger number of point sources from self-calibration observations instead of an EWS field, a larger number of basis vectors (hundreds instead of tens), and in $\ang{;;7}\times\ang{;;7}$ stamps instead of $\ang{;;3}\times\ang{;;3}$ for each source. These updates require changes in the {\tt PSFex} configuration that will be implemented for the next data release.

\subsubsection{Astrometric calibration}
\label{sec:astrometry}

Astrometric calibration accounts for geometric distortions from instrument optics and focal-plane metrology, which define detector positions and orientations. For each pointing and band, this is done by minimising positional differences between sources detected in dithered exposures and those in an external reference catalogue. Currently, {\it Gaia} DR3 \citep{GaiaDR3} is preferred over the VIS catalogue \citep{Q1-TP002}, since its bright stars -- saturated in VIS  -- remain unsaturated in NISP, providing more reliable reference sources.

Astrometric calibration is performed using {\tt SCAMP} \citep{Bertin2006}, which operates on catalogues rather than images. This requires both a reference catalogue and source catalogues for the dithered exposures. Source catalogues are provided by {\tt SourceExtractor} with a configuration optimised for astrometric calibration. The astrometric solution -- modelled in \ac{NIR PF} as a 3rd-order polynomial per detector -- is derived by minimising the $\chi^2$-sum of positional differences between overlapping detections.

In some cases, particularly when too few stars are available in common with the external catalogue for certain detectors, {\tt SCAMP} may fail to converge unless its parameters are fine tuned. However, the optimal parameters can vary across fields with different stellar densities. To address this, we implemented an astrometric pre-solution, that is an initial model for geometric distortion along with the relative positions and orientations of the detectors. This pre-solution helps reduce sensitivity to the initial matching radius with the external catalogue. If the radius is too large, it can cause false matches. If it is too small, it may result in no matches, especially near the field edges, where distortion can shift source positions by up to \ang{;;10}. As a result, the robustness of the distortion-model fitting is significantly improved.

The astrometric pre-solution is created by the distortion-model pipeline in \ac{NIR PF}, where {\tt SCAMP} is run on the 76 dithered exposures, per band, of the self-calibration field. {\tt SCAMP}, following the polynomial PV\footnote{Here, PV refers to a polynomial convention, not \Euclid's performance-verification phase.} convention \citep{CalabrettaGreisen2022}, applies the distortion polynomial to the world coordinates. Therefore, when transferring the pre-solution to a different dither, adjusting for the different dither's rotation on the sky is required. This is is done by first converting the pre-solution to the Simple Imaging Polynomial (SIP) convention \citep{Shupe2005}, where the distortion is instead applied to pixel coordinates. It is then translated back to the PV convention using the dither's centre and rotation, as encoded in the \ac{WCS} keywords. Currently, these transformations use  polynomial fitting functions. An analytic conversion method \citep[see e.g.,][]{Shupe2012} will be used for the next data releases.


\subsubsection{Photometric calibration}
\label{sect:photom_calib}

For \Euclid's cosmology science aims, we have an accuracy goal of 5\% on the absolute flux calibration of NISP. This comes easily, since we calibrate against spectrophotometric \acp{WD} in \ac{HST} CALSPEC, which have an estimated accuracy of 1\% \citep{bohlin2014,bohlin2017,bohlin2020}.

For NISP imaging we could use the all-sky faint \acp{WD} network \citep{narayan2019,bohlin2024} in CALSPEC. However, none of them are visible in \Euclid's continuous viewing zones that extend \ang{2.5;;} from the ecliptic poles. We therefore established a faint \ac{WD}, WDJ175318.65+644502.15 ({\it Gaia} SourceID 1440758225532172032) in the self-calibration field near the \ac{NEP}. This was tied to CALSPEC using photometric and spectroscopic \ac{HST} observations with WFC3/IR and STIS \citep[programme IDs 16702 and 17442,][]{appleton2021,deustua2023}. This star is stable to within $1\%$ over timescales ranging from 100 seconds to several years. With $\HE = 18.36$, it is faint enough for NISP, which saturates at around 16.5\,AB\,mag. 

Flux calibration is established by comparing the observed flux $F_\mathrm{obs}$ in instrumental units (e$^-$\,s$^{-1}$) with the true flux $F_\mathrm{true}$ in $\mathrm{\muup}$Jy. The true flux integrated over a passband is obtained by convolving the \ac{WD}'s model spectral flux density $S_\nu$ with the passbands' spectral response $T(\nu)$, 
\begin{equation}
    F_\mathrm{true} = \frac{ \int \dfrac{T(\nu) \,S_{\nu}}{\nu} \, {\rm d}\nu }{\int \dfrac{T(\nu)}{\nu} \, {\rm d}\nu }\,.
\end{equation}
Here, $\nu$ is the frequency, and the spectral response $T(\nu)$ for passbands \YE, \JE, and \HE is taken from \citet{Schirmer-EP18}. 

$F_\mathrm{obs}$ is measured with {\tt SourceExtractor} within a fixed aperture of diameter 6 pixels. We use our \ac{PSF} model (\cref{sec:psf}) to compute the fraction of the flux that falls outside this aperture, and correct $F_\mathrm{obs}$ accordingly. Finally, the absolute flux calibration factor, or \ac{IS} in \ac{NIR PF}, is defined as
\begin{equation}
    \mathrm{IS} = \frac{F_\mathrm{true}\,[\mathrm{\muup Jy} ]}{F_\mathrm{obs} \,[\mathrm{e}^{-}\,\mathrm{s}^{-1}]}\,. 
\end{equation}
 
\begin{figure}
\centering
\resizebox{\hsize}{!}{\includegraphics{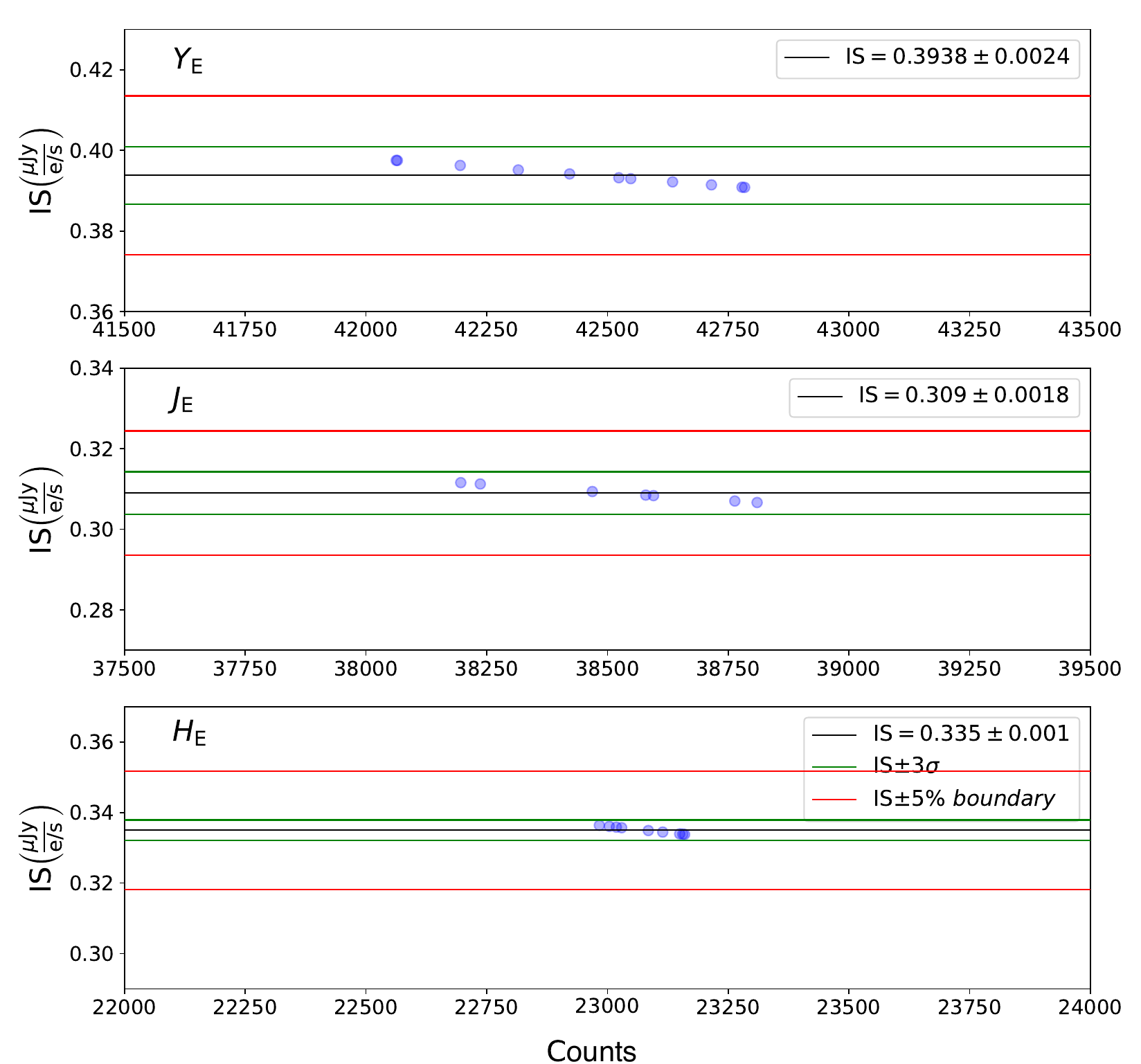}}
\caption{Absolute flux calibration factor \ac{IS} as a function of the observed counts, from the June 2024 \ac{WD} calibration observations used for Q1. The observed counts were corrected for aperture losses. The black and green lines indicate the respective sigma-clipped mean of the individual \ac{IS} measurements and the sample's 3\,$\sigma$ uncertainty. The red lines denote the 5\% requirement.}
\label{figure:IS-counts}
\end{figure}

Flux calibration of the Q1 data is based on the self-calibration observations -- that contain the \ac{WD} -- in June 2024 (see \cref{subsection:zeropoint}). \Cref{figure:IS-counts} presents the individual measurements of the WD in two NISP detectors, together with the corresponding \ac{IS}. As can be seen, the 3\,$\sigma$ uncertainty of the \ac{IS} is well within the required $5\%$ boundary.

\subsubsection{Ghost masking}
\label{sec:ghosts}

Although a number of ghosts are seen within the NISP-P data \citep[see][]{EuclidSkyNISP}, only the dichroic and filter ghosts are masked in Q1. The ghosts' size, shape, and position relative to their bright source stars are described in detail in Paterson et al. (in prep.). There, we also discuss the dependence on the filter used, detection methods, and lower flux limits that require masking. For the masking itself, we first select all stars from {\it Gaia} brighter than the magnitude limit set by Paterson et al. (in prep.). Pixels that fall within the computed masking area of these stars are then flagged with {\tt GHOST} (bit 18) and {\tt INVALID} (bit 0) in the \ac{DQ} layer \cref{table:flag_descriptions}. 

Stars brighter than 11\,mag also exhibit an arc with a curvature radius of about \ang{;1.7;} \citep[see figure 19 of][]{EuclidSkyNISP}. The arcs extend up to $\pm\,45\degree$ from the $x$-axis in pixel coordinates. These ghosts are the result of charge persistence that accumulates when NISP switches between filters and grisms \citep{EuclidSkyNISP}. The procedure to mask these arcs is similar to that of the ghosts. First, bright stars expected to produce arcs are selected from {\it Gaia}. Second, affected pixels are flagged with both {\tt PERSIST} (bit 13) and {\tt INVALID} in the \ac{DQ} layer.


\subsubsection{Catalogue production}
\label{sec:catalog}

The final step of \ac{NIR PF} is catalogue extraction, producing primarily single-band catalogues for the calibrated individual and stacked images. Stacked images and their catalogues, however, are beyond the scope of this Q1 paper and we refer to \ac{MER PF} instead \citep{Q1-TP004}. In the following, we describe {\tt NIR\_CatalogExtraction} that operates on individual calibrated images, processing one dither at a time on a detector-by-detector basis. The module consists of two main steps, the first generates the source map, which is subsequently passed to the second step as a detection constraint for extracting source photometry.

The identification process begins by computing the second derivative of the image along four distinct directions. This technique is based on the {\tt CuTEx} tool \citep{Molinari2011}. Since two-dimensional images represent discrete data sets, we employ Lagrangian methods for numerical differentiation, specifically using the 3-point formula \citep[detailed in][]{hildebrand1956}. For each pixel in the input image, the algorithm applies a convolution with a kernel derived from the second-order derivative formulas along four directions ($x$, $y$, and the two $45\degree$ bisectors). This ensures that the results are independent of the derivation direction. The resulting second derivative map is then used to detect objects and construct the source map, a technique that proves highly effective in crowded regions as well as in the presence of variable background without the need for dedicated tunings.

A dynamic threshold is applied for source identification, with its value automatically adjusted by the code through iterative refinement. Starting from an initial threshold, the algorithm calculates the number of detected sources and evaluates how many of these allow reliable PSF photometry. The threshold is iteratively optimised to balance source detection with photometric reliability. Once the optimal threshold is determined, a mask is generated to mark the source positions. This mask is subsequently used as input for the photometry step, where {\tt SourceExtractor} processes the calibrated dither, constrained by the detection map for source positions.

The version of the code used for Q1 simultaneously performs three types of photometry for each detected object: aperture with a fixed radius of 3 pixels; automatic; and \ac{PSF} fitting. Consequently, the module requires the \ac{PSF} model computed by {\tt NIR\_PointSpreadFunction} as additional input. As discussed in \cref{sec:psf}, the fixed aperture adopted for catalogue extraction includes most of the flux for compact objects: 99.99\% for \YE; 98.8\% for \JE; and 99.7\% for \HE.

The final output catalogue is stored as a FITS file containing 16 layers, one for each detector, to maintain correspondence with the structure of the calibrated image (\cref{sec:catalog}).
This approach ensures efficient and accurate extraction of sources and their photometric properties.

\subsection{DQC pipeline}
\label{sec:dqc}

The \ac{LE1} data from the selected sky patches in this release are processed using \ac{NIR PF}, generating a set of \acp{DP} per pipeline run (see \cref{sec:products}). These \acp{DP} must be verified to ensure data quality meets the required standards.

The \ac{DQC} pipeline inspects \ac{NIR PF} \acp{DP} by checking the \ac{DQC} parameters computed during processing. These parameters are stored as metadata and ingested into the \ac{EAS} at the end of each run.

Each \ac{DQC} pipeline run produces a PDF report for each input \ac{PPO} ID. The report contains test results, with each test consisting of multiple checks. In each check, selected \ac{DQC} parameters from a given \ac{DP} are compared against predefined thresholds to determine compliance. Test results fall into four categories: PASS, FAIL, WARN, or N/A. We maintain the list of parameter thresholds for verification.

Each test result is recorded at the beginning of the PDF report and stored in a local database. We review a summary table of all test results for a sky patch before approving them for publication. The available tests include specific and shared tests for both \ac{NIR PF} and SIR PF, which check technical aspects such as input/output products and profiling. For this release, the tests use calibrated images and catalogues to evaluate PSF quality, flagged pixel percentage, astrometric RMS, background level, and photometric accuracy.


\section{NIR Data Products}
\label{sec:products}

The \Euclid Common Data Model is an XML schema (XSD) that defines all \Euclid Data Products, along with a dictionary of all data types, interfaces, and FITS file formats. A Data Product is therefore represented as an XML file containing the names of data files and their corresponding metadata, which can be stored in and retrieved from the \ac{EAS}. 
 
A detailed description of the metadata for all NIR \acp{DP} is available in the Data Product Description Document \citep[DPDD,][]{EuclidDpdd}. The \acp{DP} produced by \ac{NIR PF} for Q1 are the {\tt DpdNirCalibratedFrame} and the corresponding {\tt DpdNirCalibratedFrameCatalog}.

\subsection{Calibrated exposures}

The {\tt DpdNirCalibratedFrame} is a fully astrometrically and photometrically calibrated exposure, obtained by applying calibration products and corrections to each individual LE1  photometric image. Image data are stored as FITS files referenced in the \ac{DP} and are organised into three main sections: \texttt{DataStorage}; \texttt{BackgroundStorage}; and  \texttt{PsfStorage}.

\texttt{DataStorage} references the main \ac{MEF} file that contains the calibrated image data. This file consists of a total of 49 \acp{HDU}. The first \ac{HDU} or primary header  stores the main metadata keywords of the observation sequence that acquired the exposure. 
For each detector, three HDUs are provided: (i) \texttt{SCI}, containing the scientific image data in electrons; (ii) \texttt{RMS} with the associated root mean square; and (iii)  \texttt{DQ} with the data-quality mask for each pixel (\cref{table:flag_descriptions}). The header of each SCI \ac{HDU} contains the detector \ac{WCS} and the zero point. 

\texttt{BackgroundStorage} references the FITS file of the background image, computed during the background-estimation step (\cref{sec:background}). Its file structure mirrors that of the scientific image described above. 

The \texttt{PsfStorage} section of the \ac{DP} refers to two FITS files: the \ac{PSF} model; and the corresponding \ac{PSF} image (\cref{sec:psf}).

Another key metadata tag is \texttt{QualityParams}, which contains a structure to store 
image-quality parameters and statistics. They are used as input for the \ac{DQC} pipeline  (\cref{sec:dqc}) and are computed by specific tasks in \ac{NIR PF}. 

\subsection{Calibrated catalogues}
The \texttt{DpdNirCalibratedFrameCatalog} is the source catalogue extracted from the calibrated exposure, as described in \cref{sec:catalog}. Its global metadata structure mirrors that of \texttt{DpdNirCalibratedFrame}. The \ac{MEF} catalogue produced is referenced under the \texttt{DataStorage} tag of the \ac{DP} and is structured into 17 \acp{HDU}: a primary header followed by a data \ac{HDU} for each detector.

    The data \acp{HDU} store the source catalogue tables, which include coordinates, size, ellipticity, fluxes, and magnitudes extracted using three different methods by {\tt SourceExtractor}: \texttt{APER}, \texttt{AUTO}, and \texttt{PSF}, as described in \cref{sec:catalog}.


\section{Results for \Euclid Quick Release 1}
\label{sec:imageaccuracy}

Q1 includes one visit for each of the three \Euclid Deep fields, North, South and Fornax, as well as observations of the Lynd's Dark Nebula LDN1641, covering a total of 63.1\,${\rm deg}^2$ \citep{Q1-TP001}.
From an initial set of 1620 NISP raw exposures, we rejected seven images that did not pass the DQC pipeline (0.4\%), while the remaining 1613 images are available from the \Euclid Science Archive.

In the following sections we provide a characterisation of the astrometric and photometric accuracy of the NIR Q1 data set.


\subsection{Accuracy of astrometric and photometric calibration based on Gaia DR3}

\begin{figure*}
\centering
\begin{minipage}[t]{0.98\textwidth}
\centering
\resizebox{\hsize}{!}{\includegraphics{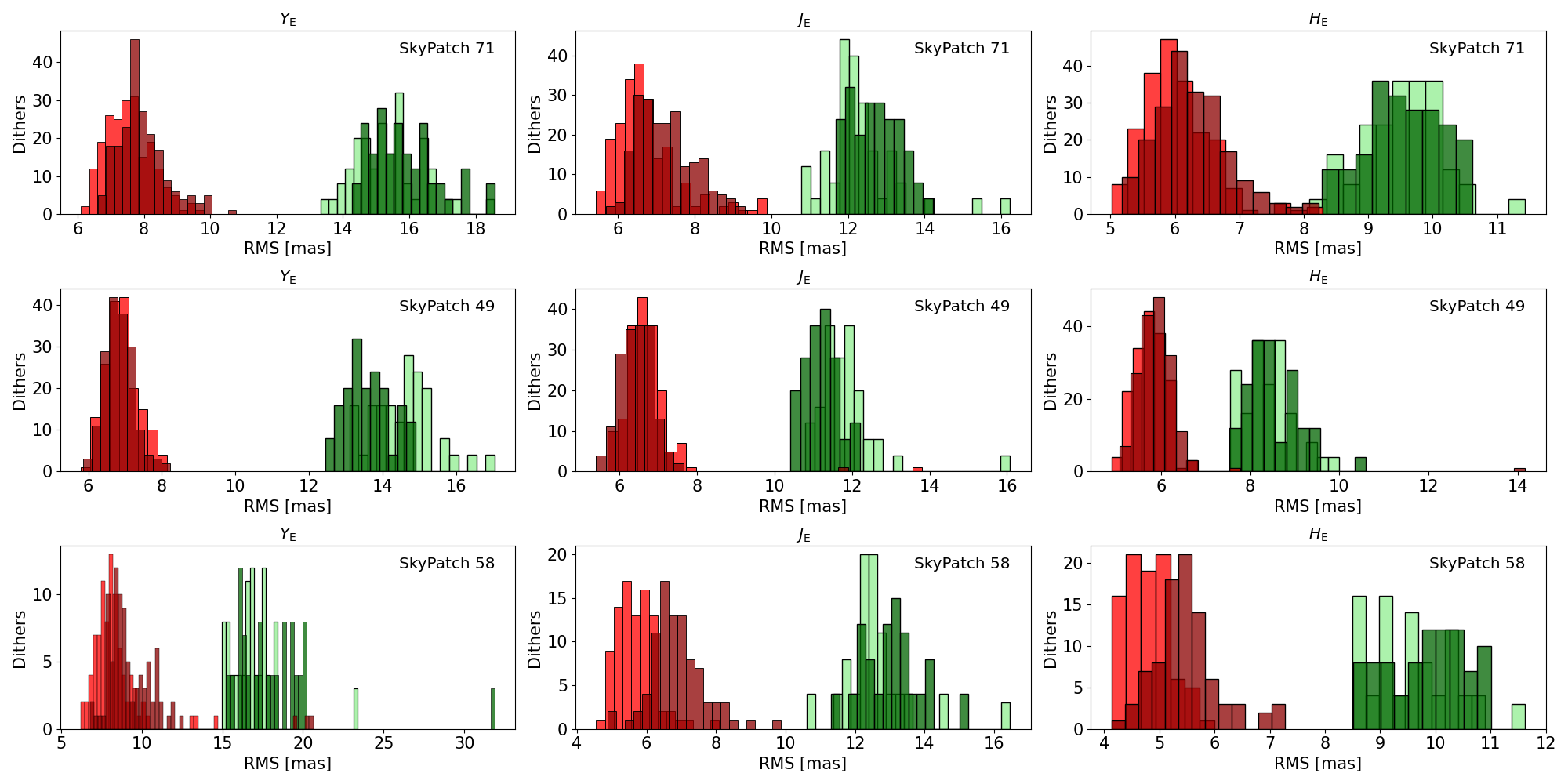}}
\end{minipage}
\caption{Astrometric \ac{RMS} residuals for dithers in the three Q1 sky patches. The red histograms are the residuals obtained from overlapping detections (required accuracy: 100\,mas), while the green histograms illustrate residuals obtained through comparison with {\it Gaia} positions (required accuracy: 200\,mas). The \ac{RMS} for $\Delta \alpha \cos(\delta)$ and $\Delta \delta$ are represented by light and dark shading, respectively. The residuals are always well within the required accuracies. } 
\label{figure:ast_rms}
\end{figure*}

As discussed in \cref{sec:astrometry}, the astrometric calibration is referenced against 
{\it Gaia} DR3 \citep{GaiaDR3} and accounts for proper motion. Astrometric residuals are about 15, 12, and 9\,mas, respectively, for the \Euclid \YE, \JE, and \HE\ bands as shown in \cref{figure:ast_rms}. The residuals are well within the required accuracy of 100\,mas.

\begin{figure*}
\centering
\begin{minipage}[t]{0.98\textwidth}
\centering
\includegraphics[width=5.9cm]{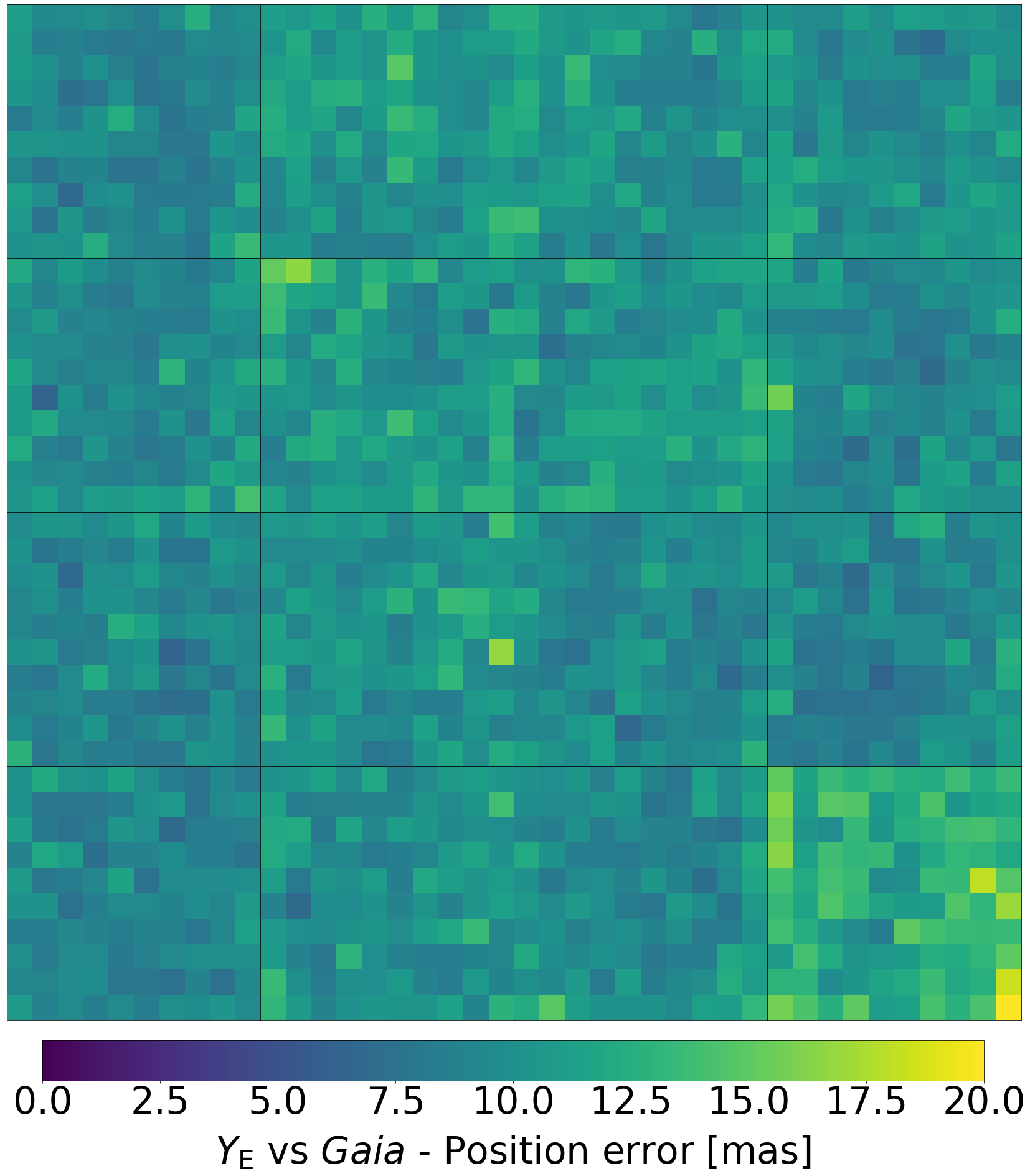}
\includegraphics[width=5.9cm]{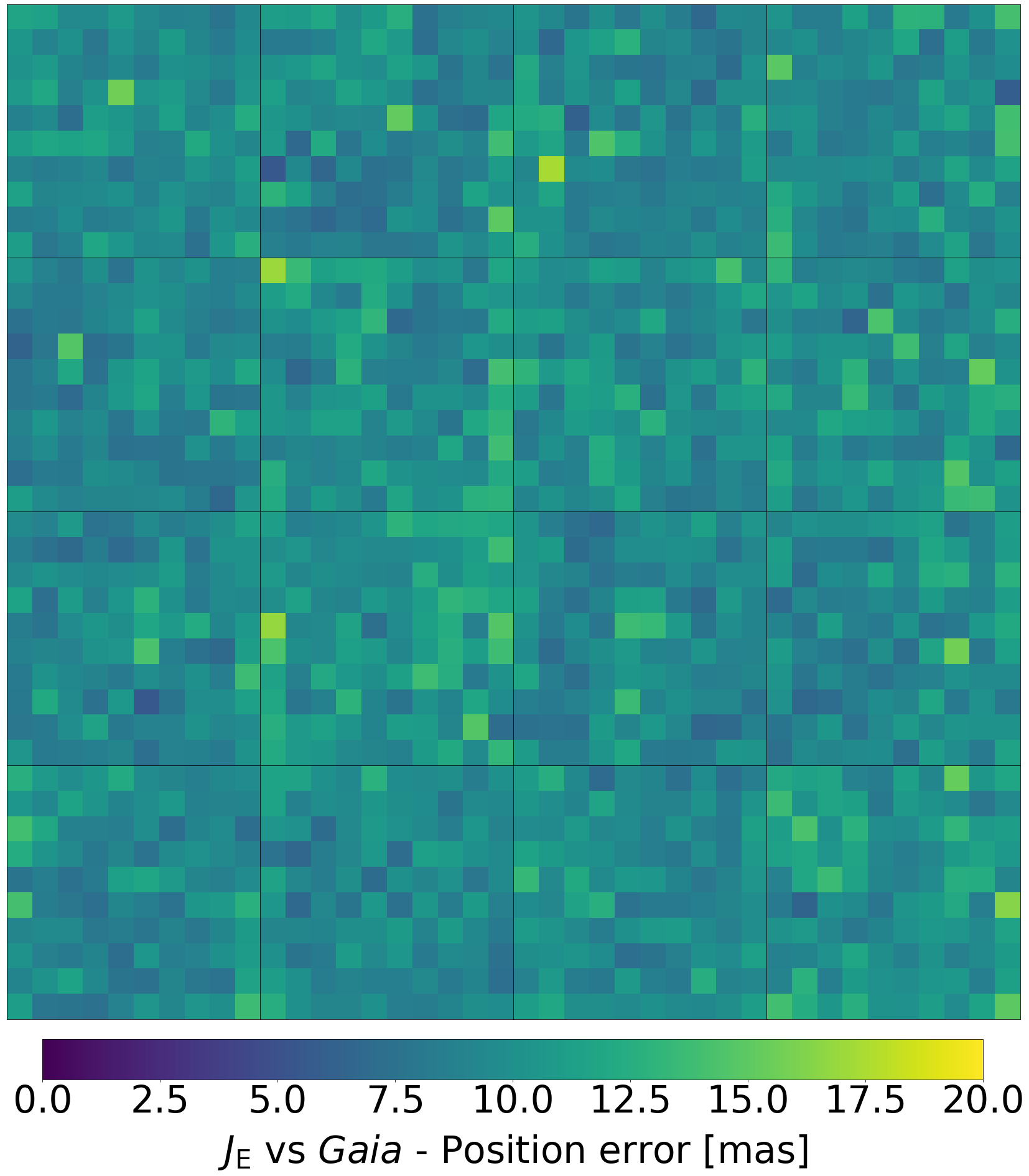}
\includegraphics[width=5.9cm]{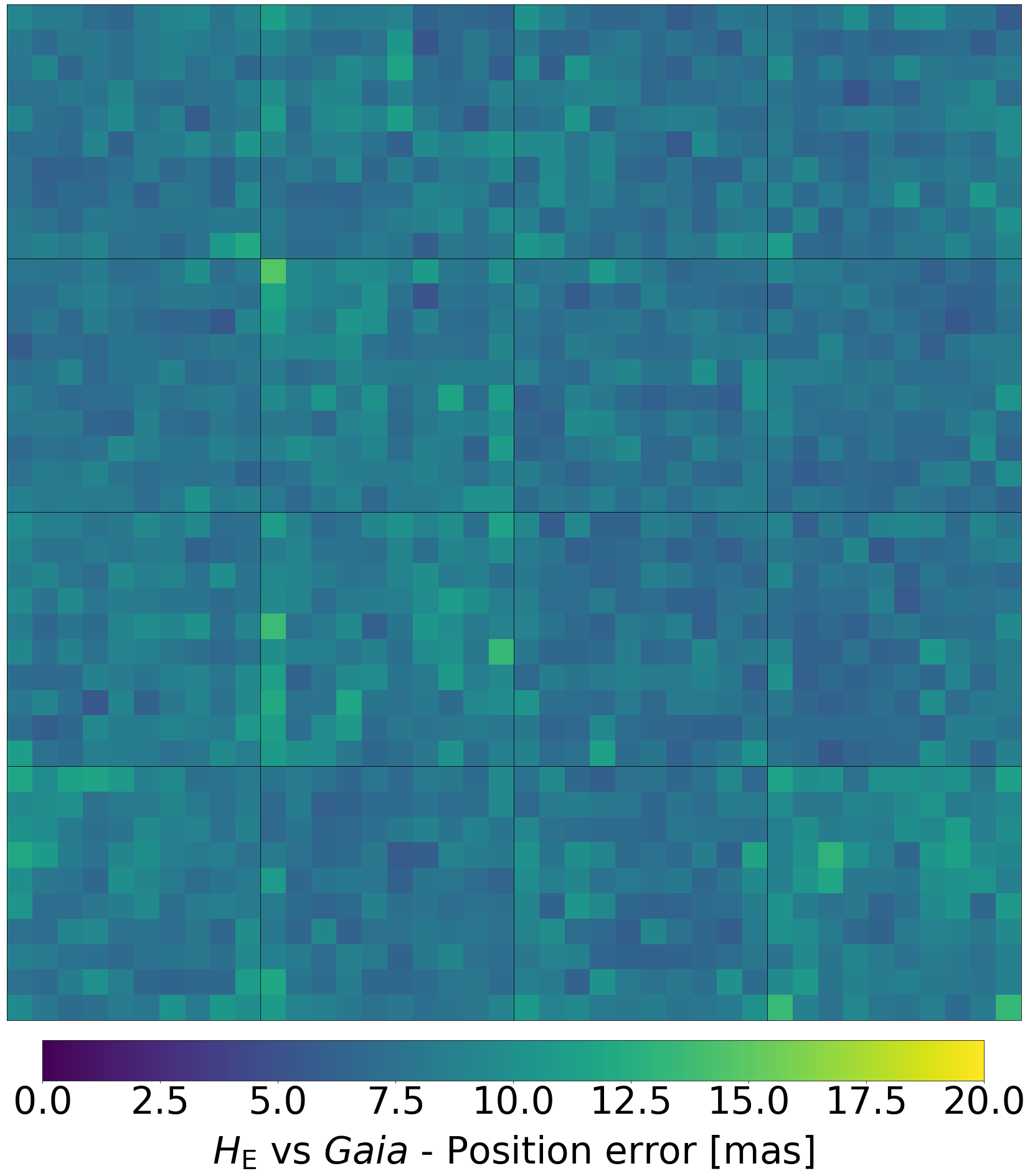}
\end{minipage}
\caption{Focal plane distribution of position error with respect to {\it Gaia}, combining all sources for sky patch 71.} 
\label{figure:ast_rms_FPA}
\end{figure*}

We also evaluated possible systematic effects related to the source position in the focal plane. For a given Q1 sky patch, the \ac{NIR PF} source catalogues were cross-matched against {\it Gaia} DR3. Then we binned the angular distances between \Euclid and {\it Gaia} positions as a function of focal-plane coordinates, taking the median value for each bin. \Cref{figure:ast_rms_FPA} shows that the astrometric accuracy has no major spatial dependence apart from a slight increase for one detector in the \YE\ band, still well below 1/10th of a pixel. The other Q1 sky patches behave similarly.

\begin{figure*}
\centering
\begin{minipage}[t]{0.98\textwidth}
\centering
\includegraphics[width=5.9cm]{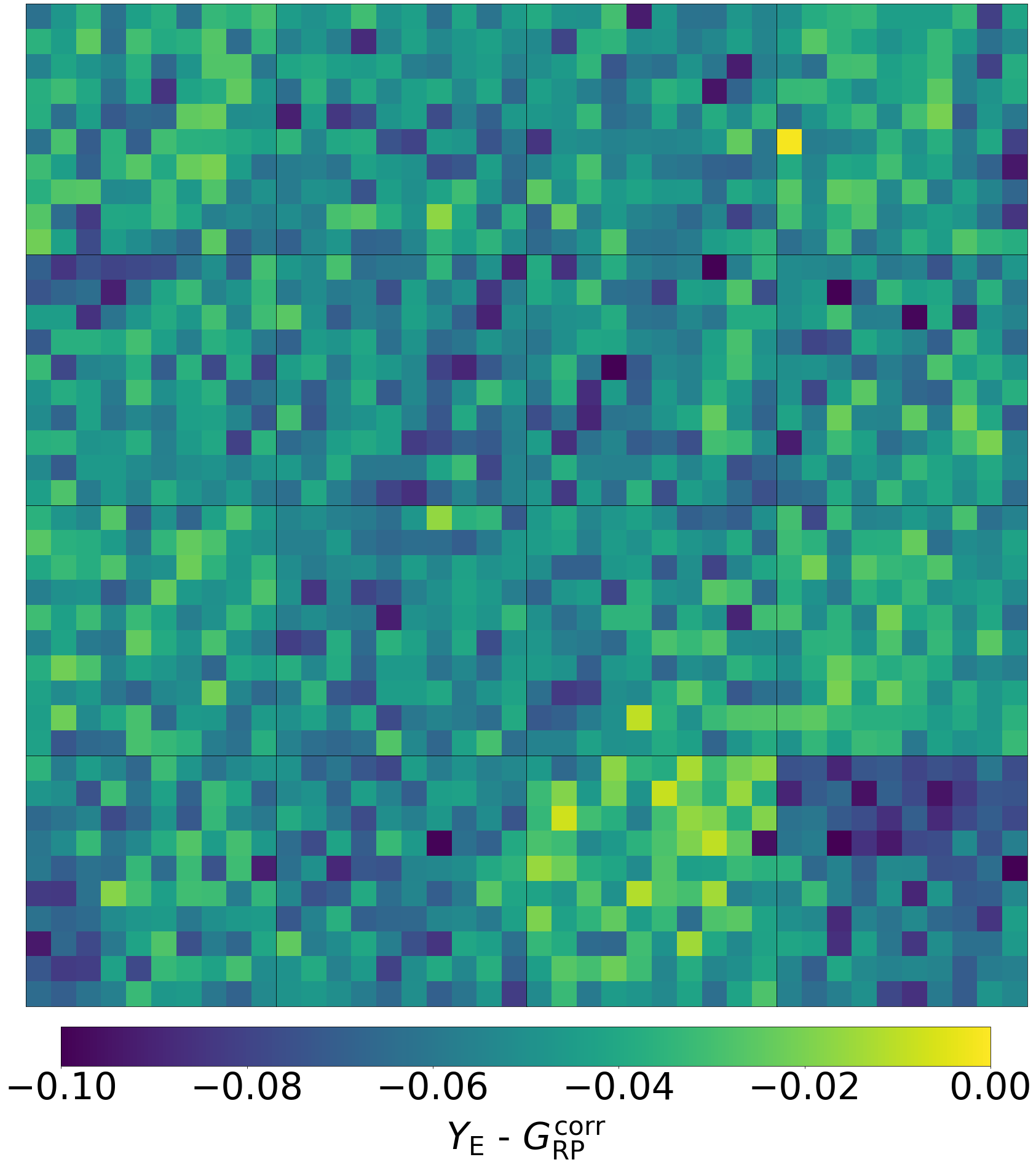}
\includegraphics[width=5.9cm]{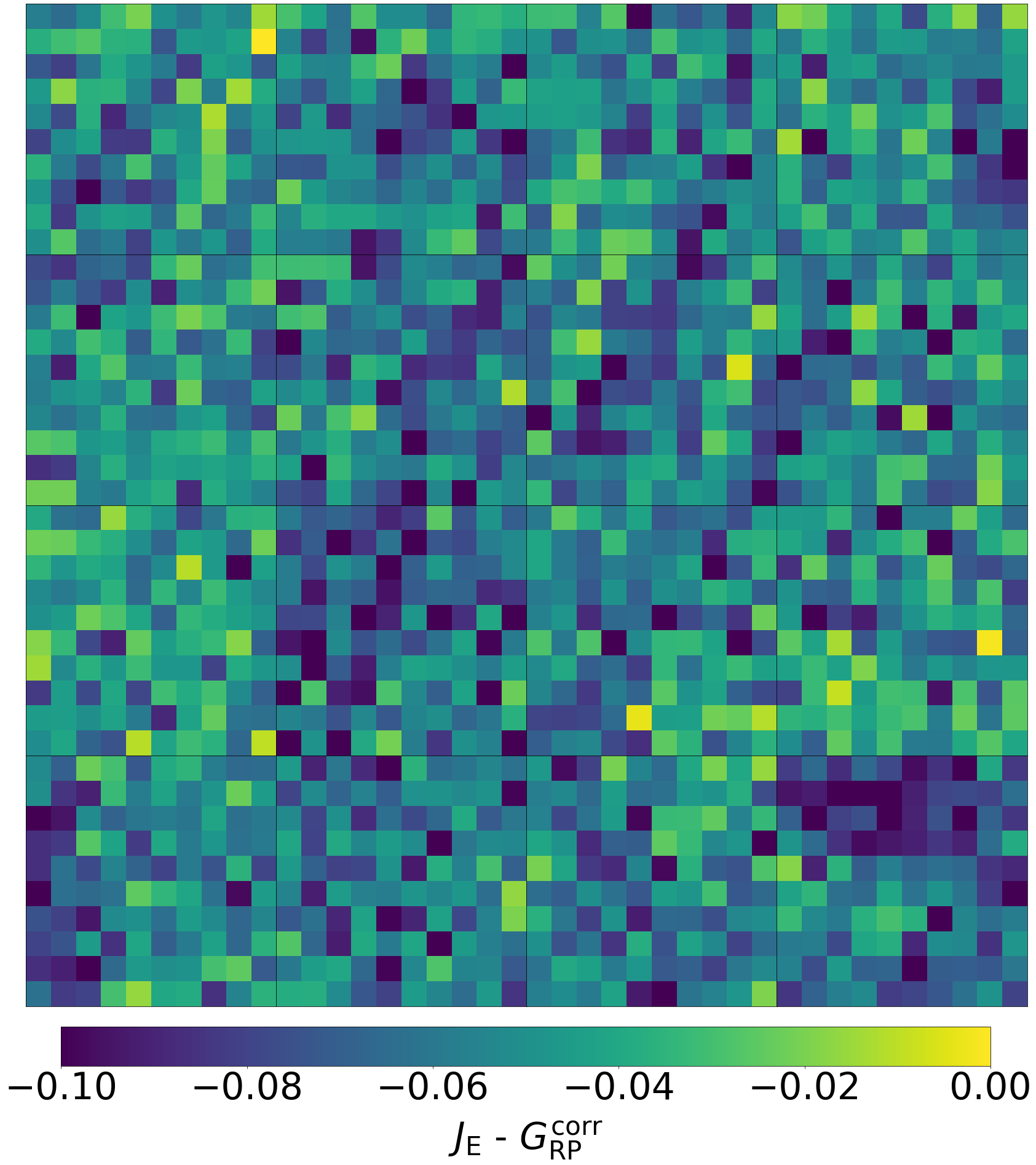}
\includegraphics[width=5.9cm]{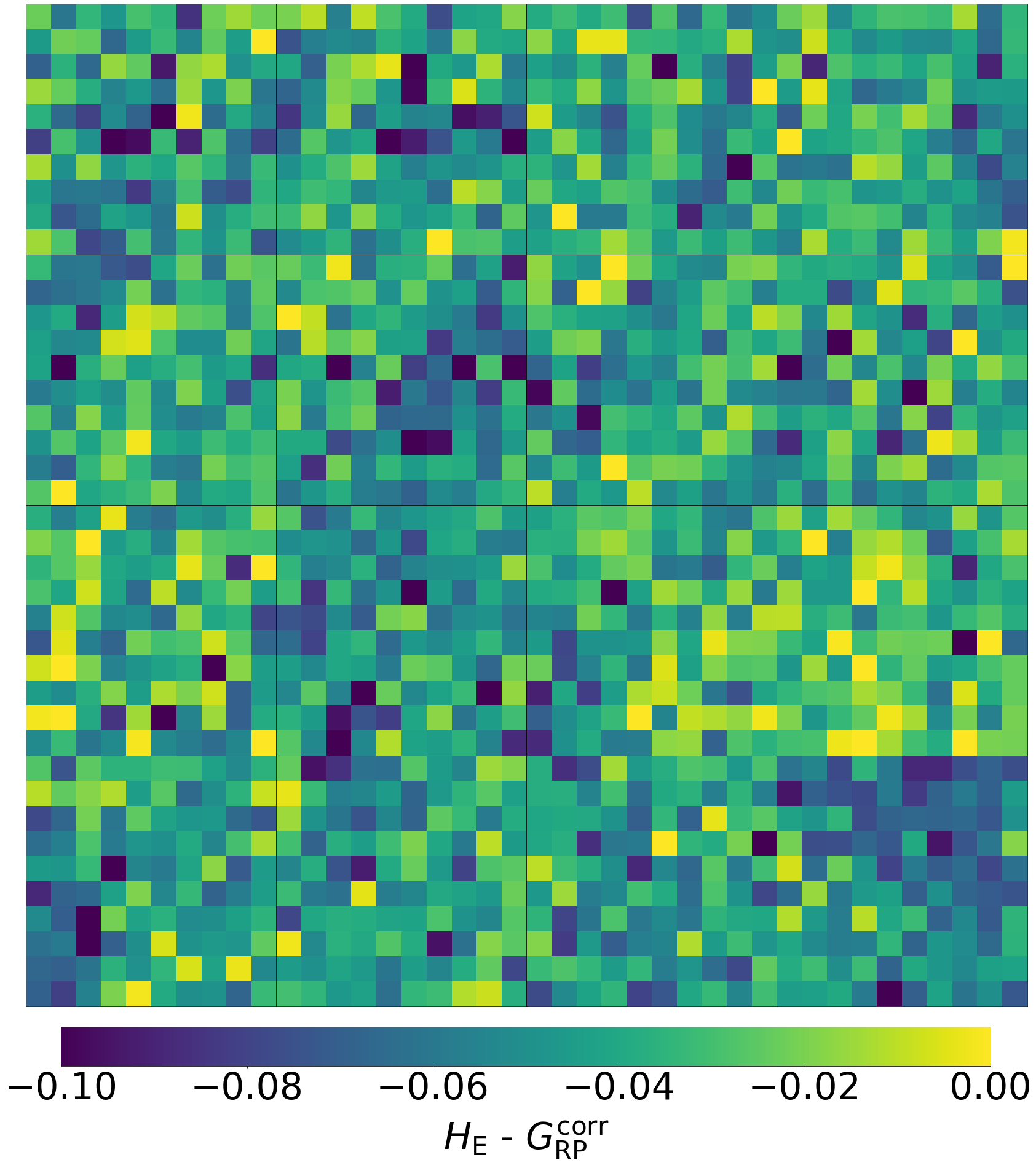}
\end{minipage}
\caption{\ac{FPA} distribution of photometric residuals for sky patch 71 between \YE, \JE, \HE, and {\it Gaia} photometry, after correcting for the colour term.} 
\label{figure:photometry_FPA}
\end{figure*}

In much the same way we analyse the positional dependence of the photometric zero points. After cross-matching with {\it Gaia} DR3, we consider the difference between \Euclid and {\it Gaia} magnitudes after correcting {\it Gaia} $G_\mathrm{RP}$ for a colour term,
\begin{equation}
G_\mathrm{RP}^\mathrm{Corr} = G_\mathrm{RP} + A + B\,(G_\mathrm{BP} - G_\mathrm{RP})\;,
\end{equation}
where $A$ and $B$ are computed for each \Euclid filter. Results for sky patch 71 are reported in \cref{figure:photometry_FPA}, showing that the photometric accuracy has no major spatial dependence. The only visible effect is an increase by about 0.02\,mag for detector DET13 in the \YE\ band. We note that the photometric residuals for the same detector for sky patch 49 are much smaller, suggesting a slow drift in the photometric calibration for this detector in \YE\ band. Indeed, each detector shows a slightly different evolution of the photometric zero point over time. In this respect, DET13 has the strongest evolution, which is not perfectly modelled by the relative calibration in the Q1 data set. Further refinements will be developed and applied for DR1 and later data releases.

\subsection{Zero-point stability from self-calibration observations}
\label{subsection:zeropoint}

A complementary way to assess calibration accuracy and photometric stability -- specifically, variations in filter throughput and zero points -- is to analyse the monthly self-calibration observations. This field is visited with a 76-point dither pattern, providing -- among many other purposes \citep{EuclidSkyOverview} -- a baseline for monitoring \Euclid’s spectrophotometric response.

The key advantage of this approach is the ability to compare fluxes across thousands of sources distributed over different detectors, within and between visits. This minimises uncertainties compared to absolute calibration based on the single \ac{WD} spectrophotometric standard.

For this comparison, we used two self-calibration visits in July and August 2024 that were close in time to the Q1 observations. We take the July visit as a reference and compute magnitude differences with respect to the August visit. Both data sets were processed with the same calibration products, so that any differences would originate from actual throughput variations.

\begin{figure}
\centering
\resizebox{\hsize}{!}{\includegraphics{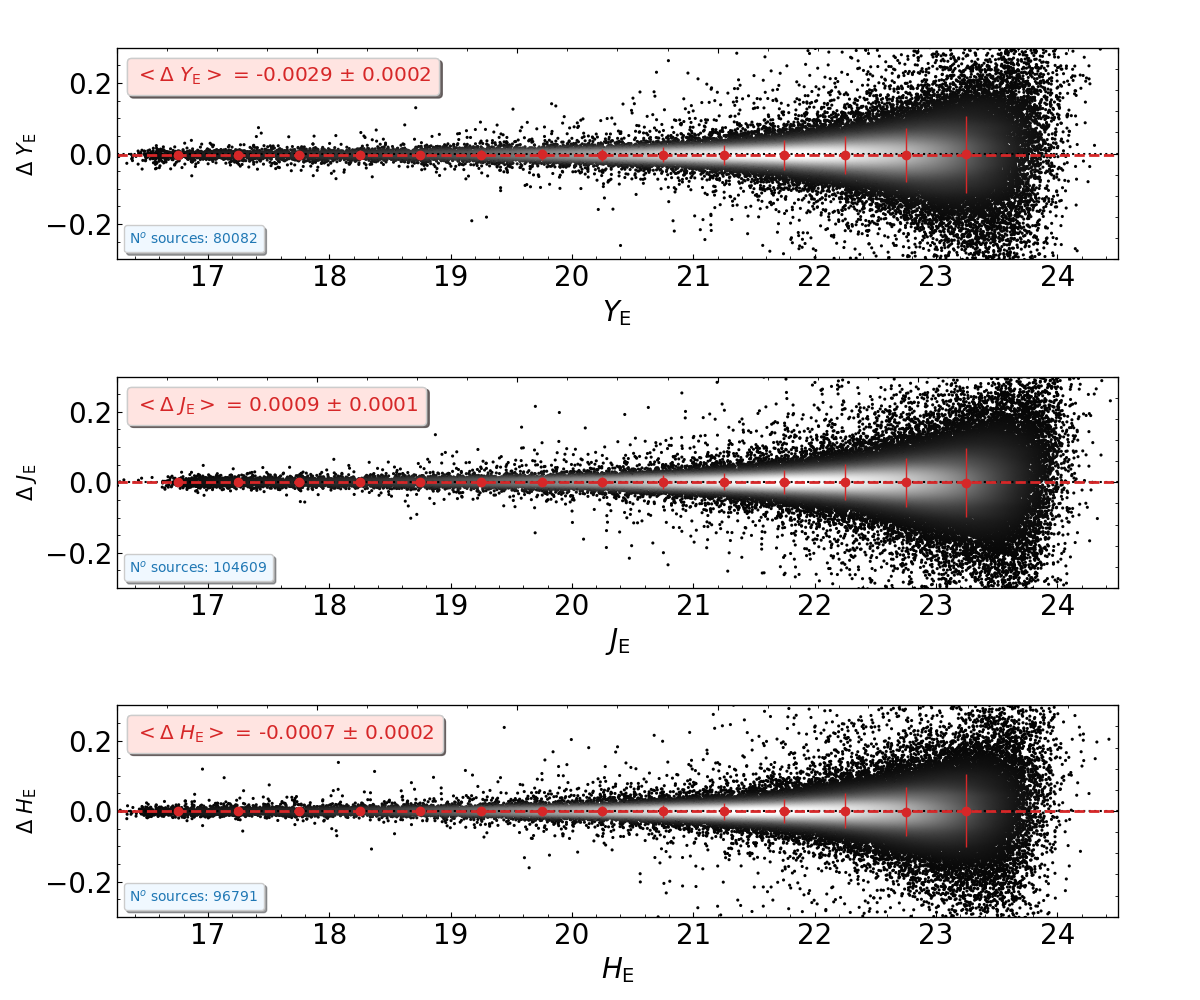}}
\caption{Photometric differences between July and August observations of the self-calibration field for the \YE, \JE, and \HE bands. The plots illustrate the photometric stability.}
\label{figure:rel_phot}
\end{figure}

Specifically, we took the stacked-frame catalogue from July and cross-matched it with the 76 catalogues from the individual exposures of that month, using a matching radius of \ang{;;0.15} or 0.5\,pixel. In this way, for each detected source in the stacked frame we have up to 76 individual measurements. This resulted in three master catalogues, one per band, with 2.0 to 2.5 million sources, each. The same procedure was repeated for the August visit. We then retained only sources in the master catalogues that had at least four flux measurements, and computed the mean flux and its error per source. The July and August master catalogues were then cross-matched with a matching radius of \ang{;;0.03} or 0.1\,pixel, and we computed the individual magnitude differences ($\Delta$mag) per source and filter. These are shown in  \cref{figure:rel_phot}.

Next, we computed the mean $\Delta$mag and its error in 0.5\,mag bins (red dots in \cref{figure:rel_phot}) to check for a flux-dependence; no such dependence was found. Mean- and median-combined results were indistinguishable. The final magnitude difference was obtained using an inverse-variance weighted mean of the individual binned means. We found $\langle\Delta{\rm mag}\rangle$ of 2.9, 0.9, and 0.7\,mmag, for \YE, \JE, and \HE, respectively. We conclude that the photometry has been stable to better than 0.3\% over this period, and that the zero-points obtained from the self-calibration observations are applicable to the Q1 sky patches without corrections.

\section{Discussion and conclusions}
\label{sec:conclusions}

This paper presents the \ac{NIR PF} pipeline used to process NISP raw photometric exposures for Q1. While the calibrated images meet the key requirements for astrometric and photometric calibration, the \ac{DQC} analysis and in-depth validation of the calibration products and images revealed certain limitations.

\begin{itemize}

\item Persistence signal. Unmasked residual streaks remain due to several factors, including the balance between masking unwanted features while preserving valid pixels and the accuracy of the model used to capture the complexities of detector physics behind persistence charges. Additionally, while masking can mitigate contamination from spectroscopic exposures on photometric ones, it cannot address persistence between consecutive photometric exposures taken at the same dither position. Persistence also impacts calibration products, such as small-scale and large-scale flats. An improved model is needed to enable an effective correction procedure, and is being worked on.

\item \ac{PSF}. An improved \ac{PSF} model has been derived from observations of the self-calibration field, extending to larger radii and therefore providing a better characterisation of the extended \ac{PSF} wings. The smaller \ac{PSF} used for the Q1 processing reflects mainly on the quality of the \ac{PSF} photometry in the Q1 catalogues, which is only available for bright sources.

\item Background estimation. We noticed a positive correlation between the estimated background signal and the position of very bright objects, resulting in an over-subtraction in the outskirts of extended objects. This limits the accuracy of low-surface-brightness measurements on such targets, and also their detection.

\item Some of the calibration products used for Q1 are based on ground calibration, and will be superseded by in-flight data in upcoming data releases.

\item Unmodelled instrumental effects that are still to be fully treated. The DQC pipeline is based on metadata and statistics, and cannot fully capture the complexity of NIR images. Indeed a visual inspection performed on a subsample of the Q1 images revealed the presence of additional instrumental effects, such as diffuse arcs produced by internal reflections, affecting a very limited number of exposures. Due to the very low incidence of such artifacts, the development of a reliable model for these effects will take place on a timescale beyond Q1.
\end{itemize}

These issues are addressed in a new release of \ac{NIR PF} for the preparation of DR1. This is scheduled for the end of 2025 within the Euclid Consortium, followed by a public release one year later.

\begin{acknowledgements}
\AckQone\\

\AckEC\\
This work has made use of data from the European Space Agency (ESA) mission
{\it Gaia} (\url{https://www.cosmos.esa.int/gaia}), processed by the {\it Gaia}
Data Processing and Analysis Consortium (DPAC,
\url{https://www.cosmos.esa.int/web/gaia/dpac/consortium}). Funding for the DPAC
has been provided by national institutions, in particular the institutions
participating in the {\it Gaia} Multilateral Agreement.
\end{acknowledgements}

\bibliography{main,Euclid,Q1}

\begin{appendix}
  
\section{Constructing filter flats from LED flats} \label{app:filter_flats}

In general, a flat field at a 2-dimensional pixel position $\mathbf{x}$ and time $t$ can be written as
\begin{equation}
  {\rm FLAT_{Filter}}(\mathbf{x},t) = \int_{\lambda_1}^{\lambda_2} \mathrm{QE}(\lambda,\mathbf{x},t)\;T(\lambda,\mathbf{x},t)\;f_\lambda(\lambda,\mathbf{x},t)\;{\rm d}\lambda\;,
\end{equation}
where QE is the quantum efficiency, $T$ is the filter (or dispersive element) transmission, and $f_\lambda$ is the spectral flux density of a flat-field lamp expressed per unit wavelength. While in theory the integration is over all wavelengths, in practice it can be bounded by some interval $[\lambda_1,\lambda_2]$ that contains the bulk of the flux.

In the case of NISP, the QE varies little over the spectral range of interest, 0.9--2.2\,\micron. The ideal lamp spectrum does not favour a particular wavelength, i.e.\ it produces the same number of photons per wavelength interval. In practice, as long as the S/N is sufficient at each wavelength while avoiding saturation, the exact shape of the lamp spectrum does not matter.

For NISP photometry, a lamp emitting the same
number of photons per wavelength interval is desirable. The spectral flux density of the lamp is then
\begin{equation}
  f_\lambda(\lambda,\mathbf{x},t) = f_\lambda^0(\mathbf{x},t)
\end{equation}
with units photons s$^{-1}$\,cm$^{-2}$\,nm$^{-1}$. We chose these units
because the provided spectral flux densities for the NISP \acp{LED} have these units.
The flat field itself then is
\begin{equation}
  {\rm FLAT_{Filter}}(\mathbf{x},t) = f_\lambda^0(\mathbf{x},t)\;\int_{\lambda_1}^{\lambda_2} \mathrm{QE}(\lambda,\mathbf{x},t)\;T(\lambda,\mathbf{x},t)\;{\rm d}\lambda\;.
\label{flatfilter}
\end{equation}

The \acp{LED} in NISP illuminate the focal plane array directly without passing through the filters or grisms. In the absence of these optical elements, the \ac{LED} flat field can be written as
\begin{equation}
  {\rm FLAT_{\rm LED}}(\mathbf{x},t) = \int_{\lambda_3}^{\lambda_4} \mathrm{QE}(\lambda,\mathbf{x},t)\;f^{\rm LED}_\lambda(\lambda,\mathbf{x},t)\;{\rm d}\lambda\;,
\label{flatled1}
\end{equation}
where ${\rm FLAT_{\rm LED}}(\mathbf{x},t)$ is the observed \ac{LED} flat, and $f^{\rm LED}_\lambda(\lambda,\mathbf{x},t)$ is the \ac{LED} spectrum measured from the laboratory. The Cauchy mean value theorem states that for two continuous functions $f(x)$ and
$g(x)$, with $g(x)\geq0$ or $g(x)\leq0$, one can always find a $x_0$ in the interval
$[a,b]$ such that
\begin{equation}
  \int_{a}^{b} g(x)\;f(x)\;{\rm d}x = f(x_0)\;\int_a^b g(x)\;{\rm d}x\;.
\end{equation}
This can be readily applied to \cref{flatled1} because QE is always positive
and $f^{\rm LED}_\lambda$ is continuous (at least within the resolution limits of the
measurements):
\begin{equation}
  {\rm FLAT_{\rm LED}}(\mathbf{x},t) = f^{\rm LED}_\lambda(\lambda_0,\mathbf{x},t)\;\int_{\lambda_3}^{\lambda_4} \mathrm{QE}(\lambda,\mathbf{x},t)\;{\rm d}\lambda\;.
\label{flatled2}
\end{equation}
We can now solve \cref{flatled2} for $f^{\rm LED}_\lambda(\lambda_0,\mathbf{x},t)$
and use it to replace $f_\lambda^0(\mathbf{x},t)$ in \cref{flatfilter}. This
can be done because ${\rm FLAT_{Filter}}$ (and ${\rm FLAT_{LED}}$) are in the end
subject to a global normalisation (chosen lamp brightness, normalisation to average
unity, or else). 
We then have
\begin{equation}
  {\rm FLAT_{Filter}}(\mathbf{x},t) = {\rm FLAT_{\rm LED}}(\mathbf{x},t)\; P(\mathbf{x},t)\;,
\label{flatfilter2}
\end{equation}
where
\begin{equation}
  P(\mathbf{x},t) = \frac{\int_{\lambda_1}^{\lambda_2} \mathrm{QE}(\lambda,\mathbf{x},t)\;T(\lambda,\mathbf{x},t)\;{\rm d}\lambda}
  {\int_{\lambda_3}^{\lambda_4} \mathrm{QE}(\lambda,\mathbf{x},t)\;{\rm d}\lambda}
  \label{propagator}
\end{equation}
propagates the \ac{LED} flat in wavelength to the bandpass $T$ of the filter in
question. This propagation depends on pixel position $\mathbf{x}$,
reflecting the spatially variably quantum efficiency as well as potential
bandpass variations. 

\section{Description of NIR PF data quality flags}

We report for convenience the bit mask convention adopted for the NIR images from the \citet{EuclidDpdd}\footnote{\url{https://st-dm.pages.euclid-sgs.uk/data-product-doc/dmq1/nirdpd/dpcards/nir_calibratedframe.html}}.
\begin{table}[h!]
\begin{tabular}{c l p{4cm} c}
\hline\hline      
\noalign{\vskip 1pt}

\textbf{Bit} & \textbf{Flag Name} & \textbf{Description} & \textbf{Invalid} \\ \hline \noalign{\vskip 1pt}
0 & INVALID & Convenience common flag & \\ \hline \noalign{\vskip 1pt}
1 & OBMASK & On-board flags & \\ \hline \noalign{\vskip 1pt}
2 & DISCONNECTED & Disconnected & Yes \\ \hline \noalign{\vskip 1pt}
3 & ZEROQE & Zero QE & Yes \\ \hline \noalign{\vskip 1pt}
4 & BADBASE & High or Low Baseline & Yes \\ \hline \noalign{\vskip 1pt}
5 & LOWQE & QE < 74\% at 1120 $\le$ $\lambda$/nm $\le$ 2020 \newline QE < (64 + ($\lambda$/nm -- 920)/20)\% at 920 $\le$ $\lambda$/nm $\le$ 1120. & \\ \hline \noalign{\vskip 1pt}
6 & SUPERQE & Pixel QE > 110\% & Yes \\ \hline \noalign{\vskip 1pt}
7 & HOT & Pixels with dark current signal falling 3\,$\sigma$ above the detector median & \\ \hline \noalign{\vskip 1pt}
8 & RTN & Random Telegraph Noise & Yes \\ \hline \noalign{\vskip 1pt}
9 & SNOWBALL & Very energetic internal deposit of signal in pixels & Yes \\ \hline \noalign{\vskip 1pt}
10 & SATUR & Saturated Pixel & Yes \\ \hline \noalign{\vskip 1pt}
11 & NLINEAR & Pixels whose signal in electrons are below or above the applicable signal limits & \\ \hline \noalign{\vskip 1pt}
12 & NLMODFAIL & Pixels whose linear correction model failed & Yes \\ \hline \noalign{\vskip 1pt}
13 & PERSIST & Pixels affected by persistence charge from previous sources & Yes \\ \hline \noalign{\vskip 1pt}
14 & PERMODFAIL & Pixels with persistence calibration procedure failed & \\ \hline \noalign{\vskip 1pt}
15 & DARKNODET & Pixels for which the dark current is not detected to within a maximum noise threshold & \\ \hline \noalign{\vskip 1pt}
16 & COSMIC & Cosmic ray hits & Yes \\ \hline \noalign{\vskip 1pt}
17 & FLATLH & Pixels in the computed flat that have too low or too high response values & \\ \hline \noalign{\vskip 1pt}
18 & GHOST & Ghosts & Yes \\ \hline \noalign{\vskip 1pt}
19 & SCATTER & Scattered Light & \\ \hline \noalign{\vskip 1pt}
20 & MOVING & Moving objects & \\ \hline \noalign{\vskip 1pt}
21 & TRANS & Transients & \\ \hline \noalign{\vskip 1pt}
22 & CROSSTALK & Cross Talk & \\ \hline \noalign{\vskip 1pt}
23 & FLOWER & `Flower pixel' found on MasterFlat & Yes \\ \hline \noalign{\vskip 1pt}
24 & VIGNET & Pixels affected by vignetting on \ac{LED} exposures & \\ \hline
\end{tabular}
\label{table:flag_descriptions}
\end{table}

\end{appendix}
\end{document}